\documentclass[12pt,a4paper]{article}
\usepackage[T1]{fontenc}
\usepackage{fancyhdr}
\usepackage{amsmath,amsthm,amsfonts,amssymb}
\usepackage{epic,eepic}
\usepackage{graphicx}
\usepackage{authblk}
\usepackage{hyperref}
\usepackage{fullpage}
\usepackage[active]{srcltx}
\usepackage{enumitem}
\usepackage{booktabs}

\usepackage{float}%
\floatstyle{plaintop}%
\restylefloat{table}

\usepackage[font={small,it}]{caption}
\usepackage[toc,page]{appendix}
\graphicspath{{Figures/}}

\usepackage[style=chicago-authordate,backend=bibtex,maxcitenames=2,maxbibnames=99]{biblatex}
\bibliography{biblio}

\makeatletter
\AtEveryBibitem{%
  \global\undef\bbx@lasthash%
  \clearfield{}}
\makeatother

\theoremstyle{plain}
\newtheorem{theorem}[]{Theorem}
\newtheorem{proposition}[]{Proposition}
\newtheorem{lemma}[]{Lemma}
\newtheorem{corollary}[]{Corollary}
\theoremstyle{definition}
\newtheorem{definition}{Definition}[subsection]
\theoremstyle{remark}
\newtheorem*{remark}{Remark}

\numberwithin{equation}{section}
\numberwithin{figure}{section}

\date{\today}
\title{\textbf{Properties of Latent Variable Network Models}}
\author[1,2,*]{Riccardo Rastelli}
\author[1,2]{Nial Friel}
\author[3]{Adrian E. Raftery}
\affil[*]{\footnotesize \href{mailto:riccardo.rastelli@ucdconnect.ie}{riccardo.rastelli@ucdconnect.ie}}
\affil[1]{\footnotesize School of Mathematical Sciences, University College Dublin, Ireland;}
\affil[2]{\footnotesize Insight: Centre for Data Analytics, University College Dublin, Ireland;}
\affil[3]{\footnotesize Department of Statistics, University of Washington, Seattle, USA.}

\begin{document}

\maketitle
\begin{abstract}
We derive properties of Latent Variable Models for 
networks, a broad class of models that includes the widely-used
Latent Position Models. These include the average degree distribution, 
clustering coefficient, average path length and degree correlations.
We introduce the Gaussian Latent Position Model, and derive
analytic expressions and asymptotic approximations for its
network properties.
We pay particular attention to one special case, 
the Gaussian Latent Position Models
with Random Effects, and show that it can represent the heavy-tailed
degree distributions, positive asymptotic clustering coefficients and
small-world behaviours that are often observed in social networks.
Several real and simulated examples illustrate the ability of the models 
to capture important features of observed networks.\\

\end{abstract}

\baselineskip=18pt
\section{Introduction}
Networks are tools for representing relations between entities.
Examples include social networks, such as acquaintance networks \parencite{amaral2000classes}, collaboration networks \parencite{newman2001structure} and interaction networks \parencite{perry2013point},
technological networks such as the World Wide Web \parencite{albert1999internet}, and biological networks such as neural networks \parencite{watts1998collective}, food webs \parencite{williams2000simple}, 
and protein-protein interaction networks \parencite{raftery2012fast}.

Social networks, specifically, tend to exhibit
transitivity \parencite{newman2003properties}, 
clustering, homophily \parencite{newman2003social}, 
the scale-free property \parencite{newman2002random} 
and small-world behaviours \parencite{watts1998collective}.

Networks are typically modelled in terms of random graphs. 
The set of nodes is fixed, and a probability distribution 
is defined over the space of all possible sets of edges, thereby
considering the observed network as a realisation of a random variable. 

One way to study networks is to define a simple generative mechanism that captures some important basic properties, such as the degree distribution \parencite{newman2001random},
clustering \parencite{newman2009random}, or small-world behaviour \parencite{watts1998collective}.
These models are deliberately made simple so to be easily fitted and studied. 
Theoretical tractability can allow the asymptotic properties 
of the fitted models to be assessed, and this can give help to determine
how well the models might fit real large networks.
It can also allow the relationships between statistics measuring
clustering, power-law behaviour and small-world behaviour to be assessed
\parencite{kiss2008comment,newman2009random,watts1998collective}. 

On the other hand, various statistical models have been proposed, 
including Exponential Random Graph Models \parencite{frank1986markov,caimo2011bayesian,krivitsky2014separable}, 
Latent Stochastic Blockmodels \parencite{nowicki2001estimation,latouche2011overlapping,airoldi2009mixed}, and Latent Position Models \parencite{hoff2002latent,raftery2012fast}.
These try to capture all the main features of observed networks within a unified framework.
However, due to their more complicated structure, only limited research has been
carried out to assess their properties \parencite{daudin2008mixture,channarond2012classification,ambroise2012new,mariadassou2015convergence}.
Moreover, recent developments \parencite{chatterjee2013estimating,shalizi2013consistency,schweinberger2015local} 
have shed light on some important limitations of ERGMs, questioning their suitability as statistical models for networks.

In this paper, we attempt to fill this gap by deriving theoretical
properties of a wide family of network models, which we call 
Latent Variable Models (LVMs). This family includes 
one well-known class of statistical network models as a special case, namely
the Latent Position Models (LPM)  \parencite{hoff2002latent,handcock2007model,krivitsky2009representing}.
These are defined by associating an observed latent position in Euclidean
space with each node, and postulating that nodes that are closer are more
likely to be linked, with the probability of connection depending on the 
distance, typically through a logistic regression model. 
In the last decade, LPMs and their extensions have been widely used for 
applications such as the analysis of international investment 
\parencite{CaoWard2014},
trophic food webs \parencite{ChiuWestveld2011,ChiuWestveld2014},
signal processing \parencite{Wang&2014},
and education research \parencite{Sweet&2013}.

Analytic expressions for the clustering properties of this model 
in its original form are hard to derive.
Because of this, we propose a new but closely related model,
the Gaussian Latent Position Model. This yields simple analytic
expressions or asymptotic approximations 
for several important clustering properties,
including a complete characterisation of the degree distribution,
the clustering coefficient, and the distribution of path lengths.
The availability of analytic expressions facilitates the analysis of very large
graphs since, for example, simulation is not required.

One result is that the Gaussian LPM can represent transitivity asymptotically,
because its clustering coefficient can be asymptotically non-zero,
unlike the Erd\H{o}s-R\'{e}nyi and Exponential Random Graph Models,
whose clustering coefficient converges to zero.

One implication of our results is that the Latent Position Model
in its original form cannot represent heavy-tailed degree distributions,
such as power-law behaviour, or small-world behaviour, as measured
by the average path length. As a result, we introduce the Gaussian
Latent Position Model with Random Effects (LPMRE), and show that it 
can overcome these limitations and capture important features of large-size real networks.
These results suggest that the Gaussian LPMRE may be a good model for social networks.


The paper is organised as follows. In Section \ref{sec:Gaussian Latent Position Models} the notation is set and the main models of interest are defined. 
Section \ref{sec:Results} gives the core theoretical results used in the paper. 
Section \ref{ref:RealisedNetworks} makes use of such results to further analyse important features of LPMs, such as transitivity, homophily, scale-free properties and small-world behaviours.
In Section \ref{sec:LPMRE}, the appealing properties of Gaussian LPMREs are illustrated through empirical studies and examples.
Section \ref{sec:Real data examples} provides several real data studies, 
while Section \ref{sec:Conclusions} concludes the paper with some final remarks.

\section{Latent Variable Network Models}\label{sec:Gaussian Latent Position Models}
\subsection{Notation and model assumptions}
Here we introduce our notation and define the various latent variable models for networks that we consider.

\paragraph{A1.} 
$\mathcal{G}=\left( V,E \right)$ is a binary random graph where $V$ is the set of node labels and $E$ is the set of random edges. 
The observed data consist of a realisation of $\mathcal{G}$. We denote $V=\left\{ 1,\dots,n\right\}$ and represent the observed edges through the adjacency matrix $\textbf{Y}=\left\{ y_{ij}\right\}_{(i,j)\in V\times V}$, where:
\begin{equation}
 y_{ij}=\begin{cases}
         1, &\mbox{ if an edge from $i$ to $j$ appears in the graph, }\\
         0, &\mbox{ otherwise. }\\
        \end{cases}
\end{equation}
Furthermore we assume that edges are undirected and self-edges are not allowed, i.e. 
$y_{ij}=y_{ji},\ \forall (i,j)\in \tilde{V}:=\lbrace (i,j): 1 \leq i < j \leq n \rbrace$ and $y_{ii}=0,\ \forall i\in V$, respectively. Our analysis can easily be extended to the case of directed edges, however.\\

A Latent Variable Model (LVM) for networks is defined by associating an unobserved random variable $\textbf{Z}_i\in \mathcal{Z}$ to actor $i$, $\forall i\in V$, for some discrete or continuous set $\mathcal{Z}$.
The set of quantities $P=\{z_1,\dots,z_n\}$ denotes a realisation of the corresponding random process.

\paragraph{A2.} The latent variables $\textbf{Z}_1,\dots,\textbf{Z}_n$ are independent and identically distributed, where each $\textbf{Z}$ is distributed according to the probability measure $p(\ \cdot\ )$.

\paragraph{A3.} Edges are assumed to be conditionally independent given the latent variables. Thus $\forall (i,j)\in\tilde{V}$, $Y_{ij}$ is a Bernoulli random variable such that
\begin{equation}
\begin{split}
 Pr\left(Y_{ij}=1\middle\vert \textbf{z}_i, \textbf{z}_j\right)&=1-Pr\left(Y_{ij}=0\middle\vert \textbf{z}_i, \textbf{z}_j\right)=r\left( \textbf{z}_i, \textbf{z}_j \right).
\end{split}
\end{equation}

The modelling assumptions \textbf{A1}-\textbf{A3} are very general, and in fact various models of interest satisfy these, including the Random Connection Models of \textcite{meester1996continuum},
the Fitness models of \textcite{caldarelli2002scale,soderberg2002general}, the LPMs of 
\textcite{hoff2002latent,handcock2007model,krivitsky2009representing},
and the Stochastic Blockmodel of \textcite{nowicki2001estimation}, 
among others.  We now give more specific modelling assumptions 
that characterise Latent Position Models.

\paragraph{A4.} In the LPM, the realised latent variables $\textbf{Z}_i$ in \textbf{A2} are points in the Euclidean space $\mathbb{R}^d$, for a fixed $d$, and they are normally distributed:
\begin{equation}\label{P}
 p\left(P\middle\vert\gamma\right)= \prod_{i=1}^{n}f_d\left(\textbf{z}_i;\textbf{0},\gamma\right)=\prod_{i=1}^{n}(2\pi\gamma)^{-\frac{d}{2}}\exp\left\{-\frac{1}{2\gamma}\textbf{z}_i^t\textbf{z}_i\right\} .
\end{equation}
In (\ref{P}), $\gamma$ is a positive real parameter and $f_d\left(\ \cdot\ ;\boldsymbol{\mu},\gamma\right)$ is the multivariate Gaussian density function with parameters $\boldsymbol{\mu}$ (mean) and 
$\gamma \mathbb{I}_d$ (covariance), where $\mathbb{I}_d$ is the $d\times d$ identity matrix and $\textbf{A}^t$ denotes the transpose of the matrix or vector $\textbf{A}$.\\

\paragraph{A5.} In our specification of the LPM, the Gaussian LPM, 
the Bernoulli parameters in \textbf{A3} are given by:
\begin{equation}\label{gaussian1}
\begin{split}
 r\left( \textbf{z}_i, \textbf{z}_j \right)
 &=\tau\exp{\left\{-\frac{\left( \textbf{z}_i-\textbf{z}_j \right)^t\left( \textbf{z}_i-\textbf{z}_j \right)}{2\varphi}\right\}},
\end{split}
\end{equation}
where $\varphi>0$, $\tau\in[0,1]$.\\

Assumption \textbf{A5} is slightly different from the original formulation of the LPM of \textcite{hoff2002latent}, in that the logistic connection function for the edges 
has been replaced by a non-normalised Gaussian density. The reasoning behind this choice will be addressed in Section \ref{sec:MotivationGaussian}.
\paragraph{A6.}
In the Logistic LPM of \textcite{hoff2002latent}, the Bernoulli parameters 
in \textbf{A3} are given by:
\begin{equation}\label{logistic1}
\begin{split}
 r\left( \textbf{z}_i, \textbf{z}_j \right)=
\frac{\exp\left\{\alpha-\beta d\left( \textbf{z}_i,\textbf{z}_j \right)\right\}}{1+\exp\left\{\alpha-\beta d\left( \textbf{z}_i,\textbf{z}_j \right)\right\}} ,
\end{split}
\end{equation}
where $\alpha\in\mathbb{R},\ \beta>0$ and $d\left( \textbf{z}_i,\textbf{z}_j \right)$ is the Euclidean distance between the latent positions $\textbf{z}_i$ and $\textbf{z}_j$.

\subsubsection{Extensions of Latent Position Models}
Two major extensions of the LPMs of \textcite{hoff2002latent} are \textcite{handcock2007model} and \textcite{krivitsky2009representing}. 
In the former, clustering is introduced through a mixture distribution on the latent process for nodal positions, while in the latter, nodal random effects are introduced to 
capture degree heterogeneity. 
In a similar fashion we introduce two variations of \textbf{A4} and \textbf{A5} to characterise the two cases:
\paragraph{A7.}
The latent positions are distributed according to a finite mixture of Gaussian distributions, i.e.:
\begin{equation}
 p\left(P\middle\vert \boldsymbol{\pi},\boldsymbol{\mu},\boldsymbol{\gamma},G\right) = \prod_{i=1}^{n} \left[\sum_{g=1}^{G}\pi_gf_d\left(\textbf{z}_i;\boldsymbol{\mu}_i,\gamma_i\right)\right]
\end{equation}
where $\boldsymbol{\pi}$ are the mixture weights, $\boldsymbol{\mu}$ and $\boldsymbol{\gamma}$ are the parameters for the components and $G$ is the number of groups. 
The components are all assumed to arise from densities with circular contours, 
but possibly different volumes.\\

\paragraph{A8.}
For every node $s\in V$, the latent information $\tilde{\textbf{z}}_s$ is composed of the realisation of a random latent position $\textbf{Z}_s$, 
which is distributed according to $p\left(\ \cdot\ \right)$, 
and a random effect $\varphi_s$. This random effect is independent of $\textbf{Z}_s$ and is distributed according to an Inverse Gamma distribution with parameters $\beta_0$ and $\beta_1$.
Also, the connection probability is modified as follows:
\begin{equation}\label{randefflikelihood}
 Pr\left(Y_{ij}=1\middle\vert \textbf{z}_i, \textbf{z}_j, \varphi_i, \varphi_j, \tau\right)=\tau\exp{\left\{-\frac{1}{2\left(\varphi_i+\varphi_j\right)^2}(\textbf{z}_i-\textbf{z}_j)^t(\textbf{z}_i-\textbf{z}_j)\right\}}.
\end{equation}
We call this the Gaussian Latent Position Model with Random
Effects, or Gaussian LPMRE.

Different combinations of assumptions \textbf{A1}-\textbf{A8} generate different Latent Variable Models. The main cases considered in the present paper are summarised in Table \ref{table:models}.

\begin{table}[htb]
\footnotesize
\centering
\begin{tabular}{llcr} \toprule  
Notation & Description & Assumptions & Model parameters\\ 
\midrule
LVM& Latent Variable Model & \textbf{A1}-\textbf{A3} & unspecified\\ 
Logistic LPM & LPM of \textcite{hoff2002latent} & \textbf{A1}-\textbf{A4}, \textbf{A6} & $\alpha,\beta,\gamma$\\ 
Gaussian LPM & Gaussian connection LPM & \textbf{A1}-\textbf{A5} & $\tau,\varphi,\gamma$\\ 
Gaussian LPCM & Clustering LPM & \textbf{A1}-\textbf{A3}, \textbf{A5}, \textbf{A7} & $\tau,\varphi,\boldsymbol{\pi},\boldsymbol{\mu},\boldsymbol{\gamma},G$\\ 
Gaussian LPMRE & $1$-cluster with random effects & \textbf{A1}-\textbf{A4}, \textbf{A8} & $\tau,\boldsymbol{\varphi},\gamma,\beta_0,\beta_1$\\ 
\bottomrule \end{tabular}
\caption{Latent Variable Models considered in the present paper. Latent variables are omitted from model parameters.}
\label{table:models}
\end{table}

\subsection{Motivation for the Gaussian likelihood assumption}\label{sec:MotivationGaussian}
The Logistic LPM has been widely used in network models. Assumption \textbf{A5} introduces a new function to define the probability of edges, which is proportional 
to a non-normalised Gaussian density.
Other variations in the form of the likelihood function have been proposed 
in the statistical community \parencite{gollini2014joint}, 
but the reasoning behind the Gaussian function mainly comes from the physics literature \parencite{deprez2013poisson,penrose1991continuum,meester1996continuum}.
The main advantage of using the Gaussian function in place of the Logistic function is that it makes it easier to derive theoretical properties
without much changing the generative process of the networks.

In the Gaussian function the model parameters $\tau$ and $\varphi$ appear. The role of $\tau$ is to control the sparsity in the network, and to allow for the fact that nodes having the same latent position might not be connected. 

The parameter $\varphi$ encompasses the core idea of the LPM, 
relating the probability of edges to the distance between latent positions.
Indeed, the larger the parameter $\varphi$ the more long range edges are supported. Moreover, as $\varphi$ goes to
infinity, the model degenerates to an Erd\H{o}s-R\'enyi random graph with connection probability $\tau$.

\begin{figure}[htb]
\centering
\includegraphics[width=0.49\textwidth]{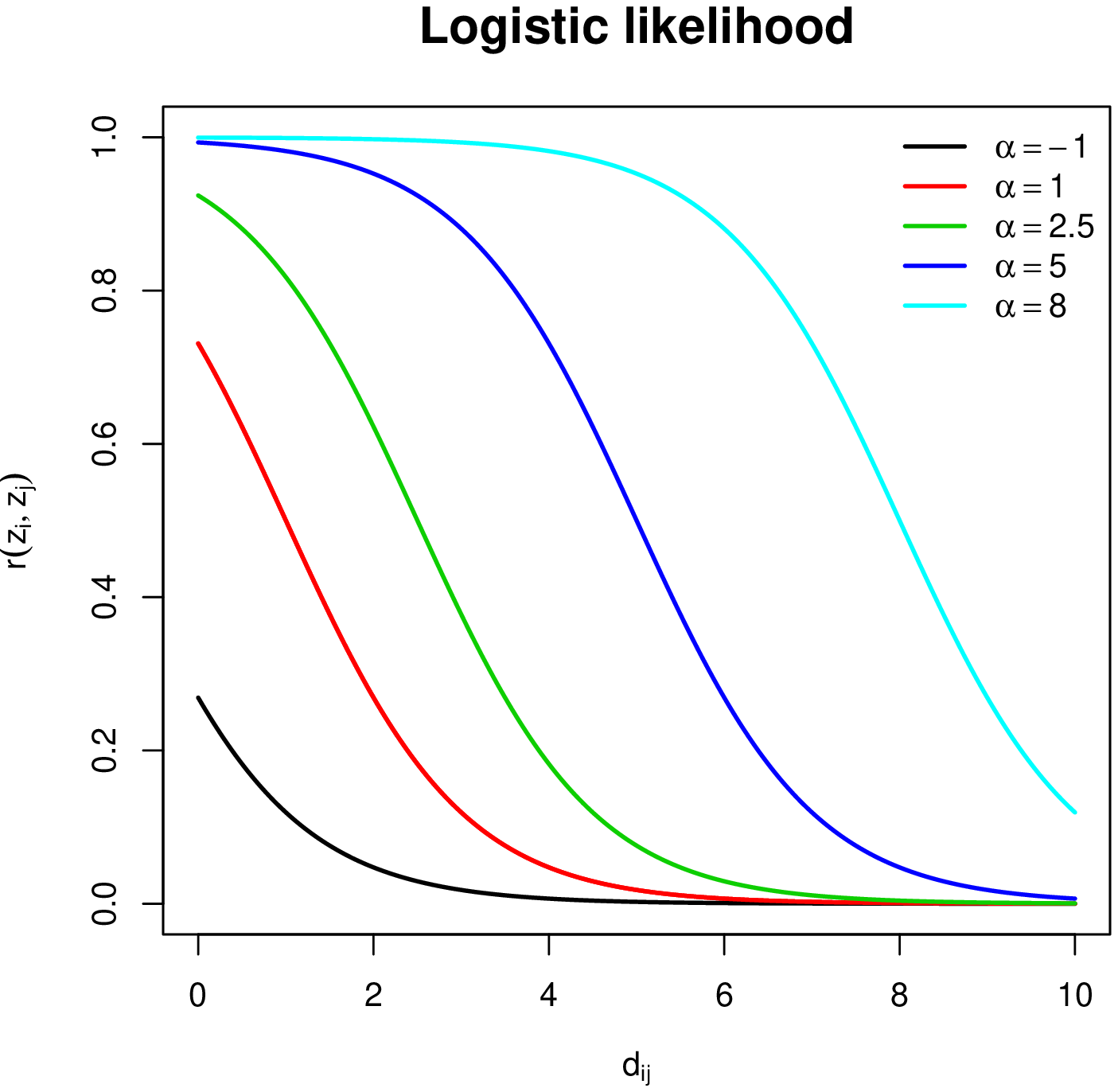}
\includegraphics[width=0.49\textwidth]{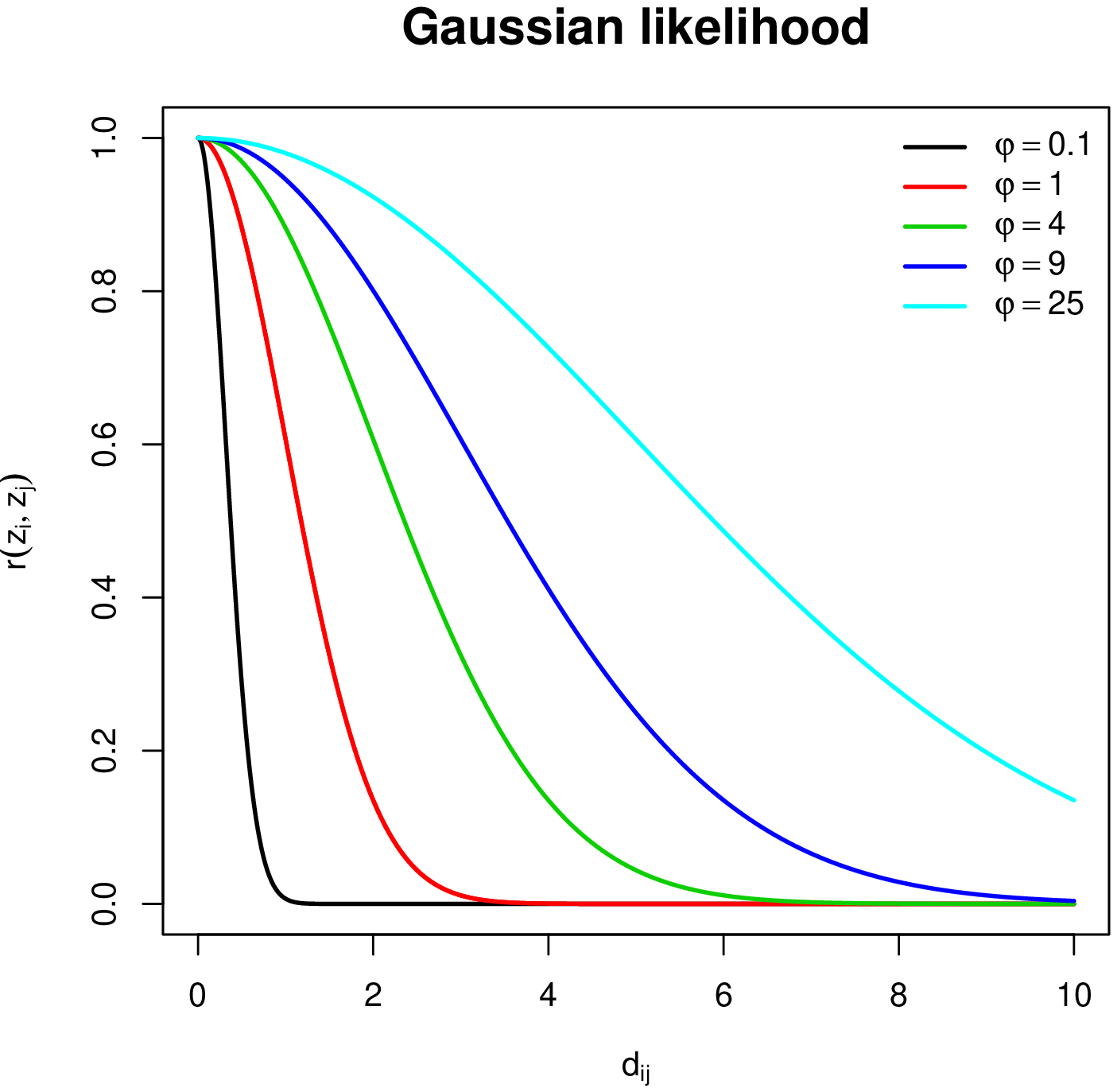}
\caption{Comparison between the Logistic and Gaussian connection functions, with $\tau=\gamma=1$. As a function of the distance between the nodes, the likelihood of a connection in both cases reaches
its maximum when the distance is null, and decreases to zero as the distance increases.}
\label{fig:2}
\end{figure}

Essentially, the difference between the two assumptions reduces to the fact that, as a function of the distance between nodes, the slopes of the curves are different 
(Figure \ref{fig:2}).
Even though an equivalence result is not provable, we argue that the properties of the Gaussian LPM are comparable and analogous to those of the Logistic LPM. 

\section{Theoretical results}\label{sec:Results}
In this section, we provide several theoretical results about LVMs, describing the distributions of features of networks realised from such models.
\subsection{Properties of the degrees}
The degree of an arbitrary actor $s$ is a discrete random variable defined by $D_s=\sum_{j\in V} Y_{sj}$. 
In this subsection, the properties of the degrees are characterised, describing their mixing behaviour and the distribution of the degree of a randomly 
chosen node, identified by the vector $\textbf{p}=\left( p_0,\dots,p_{n-1} \right)$, where $p_k=Pr\left( D=k \right),\ \forall k=0,\dots,n-1$.
To study the degree distribution of general LVMs (including LPMs), 
we propose a framework resembling that of \textcite{newman2001random}, 
which relies on the use of Probability Generating Functions (PGFs).

The study will focus on the following quantities:
\begin{itemize}
 \item \textbf{D1:} $\theta\left( \textbf{z}_s \right)$, defined as the probability that an actor chosen at random is a neighbour of a node with latent information $\textbf{z}_s$.
 \item \textbf{D2:} The PGF of the degree of a randomly chosen actor, $G(x)=\sum_{k=0}^{n-1}x^kp_k$.
 \item \textbf{D3:} The factorial moments of the degree of a randomly chosen actor. Note that central and non-central moments can be recovered iteratively from factorial moments.
 \item \textbf{D4:} The first factorial moment, i.e. the average degree of a random node: $\bar{k}$.
 \item \textbf{D5:} The values of $p_k$, for every $k=0,\dots,n-1$.
 \item \textbf{D6:} $\bar{k}(\textbf{z}_s)$, defined as the average degree of a node with latent information $\textbf{z}_s$.
 \item \textbf{D7:} $\bar{k}_{nn}(\textbf{z}_s)$, defined as the average degree of the neighbours of a node with latent information $\textbf{z}_s$.
 \item \textbf{D8:} $\bar{k}_{nn}(k)$, defined as the average degree of the neighbours of a node with degree $k$.
\end{itemize}
The following main result characterises all of the quantities listed under a very general LVM:
\begin{theorem}\label{thm1}
 Under assumptions $\textbf{A1}-\textbf{A3}$, the following results hold:
 \begin{align}
  &\textbf{D1:\ }\theta(\textbf{z}_s)= \int_{\mathcal{Z}}p\left(\textbf{z}_j\right)r\left( \textbf{z}_s,\textbf{z}_j \right)d\textbf{z}_j\label{theta1}\\
  &\textbf{D2:\ }G(x)=\int_{\mathcal{Z}}p\left(\textbf{z}_s\right)\left[x\theta(\textbf{z}_s) + 1 - \theta(\textbf{z}_s)\right]^{n-1}d\textbf{z}_s\label{pgflsm}\\
  &\textbf{D3:\ }\frac{\partial{}^rG}{\partial{x}^r}(1)  = \frac{(n-1)!}{(n-r-1)!}\int_{\mathcal{Z}}p\left(\textbf{z}_s\right)\theta(\textbf{z}_s)^rd\textbf{z}_s\label{Gder}\\
  &\textbf{D4:\ }\bar{k} = (n-1)\int_{\mathcal{Z}}p\left(\textbf{z}_s\right)\theta(\textbf{z}_s)d\textbf{z}_s\label{avgdeg}\\
  &\textbf{D5:\ }p_k= \int_{\mathcal{Z}}p\left(\textbf{z}_s\right)\binom{n-1}{k}\theta(\textbf{z}_s)^k\left[1-\theta(\textbf{z}_s)\right]^{n-k-1}d\textbf{z}_s\label{degree1}\\
  &\textbf{D6:\ }\bar{k}(\textbf{z}_s)=(n-1)\theta(\textbf{z}_s)\label{kbar1}\\
  &\textbf{D7:\ }\bar{k}_{nn}(\textbf{z}_s) = 1+\frac{(n-2)}{\theta\left( \textbf{z}_s \right)} \int_{\mathcal{Z}} p\left( \textbf{z}_j \right)r\left( \textbf{z}_s,\textbf{z}_j \right)\theta\left( \textbf{z}_j \right)d\textbf{z}_j\label{knn1}\\
  &\textbf{D8:\ }\bar{k}_{nn}(k) = \frac{1}{p_k}\int_{\mathcal{Z}} p(\textbf{z}_j)\binom{n-1}{k}\theta(\textbf{z}_j)^k\left[1-\theta(\textbf{z}_j)\right]^{n-k-1} \bar{k}_{nn}(\textbf{z}_j)d\textbf{z}_j\label{knn2}
\end{align}
\end{theorem}
The proof of Theorem \ref{thm1} is provided in Appendix \ref{proof1}.

\begin{remark}
Equation  \eqref{knn2} is a generalisation of a result from \textcite{boguna2003class}, where a general framework to study the degree correlations for the fitness model of \textcite{caldarelli2002scale} and \textcite{soderberg2002general}
is introduced. 
\end{remark}
\begin{remark}
 Particular instances of some of the results of Theorem \ref{thm1} have been already shown in \textcite{olhede2012degree} and \textcite{channarond2012classification,daudin2008mixture} 
 for Stochastic Block Models and Fitness models, without resorting to PGFs. 
 Theorem \ref{thm1} encompasses those special cases and extends the range of results offered.
\end{remark}

The results presented in Theorem \ref{thm1} are valid for all LVMs. 
Essentially, they relate the distributional assumptions about the latent 
variables and the edge probabilities to the properties of the degrees 
of the realised networks.  

We now apply these results to LPMs. 
The following Corollaries show how the formulas involved in \textbf{D1}-\textbf{D8} simplify under the Gaussian models of Table \ref{table:models}. Proofs are shown in 
Appendices \ref{proof:cor1} and \ref{proof:cor2}.
\begin{corollary}\label{cor1}
Under the Gaussian LPM, the following quantities have an explicit form:
\begin{align}
\textbf{D1:\ }&\theta(\textbf{z}_s)= \tau\left(\frac{\varphi}{\gamma+\varphi}\right)^{\frac{d}{2}}\exp\left\{-\frac{1}{2(\gamma+\varphi)}\textbf{z}_s^t\textbf{z}_s\right\}\\
\textbf{D3:\ }&\frac{\partial{}^rG}{\partial{x}^r}(1)= \frac{(n-1)!}{(n-r-1)!}\tau^r\left\{\frac{\varphi^r}{\left(\gamma+\varphi\right)^{r-1}\left[(r+1)\gamma+\varphi\right]}\right\}^\frac{d}{2}\label{Gderlpm}\\
\textbf{D4:\ }&\bar{k} = (n-1)\tau\left\{\frac{\varphi}{2\gamma+\varphi}\right\}^\frac{d}{2}\label{avgdeglpm}\\
\textbf{D7:\ }&\bar{k}_{nn}(\textbf{z}_s) = 1+\bar{k}\left(\frac{n-2}{n-1}\right)\frac{f_d\left( \textbf{z}_s;\textbf{0},\frac{\gamma^2+3\gamma\varphi+\varphi^2}{2\gamma+\varphi} \right)}{f_d\left( \textbf{z}_s;\textbf{0},\gamma+\varphi \right)}\label{knn1lpm}
\end{align}
\end{corollary}
Note that $\theta\left(\ \cdot\ \right)$ has an explicit expression, thus evaluation of the quantities in \textbf{D2}, \textbf{D5} and \textbf{D8} boils down to an approximation of 
a single integral.
\begin{corollary}\label{cor2}
Under the Gaussian LPCM, the following results hold:
\begin{align}
\textbf{D1:\ }&\theta(\textbf{z}_s)= \tau\left( 2\pi\varphi \right)^{\frac{d}{2}}\sum_{g=1}^{G}\pi_gf_d\left( \textbf{z}_s;\boldsymbol{\mu}_g,\gamma_g+\varphi \right)\\
\textbf{D4:\ }&\bar{k}=(n-1)\tau\left( 2\pi\varphi \right)^{\frac{d}{2}}\sum_{g=1}^{G}\sum_{h=1}^{G}\pi_g\pi_hf_d\left( \mu_g-\mu_h;\textbf{0},\gamma_g+\gamma_h+\varphi \right).
\end{align}
Also, the degree distribution is a continuous mixture of binomial distributions,
where the mixture weights are themselves distributed as mixtures of Gaussians:
\begin{equation}
\textbf{D7:\ }p_k= \int_{\mathbb{R}^d}\left[\sum_{g=1}^{G}\pi_gf_d\left(\textbf{z}_s;\mu_g,\gamma_g\right)\right]\binom{n-1}{k}\theta(\textbf{z}_s)^k\left[1-\theta(\textbf{z}_s)\right]^{n-k-1}d\textbf{z}_s.\label{degree2}\\ 
\end{equation}
\end{corollary}

Under the Gaussian LPMRE, none of the equations can be written explicitly, 
since the integrals over the random effects cannot be evaluated analytically. 
However, we will make use of the following two quantities, 
which will be calculated in an approximate form:
\begin{align}
&\theta(\textbf{z}_s,\varphi_s)= \int_{\mathbb{R}^d}\int_{0}^{\infty}f_d\left(\textbf{z}_j;\textbf{0},\gamma\right)p\left(\varphi_j\middle\vert\beta_0,\beta_1\right)r\left( \textbf{z}_s,\textbf{z}_j \right)d\varphi_jd\textbf{z}_j\label{thetaLPMRE},\\
&G^{(r)}(1) = \frac{(n-1)!}{(n-r-1)!}\int_{\mathbb{R}^d}\int_{0}^{\infty}f_d\left(\textbf{z}_s;\textbf{0},\gamma\right)p\left(\varphi_j\middle\vert\beta_0,\beta_1\right)\theta(\textbf{z}_s,\varphi_s)^rd\textbf{z}_sd\varphi_s,\label{Gder2}
\end{align}
\begin{remark}
The advantage of using the Gaussian function rather than the Logistic function of \textcite{hoff2002latent,handcock2007model,krivitsky2009representing} is mainly highlighted in 
Corollary \ref{cor1}: under the Gaussian hypothesis most of the integrals of Equations \ref{theta1}-\ref{knn2} can be evaluated analytically since they become a 
convolution of two Gaussian densities, which is solvable for any $d$. Also, quantities that do not have an exact expression, such as $p_k$ or $\bar{k}_{nn}(k)$, can 
be efficiently evaluated through numerical methods, since the number of integrals to approximate is constant (depending on $d$, but not on $n$).
\end{remark}

\begin{remark}\label{nonidentifiability}
In Gaussian LPMs, a nonidentifiability issue arises between the parameters 
$\varphi$ and $\gamma$, since the factorial moments depend only on their ratio,
$\varphi / \gamma$. We argue, however, that both parameters should be included in our study, to keep the model as close as possible to the original LPM 
 of \textcite{hoff2002latent},
 and to provide a proper basis for possible extensions, such as the 
Gaussian LPCM and the Gaussian LPMRE.
\end{remark}

\subsection{Clustering coefficient}
In this section, we take advantage of the Gaussian assumption to study the clustering coefficient value for Gaussian LPMs analytically.

Since there is more than one definition for the clustering coefficient, we clarify that the one used in this paper is the global clustering coefficient
of \textcite{newman2003properties}, 
equal to three times the number of triangles divided by the number of connected triples of nodes. 
Thanks to the exchangeability of actor labels, this quantity is an unbiased estimator of the probability that,
given two consecutive edges, the extremities of such $2$-steps path are connected themselves. 

\begin{proposition}\label{thm2}
Under assumptions \textbf{A1}-\textbf{A3}, the global clustering coefficient $\mathcal{C}$ can be written as:
\begin{equation}\label{clustering1}
\begin{split}
 \mathcal{C} 
 &= \frac{\int_{\mathcal{Z}}\int_{\mathcal{Z}}\int_{\mathcal{Z}} p(\textbf{z}_i)p(\textbf{z}_k)p(\textbf{z}_j)r\left( \textbf{z}_i,\textbf{z}_k \right)r\left( \textbf{z}_k,\textbf{z}_j \right)r\left( \textbf{z}_j,\textbf{z}_i \right)d\textbf{z}_id\textbf{z}_kd\textbf{z}_j}
 {\int_{\mathcal{Z}}\int_{\mathcal{Z}}\int_{\mathcal{Z}} p(\textbf{z}_i)p(\textbf{z}_k)p(\textbf{z}_j)r\left( \textbf{z}_i,\textbf{z}_k \right)r\left( \textbf{z}_k,\textbf{z}_j \right)d\textbf{z}_id\textbf{z}_kd\textbf{z}_j}.
\end{split}
\end{equation} 
Under the Gaussian LPM both the numerator and the denominator can be 
expressed analytically, yielding the following result:
\begin{equation}\label{trans}
 \mathcal{C} = \tau\left(\frac{\gamma+\varphi}{3\gamma+\varphi}\right)^{\frac{d}{2}}.
\end{equation}
\end{proposition}
A proof of Proposition \ref{thm2} is provided in Appendix \ref{proof:clustering}.
We note that the \eqref{trans} gives an exact result for the clustering coefficient of an LPM of any size. This is an interesting result and contrasts with many network models, 
where the clustering coefficient can only be recovered asymptotically. Some interesting consequences of \eqref{trans} will be illustrated in Section \ref{sec:Asymptotics for clustering}.

\subsection{Connectivity properties}
The study of the theoretical properties of LPMs can be further extended, characterising the connectivity structure of realised networks. To do so, we give the 
definition of a path for a random graph, and show a general result about the connection of two nodes in Gaussian LPMs, once their latent position is known.

\begin{definition}[Path]
 Under assumptions \textbf{A1}-\textbf{A3}, a $k$-step path is a sequence of $k+1$ distinct nodes $\left\{i_0,i_1,\dots,i_{k}\right\}$ such that an edge is present 
 between every two consecutive nodes, i.e. $y_{i_0i_1}=y_{i_1i_2}=\cdots=y_{i_{k-1}i_k}=1$.
\end{definition}
Under the same assumptions, the probability of a $k$-step path appearing between two nodes with latent information $\textbf{z}_i$ and $\textbf{z}_j$ can be written as:
\begin{equation}\label{WalkIntegral1}
 I_{k}(\textbf{z}_i,\textbf{z}_j) = \int_{\mathcal{Z}}\cdots\int_{\mathcal{Z}} p(\textbf{z}_1)\dots p(\textbf{z}_{k-1})r\left( \textbf{z}_i,\textbf{z}_1 \right)r\left( \textbf{z}_1,\textbf{z}_2 \right)\cdots r\left( \textbf{z}_{k-1},\textbf{z}_j \right)d\textbf{z}_1\cdots d\textbf{z}_{k-1}.
\end{equation}
For a Gaussian LPM, the integrals on the right-hand side of \eqref{WalkIntegral1} involve Gaussian kernels only, and therefore they can be evaluated exactly. 
We provide a more explicit 
formula for $I_{k}(\textbf{z}_i,\textbf{z}_j)$ in the following Proposition:
\begin{proposition}\label{thm3}
 Under the Gaussian LPM, let $I_{k}(\textbf{z}_i,\textbf{z}_j)$ be defined as in \eqref{WalkIntegral1}, for any $k=1,2,\dots,n-1$, $\textbf{z}_i\in\mathbb{R}^d$ 
 and $\textbf{z}_j\in\mathbb{R}^d$. Define the following recurrence relations:
\begin{equation}\label{recurrence}
 \begin{cases}
 h_{r+1} &= h_r\alpha_r^{-d}\tau\left( 2\pi\varphi \right)^{\frac{d}{2}} f_d\left( \textbf{z}_i;\textbf{0},\frac{\omega_r+\gamma}{\alpha_r^2} \right)\\
  \alpha_{r+1} &= \frac{\alpha_r\gamma}{\omega_r+\gamma}\\
  \omega_{r+1} &= \frac{\omega_r\varphi + \omega_r\gamma + \gamma\varphi}{\omega_r+\gamma}
 \end{cases}\hspace{0.2cm}\mbox{, with}\hspace{0.2cm}
 \begin{cases}
 h_1&=\tau\left( 2\pi\varphi \right)^{\frac{d}{2}}\\
 \alpha_1&=1\\
 \omega_1&=\varphi\\
\end{cases}.
\end{equation}
Then, the following result holds:
\begin{equation}
 I_{k}(\textbf{z}_i,\textbf{z}_j) = h_{k}f_d\left( \textbf{z}_j - \alpha_{k}\textbf{z}_i; \textbf{0},\omega_{k} \right),\ \mbox{ for }k=1,2,\dots,n-1.
\end{equation}
\end{proposition}
The proof of Proposition \ref{thm3} is provided in Appendix \ref{thm3proof}.
\begin{remark}
 Note that the previous result could be extended by integrating out the 
latent positions $\textbf{z}_i$ and $\textbf{z}_j$ as well. 
However, this is not of interest for the present work.
\end{remark}
The result of Proposition \ref{thm3} is a useful tool for studying the 
statistical properties of path lengths for Gaussian LPMs, which we develop in 
Section \ref{sec:Path lengths}.

\section{Properties of realised networks}\label{ref:RealisedNetworks}
We now use the results in the previous section to obtain properties of the Gaussian LPM.

A main drawback of all LPMs is that, given the complete set of latent positions, the evaluation of the likelihood for the corresponding realised graph requires the calculation of a 
distance matrix, with a computational and storage cost of $O(n^2)$. This cost is the main obstacle to inference for large graphs, making estimation 
impractical for networks larger than a few thousands nodes. 
The issue extends also to the generation of LPMs, which is usually performed in two sequential steps: firstly latent positions are sampled, and then edges are created with the 
Gaussian probability. The evaluation of the distance matrix is thus needed in between the two steps. This makes any empirical study of the properties of LPMs rather inefficient 
and limited to relatively small graphs, only.

By contrast, the results presented in Theorem \ref{thm1} and related Corollaries involve either exact formulas, which have negligible computational cost, 
or integral approximations whose computational cost is independent of $n$.
Hence, the analysis that we propose in this Section does not require any intensive calculation and can be performed on networks of any size.
Note that \textcite{raftery2012fast} proposed a computational approximation to overcome this difficulty, whereas here we provide exact results and analytical approximations.

\subsection{Characterisation of the degree distribution for LPMs}\label{sec:Characterisation of the degree distribution}
Empirical evaluations \parencite{newman2003structure} suggest that typically the proportion of nodes with degree greater than $k$ is expected to be proportional 
to $k^{-\alpha}$, for a positive $\alpha$ which can be as small as $2$. Networks exhibiting such behaviour are usually referred to as scale-free, and the corresponding degree 
distribution is said to follow a power-law decay. The highly connected nodes, denoted hubs, fulfil a crucial role in defining the structure of the network \parencite{albert2000error}, 
and as a result this is a feature which many network models aim to capture \parencite{barabasi1999emergence,newman2001random}.

\begin{figure}[htb]
\centering
\includegraphics[width=0.48\textwidth]{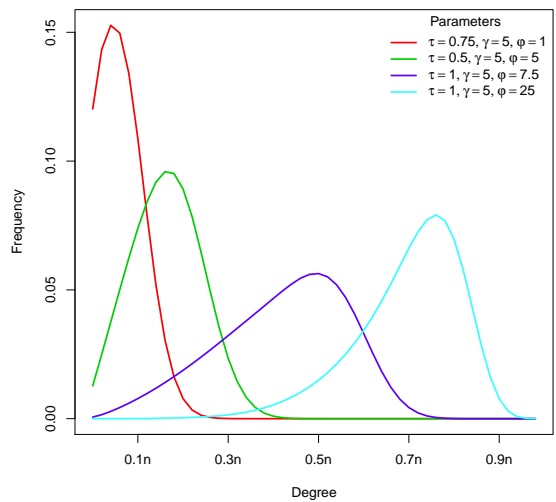}
\caption{Gaussian Latent Position Model: Approximate degree distribution 
for different sets of model parameters $\tau,\gamma,\varphi$.}
\label{fig:dd1}
\end{figure}

According to the results of the previous section, the theoretical degree distribution of a Gaussian LPM has the form of a continuous mixture of binomials, 
and can be approximated efficiently for any network size.
Figure \ref{fig:dd1} shows approximate degree distributions for various choices of model parameters. 

While the degree distributions of sparse networks often resemble Poisson distributions, denser networks tend to be associated with more left-skewed shapes.
However, the theoretical degree distribution of LPMs in Figure \ref{fig:dd1} resembles a truncated shape, suggesting that the model may not successfully represent heavy tails. 
It should be noted, however, that truncated shapes do arise in social networks: data are often collected through surveys, where each actor is asked to specify up to a fixed number of preferences, 
so that the degree distribution will exhibit an artificial truncation at the corresponding value.
Popular social datasets have been obtained using such a design, such as Sampson's monks data \parencite{sampson1968novitiate} and the Adolescent Health data \parencite{handcock2007model}.
Moreover, some important empirical evidence has been shown in \textcite{dunbar1992neocortex} demonstrating the existence of a theoretical cognitive limit on the number of stable 
relationships that social actors can maintain. Hence both power-laws and non-power-laws behaviours are of interest in statistical modelling of networks.

We now propose a more rigorous analysis of the degree distribution using the dispersion and skewness indexes, which can be evaluated through the exact formulas for 
the factorial moments in \eqref{Gderlpm}.

\begin{corollary}\label{cor:dispersion}
Under the Gaussian LPM, the dispersion index is given by:
\begin{equation}
\mathcal{D} = 1+(n-2)\tau\left(\frac{\varphi(2\gamma+\varphi)}{(\gamma+\varphi)(3\gamma+\varphi)}\right)^{\frac{d}{2}} - (n-1)\tau\left(\frac{\varphi}{2\gamma + \varphi}\right)^{\frac{d}{2}}.
\end{equation}
\end{corollary}
The proof is given in Appendix \ref{proof:dispersion}.
\begin{remark}
 The calculation of the skewness does not involve any simplification, 
and so it is omitted here.
\end{remark}

The dispersion index can be used to assess how dispersed the distribution is when compared to a Poisson, which has an index of $1$. A value greater than $1$ corresponds to 
an overdispersed distribution while a value smaller than $1$ corresponds to an underdispersed one. The Binomial distribution arising from a finite Erd\H{o}s-R\'enyi random graph has 
a dispersion index smaller than $1$, and so it qualifies as underdispersed.

Corollary \ref{cor:dispersion} allows us to study how the model parameters $\tau,\gamma$ and $\varphi$ affect the dispersion of the distribution. 
For $d=2$, our results can be summarised as follows:
\begin{itemize}
 \item When $\varphi=\gamma(\sqrt{n-1} - 2)$, the distribution has dispersion index $1$, typical of a Poisson distribution.
 \item When $\varphi<\gamma(\sqrt{n-1} - 2)$, the distribution has dispersion index greater than $1$, so that the distribution is overdispersed.
 \item When $\varphi>\gamma(\sqrt{n-1} - 2)$, the distribution has dispersion index smaller than $1$, typical of a Binomial distribution, and so is underdispersed.
\end{itemize}
Note that the characterisation does not depend on $\tau$. 

\begin{figure}[htb]
\centering
\includegraphics[width=0.48\textwidth]{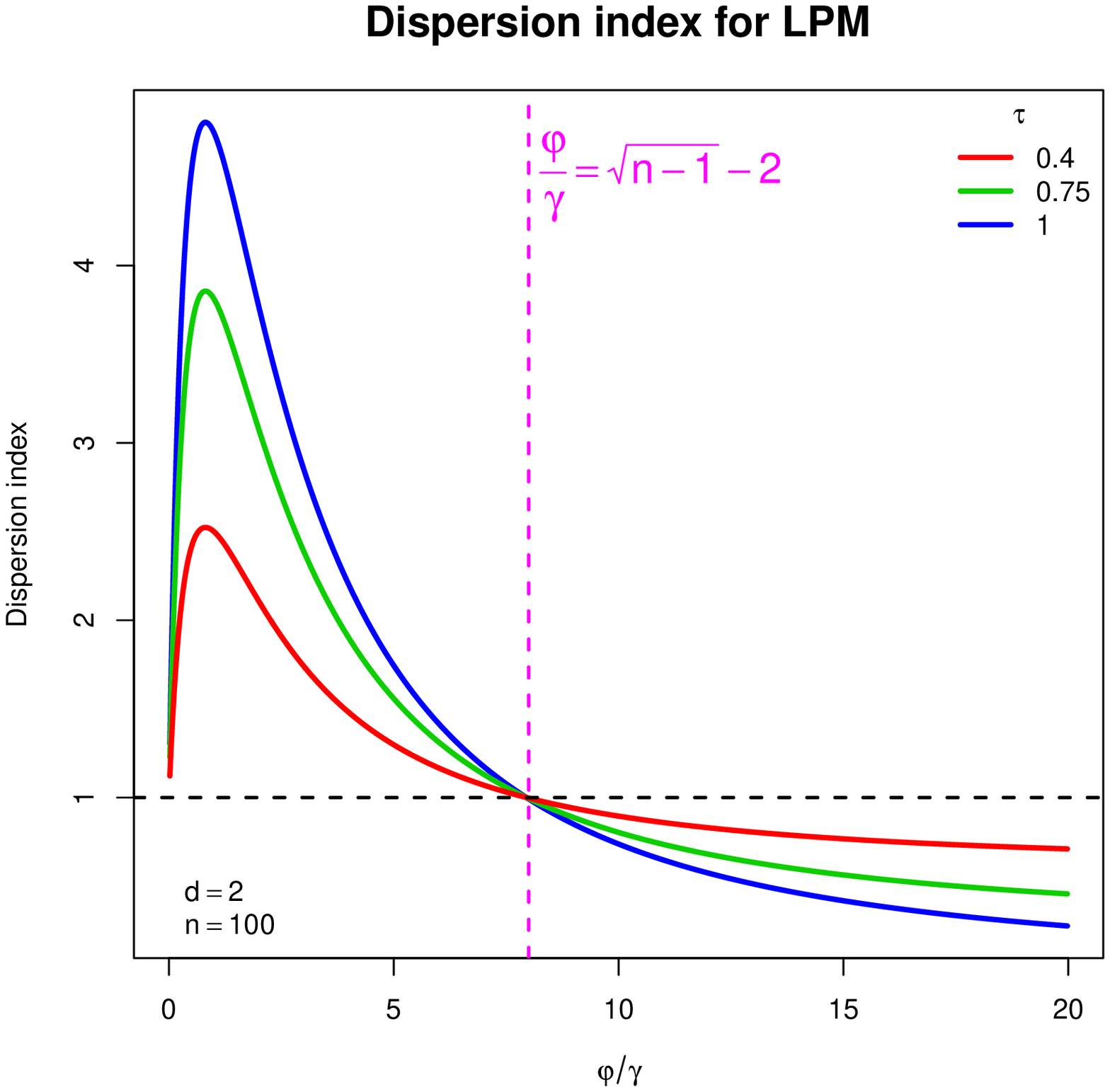}
\includegraphics[width=0.48\textwidth]{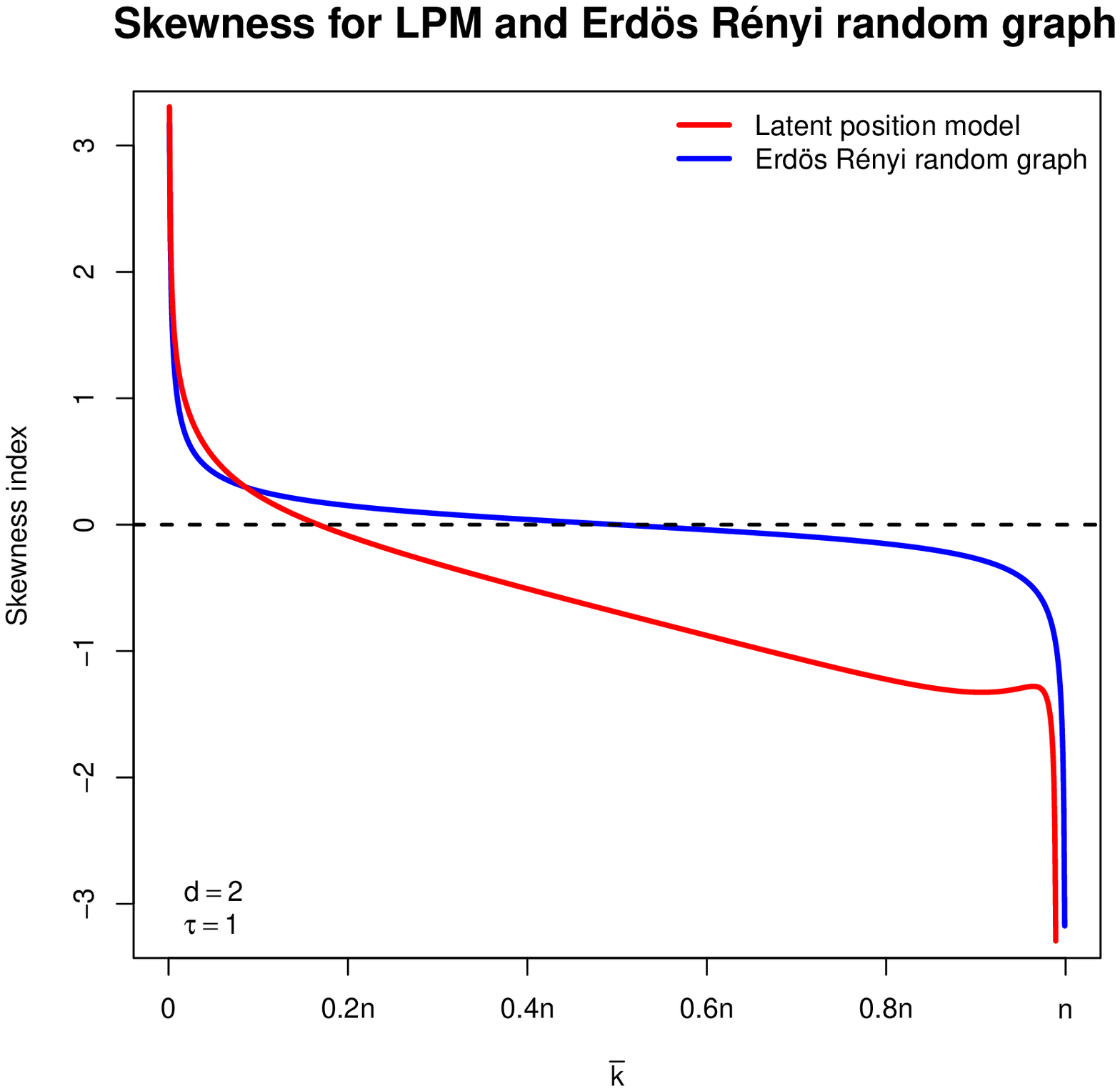}
\caption{Gaussian Latent Position Model: 
\textbf{Left:} Dispersion index versus the ratio between $\varphi$ and $\gamma$. The vertical line is the threshold corresponding to a Poisson dispersion.
For larger values of $\varphi$, the distributions arising are not more dispersed than an Erd\H{o}s-R\'enyi random graph, asymptotically degenerating to such model as $\varphi$ gets larger.
\textbf{Right:} Unless the graph is very sparse, the skewness index for Gaussian LPMs (red line) is smaller than the skewness of a Erd\H{o}s-R\'enyi random graph (blue line) with the 
same average degree.}
\label{fig:9}
\end{figure}

The left panel of Figure \ref{fig:9} shows the dispersion as a function of the model parameters.
The motivation behind this result is that the Erd\H{o}s-R\'enyi random graph model is recovered as a special case asymptotically, as $\varphi$ gets larger.
Therefore, as $\varphi$ increases, the model degenerates and the degree distribution becomes binomial and thus underdispersed, regardless of how sparse the network is. If $\varphi$ is small enough,
namely below the threshold, then the model is nondegenerate and produces networks with an overdispersed degree distribution.
Hence, Gaussian LPMs are able to represent degree heterogeneity, since for many choices of the model parameters the degree distribution is overdispersed.
However, degree heterogeneity does not imply heavy tails or power-law behaviour.

 We now analyse the skewness index,
which is useful for identifying asymmetries in overdispersed distributions.
In the case of degree distributions of networks, a negative value of the skewness index corresponds to shapes exhibiting a left tail heavier than the right one,
while a positive value corresponds to the opposite behaviour. As a tool to assess the presence of hubs, we expect a scale-free network to have a positive and relatively large skewness index. 
However, as shown in the right panel of Figure \ref{fig:9}, 
this scenario does not arise in Gaussian LPMs.

Given that in Erd\H{o}s-R\'enyi random graphs $p_k$ goes to zero at the
rate  $1/k!$ (i.e. power laws are not represented), 
the right panel in Figure \ref{fig:9} shows that, unless the graph is very 
sparse, Gaussian LPMs exhibit degree distributions that are always more skewed to the left than those of the Erd\H{o}s-R\'enyi model with the same average degree. Even for very sparse
networks, the difference is not large enough to justify the presence of a low-order power-law tail. 

This shows that Gaussian LPMs cannot capture power-law behaviour. 
They are able to represent degree heterogeneity, but in 
the sense that degrees will not be concentrated around the mean value, but will
rather have a nontrivially dispersed distribution between $0$ and a 
maximum degree value, confirming the shapes already shown in 
Figure \ref{fig:dd1}.

\subsection{Degree correlations}\label{sec:Degree correlations}
In the study of networks, one is often interested in the mixing properties of the graph. One mixing structure arises when nodes that share common features
are more likely to be linked.
In the context of social networks, this behaviour is called homophily. 

A special case is mixing according to the nodes' degrees, called degree 
correlation.  For example, one might be interested in whether the degrees 
of two random nearest neighbours are positively or negatively correlated. 
Positive correlation, or assortative mixing of the degrees, is 
a recurring feature in social networks \parencite{newman2003social,newman2002assortative}, in contrast to many other kinds of networks
(World Wide Web, protein interactions, food webs; see \textcite{newman2003structure}), which typically have negative degree correlation or dissortative mixing.

\begin{figure}[htb]
\centering
 \includegraphics[width=0.49\textwidth]{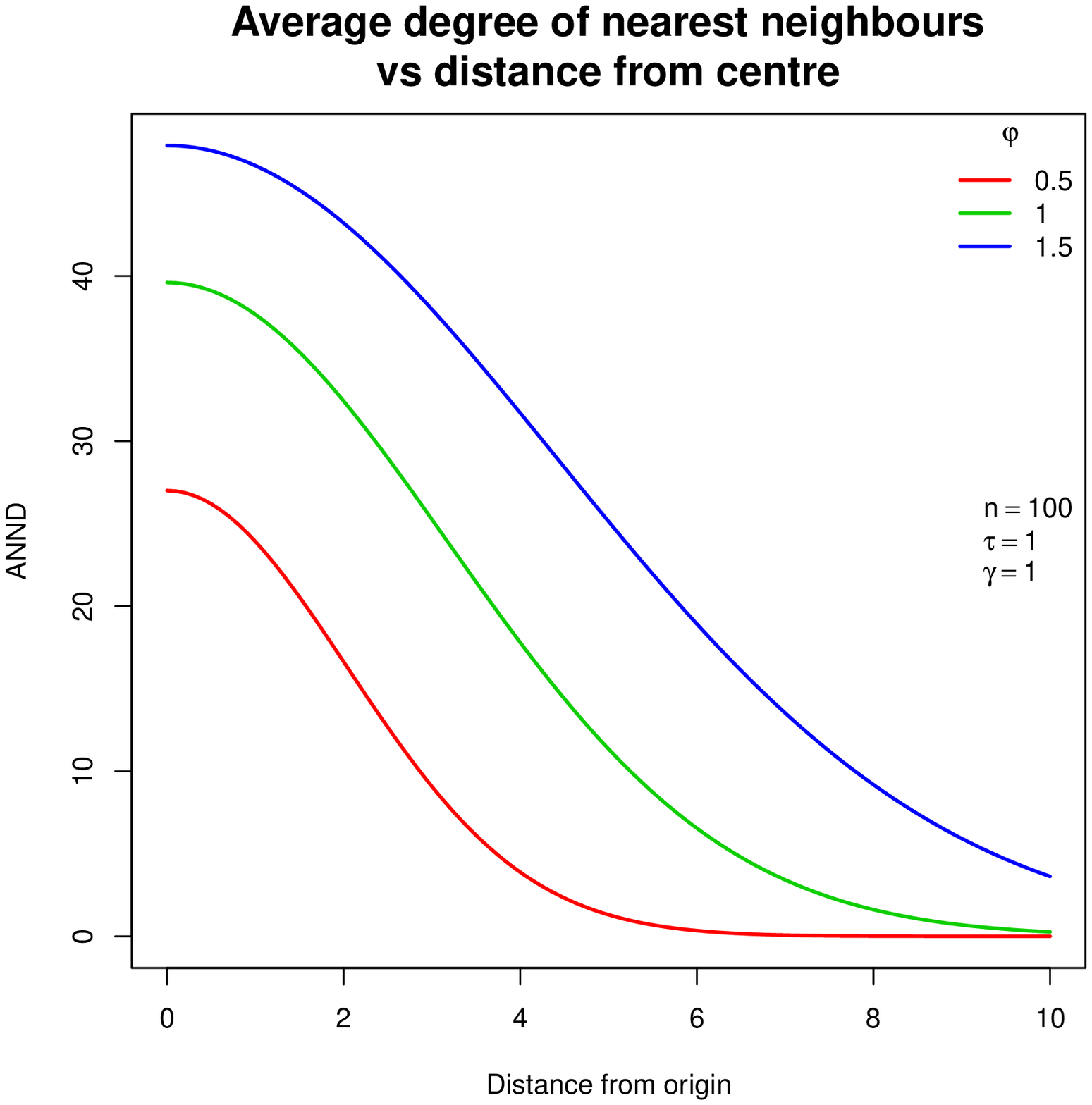}
  \includegraphics[width=0.49\textwidth]{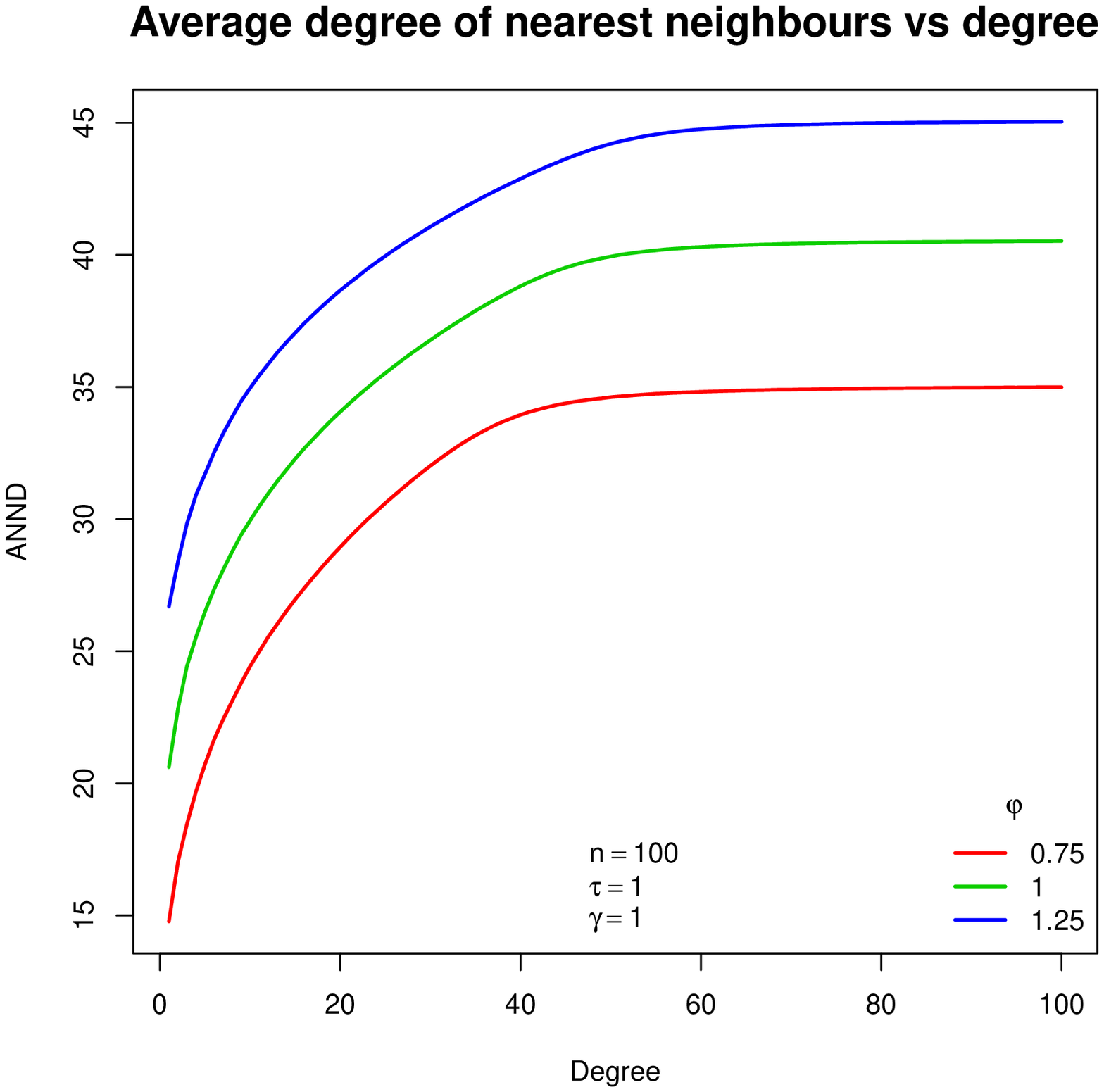}
\caption{Gaussian Latent Position Model:
\textbf{Left:} Average degree of the closest neighbours of a node as a function of its distance from the centre. Nodes located in the centre will more likely connect 
to high degree nodes. \textbf{Right:} Average degree of the closest neighbours as a function of the degree of a node. The ANND index is clearly a nondecreasing function, verifying that
Gaussian LPMs exhibit assortative mixing in the degrees of the nodes.}
 \label{fig:LPMAssortativity}
\end{figure}

Here, we illustrate the fact that Gaussian LPMs can represent assortative 
mixing in the degrees, using the results of Theorem \ref{thm1}.
Equation \eqref{knn1lpm} shows that the Average Nearest Neighbours' Degree (ANND) of an arbitrary node $i$ is an exact function of its latent position
$\textbf{z}_i$. The left panel of Figure \ref{fig:LPMAssortativity} displays 
this function in terms of the distance between $\textbf{z}_i$ and the centre of the latent space.  

It is not surprising that nodes located closer to the centre have highly connected neighbours. 
Instead, \eqref{knn2} provides a less explicit formula for the ANND index as a function of the degree of node $i$, rather than its distance from the centre.
This quantity can be efficiently approximated for every degree value.
The right panel of Figure \ref{fig:LPMAssortativity} represents this case. 
The average degree of the neighbours of a node of degree $k$,
$\bar{k}_{nn}(k)$, appears to be a nondecreasing function of the degree $k$, 
indicating the presence of assortative mixing in the degrees, 
using the same criterion as \textcite{boguna2003class}. 
It follows that realised Gaussian LPM networks exhibit assortative 
mixing of the degrees, suggesting them to be well suited for social networks
\parencite{newman2003social}.

\subsection{Asymptotics for the clustering coefficient}\label{sec:Asymptotics for clustering}
Transitivity, defined as the propensity of two neighbours of a node also to be neighbours of one another, is ubiquitous in network analysis. 
In social networks, the tendency of three or more nodes to cluster 
is a feature of interest since it has a nontrivial relation with the 
structure of path lengths, for example impacting the dynamics of the spread of diseases \parencite{newman2003properties,newman2009random,kiss2008comment}.

LPMs capture transitivity in a very natural way. 
Indeed, when two actors have a neighbour in common, 
it is expected that the three corresponding nodes will 
be close in the latent space, 
making triangles more likely. This reasoning extends to higher order configurations as well. 
In this section, we show how Proposition \ref{thm2} provides 
a more objective justification to this intuition.

One well-known drawback of the Erd\H{o}s-R\'enyi model is that
it cannot capture transitivity when the network is large.
To see this, let $p$ be the connection probability 
and $\bar{k}=p(n-1)$ be the expected average degree of the 
corresponding realised network. We focus on the realistic case where the size of the network increases ($n$ tends to infinity), while $\bar{k}$ remains constant with respect to $n$.
It follows that $p$ must tend to zero as $n$ increases, as well as $\mathcal{C}\rightarrow 0$ since $\mathcal{C}=p$.
Hence, asymptotically, the clustering coefficient for Erd\H{o}s-R\'enyi random graphs is zero.

Even more structured models such as Exponential Random Graph Models, 
have been shown to degenerate asymptotically to Erd\H{o}s-R\'enyi random 
graphs, under some nonrestrictive conditions
\parencite{chatterjee2013estimating},
thus losing the ability to represent a nontrivial transitivity structure.

In contrast, Gaussian LPMs can represent transitivity, even asymptotically.
To see this, first, recall \eqref{avgdeglpm}, which defines the average degree of 
a random node in a Gaussian LPM.
In order to have an asymptotically constant average degree $\bar{k}_0$, the parameters $\varphi$ and $\gamma$ should satisfy:
\begin{equation}
\varphi=\frac{2\bar{k}_0^{\frac{2}{d}}\gamma}{(n-1)^{\frac{2}{d}}\tau^{\frac{2}{d}} - \bar{k}_0^{\frac{2}{d}}}.
\end{equation}
In the limit of large $n$, the corresponding clustering coefficient satisfies:
\begin{equation}
\mathcal{C}=\frac{\tau}{3^{\frac{d}{2}}}.
\end{equation}
Thus the limiting clustering coefficient has a non-zero value that can be 
as large as $3^{-\frac{d}{2}}$. 
This highlights an important difference between 
the Erd\H{o}s-R\'enyi and Exponential Random Graph models on one hand,
and LPMs on the other, in that the latter are able to represent transitivity 
in large networks.

Furthermore, the non-null clustering coefficient classifies Gaussian LPMs as highly clustered networks. Such models lack the loopless tree structure which simplifies the study of 
component sizes and path lengths. A review of the main difficulties arising when dealing with highly clustered models can be found in \textcite{newman2002random}.

\subsection{Path lengths}\label{sec:Path lengths}
In a well known experiment, \textcite{milgram1967small} observed that any two strangers are connected by a chain of intermediate acquaintances of length at most
six. Later on, similar observations were made in \textcite{albert1999internet} about the connectivity of certain portions of the Internet, stating that any two web 
pages are at most 19 clicks away from one another. The small-world effect defines this behaviour exactly: given any two connected nodes, the shortest path from
one node to the other will have an average length which is very small when compared to the size of the network $n$, typically comparable to $\log(n)$ or 
smaller \parencite{newman2001structure}. The small-world property has motivated research on the connectivity of graphs, relevant to
fields such as communication systems, epidemiology and optimisation.

Hence, understanding how a statistical model relates to the small-world 
property is important. For LPMs, not much is known about the 
diameter and connectivity of the realised networks. 
Here, we use Proposition \ref{thm3} to apply a procedure similar to that of \textcite{fronczak2004average}, showing how the distribution of the 
geodesic distances can be evaluated in a Gaussian LPM. We also characterise the average path length (APL) 
for Gaussian LPM networks of any size, giving appropriate insights on the asymptotic behaviour of such an index. 

\textcite{fronczak2004average} focused on the family of fitness models for
networks, which includes Erd\H{o}s-R\'enyi random 
graphs and the preferential attachment model of \textcite{barabasi1999emergence}. 
These models satisfy assumptions \textbf{A1}-\textbf{A3}, where the latent information is coded by a fitness value $h_i$, for every $i\in V$. 
Then, edge probabilities are given by:
\begin{equation}\label{fitness1}
 r\left( h_i,h_j \right)=\frac{h_ih_j}{\beta},
\end{equation}
where $\beta$ is a suitable constant. The model degenerates to an Erd\H{o}s-R\'enyi random graph when $h_i=\bar{k}$ for every $i$, and $\beta=\bar{k}(n-1)$.

Here, we exploit the fact that fitness models and LPMs both originate from LVMs, generalising the work of \textcite{fronczak2004average} to a wider family of models.
To study the connectivity of the networks and the path lengths' distribution, we focus on the quantities $\ell_k\left( \textbf{z}_i,\textbf{z}_j \right)$, 
defined as the probability that the shortest path between two nodes located in $\textbf{z}_i$ and $\textbf{z}_j$ has length $k$. 
We also define $r_k\left(\textbf{z}_i,\textbf{z}_j\right)$ as the probability that a path of length $k$ exists between two nodes. 
In both definitions, and from now on, we condition on the fact that the two nodes are connected, i.e.~that there exists a finite-length
path that has the two nodes as extremes. Such an assumption is natural since usually statistics of path lengths are defined only for sets of connected nodes.
Note that $I_k\left(\textbf{z}_i,\textbf{z}_j\right)$ differs from 
$r_k\left(\textbf{z}_i,\textbf{z}_j\right)$ in that the latter is 
the probability that there is at least one $k$-step path between the two nodes.
We now describe a way to evaluate 
$\ell_k\left( \textbf{z}_i,\textbf{z}_j \right)$ efficiently, 
as a function of the model parameters of a Gaussian LPM.
\paragraph{A9.} The graphs considered are dense enough, such that for every $(i,j)\in \tilde{V}$, if there exists a path of length $k$ between nodes $i$ and $j$, 
then a path of length $t$ exists between the same nodes for every $t=k+1,\dots,n-1$.

\begin{proposition}\label{prop1}
Under the Gaussian LPM and assumption \textbf{A9}, let $i$ and $j$ be any two nodes. Then the following two statements are equivalent:
\begin{itemize}
 \item The geodesic distance between $i$ and $j$ is less than $k$.
 \item There exists a $k$-step path between $i$ and $j$.
\end{itemize}
\end{proposition}
The proof of Proposition \ref{prop1} relies heavily on \textbf{A9} and is straightforward.
From Proposition \ref{prop1} it follows that, for any $i$ and $j$:
\begin{equation}\label{lij0}
r_{k}\left( \textbf{z}_i,\textbf{z}_j \right) = \sum_{t=1}^{k}\ell_{t}\left( \textbf{z}_i,\textbf{z}_j \right).
\end{equation}
Moreover, since $\ell_{1}\left( \textbf{z}_i,\textbf{z}_j \right)=r_{1}\left( \textbf{z}_i,\textbf{z}_j \right)=r\left( \textbf{z}_i,\textbf{z}_j \right)$, the following holds:
\begin{equation}\label{lij1}
\ell_{k}\left( \textbf{z}_i,\textbf{z}_j \right) = r_{k}\left( \textbf{z}_i,\textbf{z}_j \right)-r_{k-1}\left( \textbf{z}_i,\textbf{z}_j \right).
\end{equation}
Hence, we aim to characterise $r_{k}\left( \textbf{z}_i,\textbf{z}_j \right)$, thereby deducing the properties of $\ell_{k}\left( \textbf{z}_i,\textbf{z}_j \right)$.

Each possible path of length $k$ from $i$ to $j$ can be thought of as a Bernoulli random variable, having a success if all the edges involved in the path appear, 
or not having a success if any of those edges fail to appear. For an Erd\H{o}s-R\'enyi random graph with average degree $\bar{k}=(n-1)p$, 
the parameter of such a random variable is $p^k$. 
For Gaussian LPMs, the success probability is $I_k\left( \textbf{z}_i, \textbf{z}_j\right)$, which has been characterised in Proposition \ref{thm3}.

However, we are interested in $r_k\left( \textbf{z}_i,\textbf{z}_j \right)$, which is the probability of the 
union of all the $k$-steps paths from $i$ to $j$. Unfortunately, these variables are not independent, since different paths will have edges in common.
We circumvent this issue by pretending that all such paths are mutually independent, following the reasoning of \textcite{fronczak2004average}. 
This assumption makes sense when $k$ is much smaller than $n$. In fact, for the purpose of the study of shortest path lengths, estimates of 
$r_k\left( \textbf{z}_i,\textbf{z}_j \right)$ will be needed only for small $k$s, since
in the general case $\ell_k\left( \textbf{z}_i,\textbf{z}_j \right)$ will drop to zero very quickly.

Using the results of Proposition \ref{thm3} and Lemma $1$ of \textcite{fronczak2004average}, we obtain:
\begin{equation}\label{lij2}
  \ell_k\left( \textbf{z}_i,\textbf{z}_j \right) \approx \exp\left\{-n^{k-1}I_{k-1}(\textbf{z}_i,\textbf{z}_j)\right\}-\exp\left\{-n^{k}I_{k}(\textbf{z}_i,\textbf{z}_j)\right\}.
\end{equation}
Equation \eqref{lij2} gives a general formula to evaluate the distribution of the geodesic distance $\ell_k\left( \textbf{z}_i,\textbf{z}_j \right)$ for every $k<<n$ for dense Gaussian LPM networks.

\begin{figure}[htb]
\centering
\includegraphics[width=0.32\textwidth]{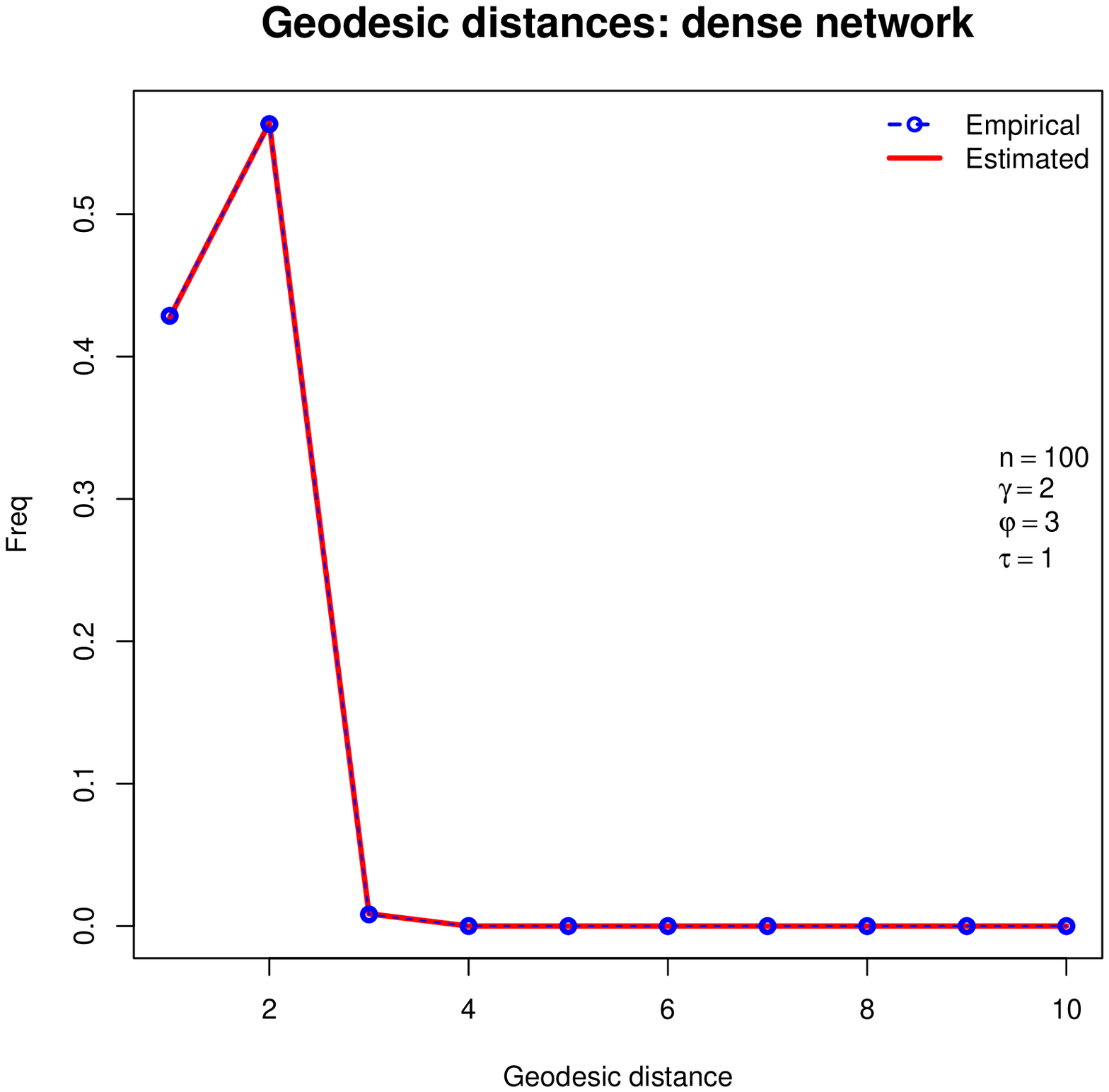}
\includegraphics[width=0.32\textwidth]{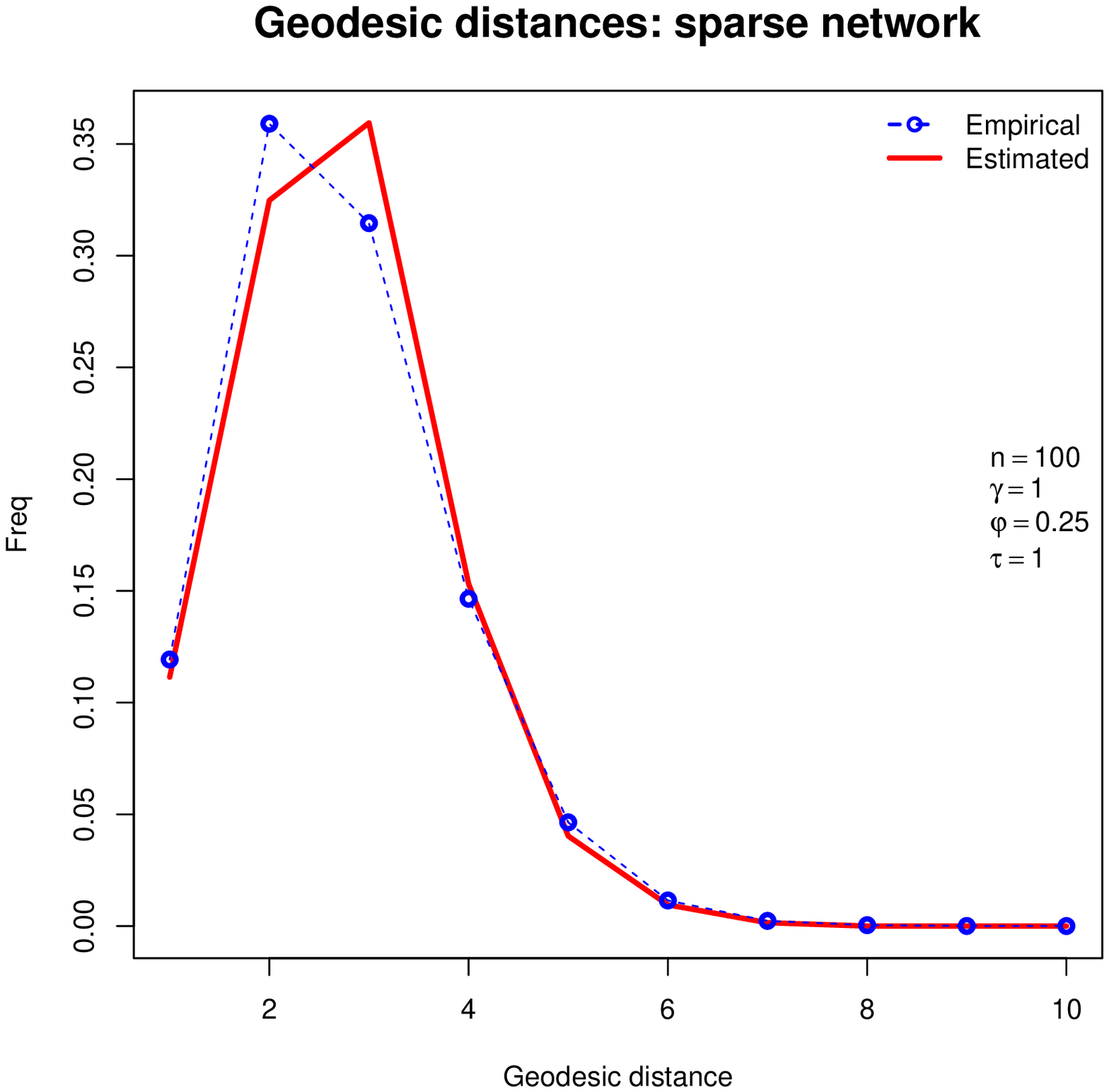}
\includegraphics[width=0.32\textwidth]{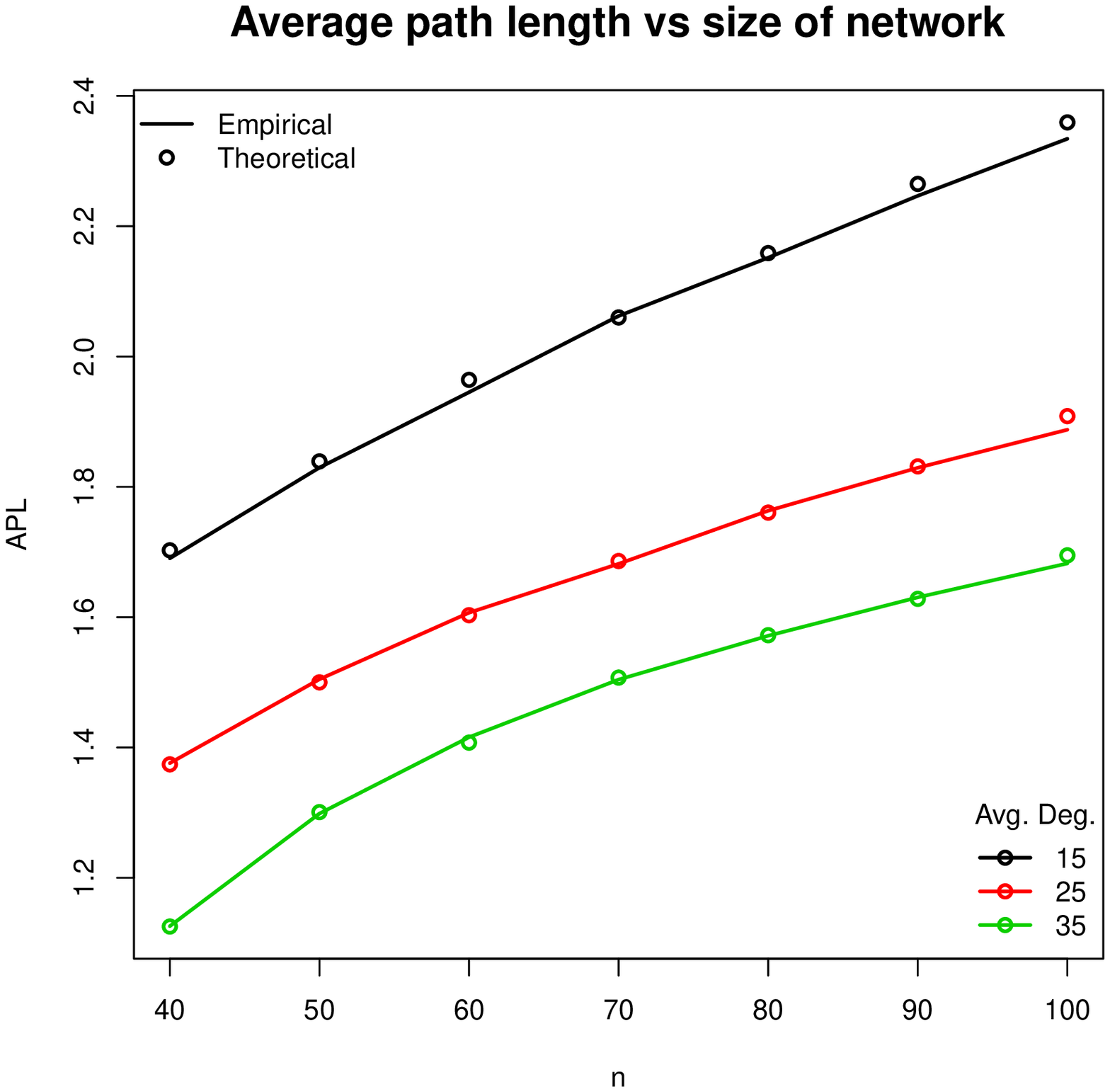}
\caption{Geodesic distances and average path lengths for 
the Gaussian LPM model. \textbf{Left and centre:} Comparison between empirical and theoretical values for the distribution of geodesic distances. 
Networks generated are composed of $100$ nodes. The left panel corresponds to a more dense graph (average degree is approximately $42$) while the 
one in the centre corresponds to a more sparse graph (average degree is approximately $11$). \textbf{Right:} Comparison between empirical (lines) and 
theoretical (dots) values of APL. 
The parameters $\tau$ and $\gamma$ are set to $1$.}
\label{fig:SW1}
\end{figure}

In Figure \ref{fig:SW1} a comparison between the empirical and theoretical values obtained is shown.
The first two panels of Figure \ref{fig:SW1} give a representation of how close the approximation of the path length distribution can be, for a dense Gaussian 
LPM network and a less dense one. Note that in less dense networks the assumption that $k<<n$ is less likely to hold because more sparsity will imply longer shortest paths.

Also, once $\ell_k\left( \textbf{z}_i,\textbf{z}_j \right)$ is known for every $k$, a straightforward evaluation of the APL can be obtained by averaging over all 
possible values of $k$, $\textbf{z}_i$ and $\textbf{z}_j$. The agreement of the estimation with the results from an empirical study is shown in the right panel of 
Figure \ref{fig:SW1}. As expected, the estimation is more accurate for graphs with a higher average degree. However, the results show that such an index is more tolerant 
when assumptions tend to be violated, possibly because the bias is limited when values are averaged.

\begin{figure}[htb]
\centering
\includegraphics[width=0.49\textwidth]{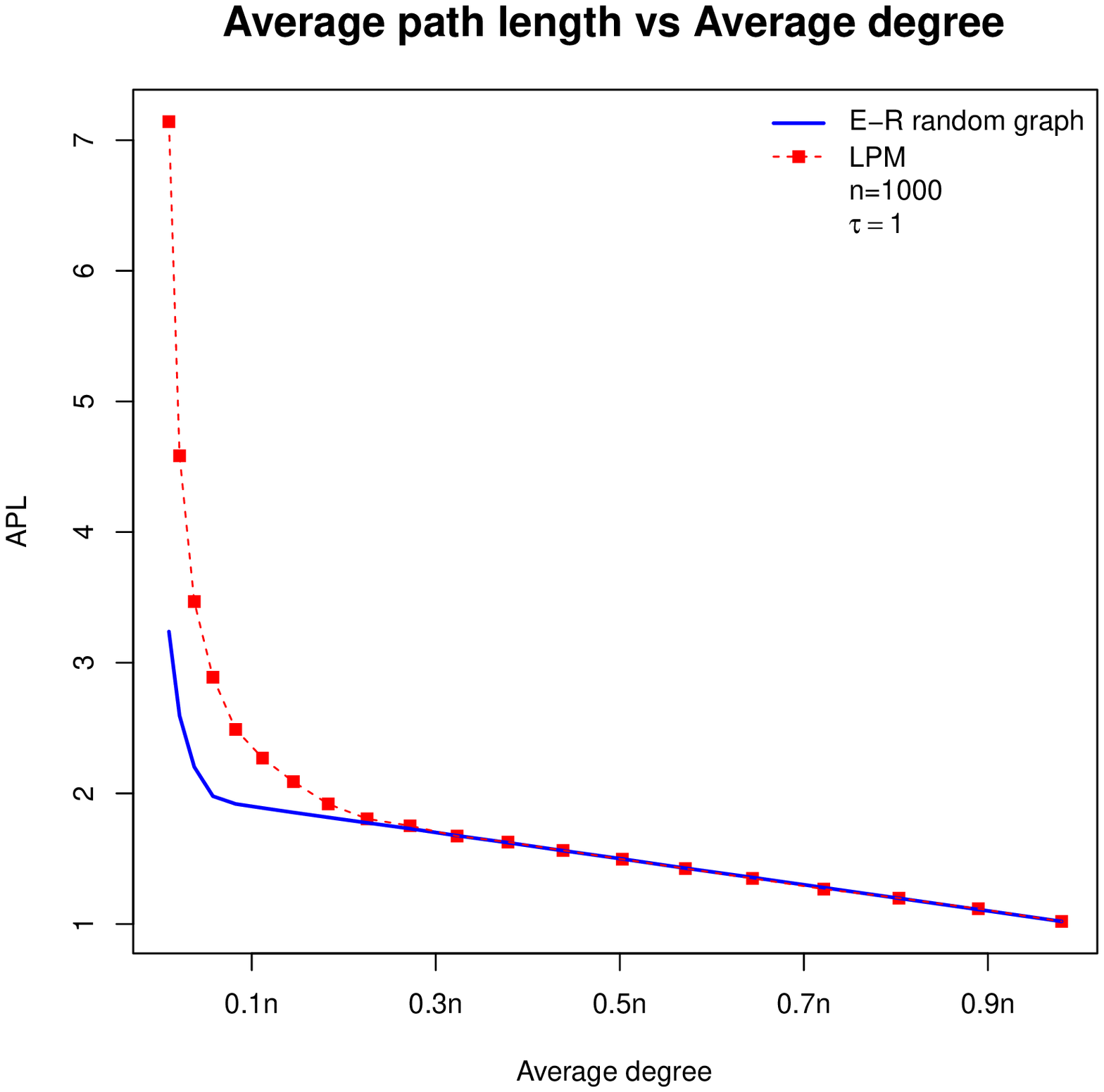}
\includegraphics[width=0.49\textwidth]{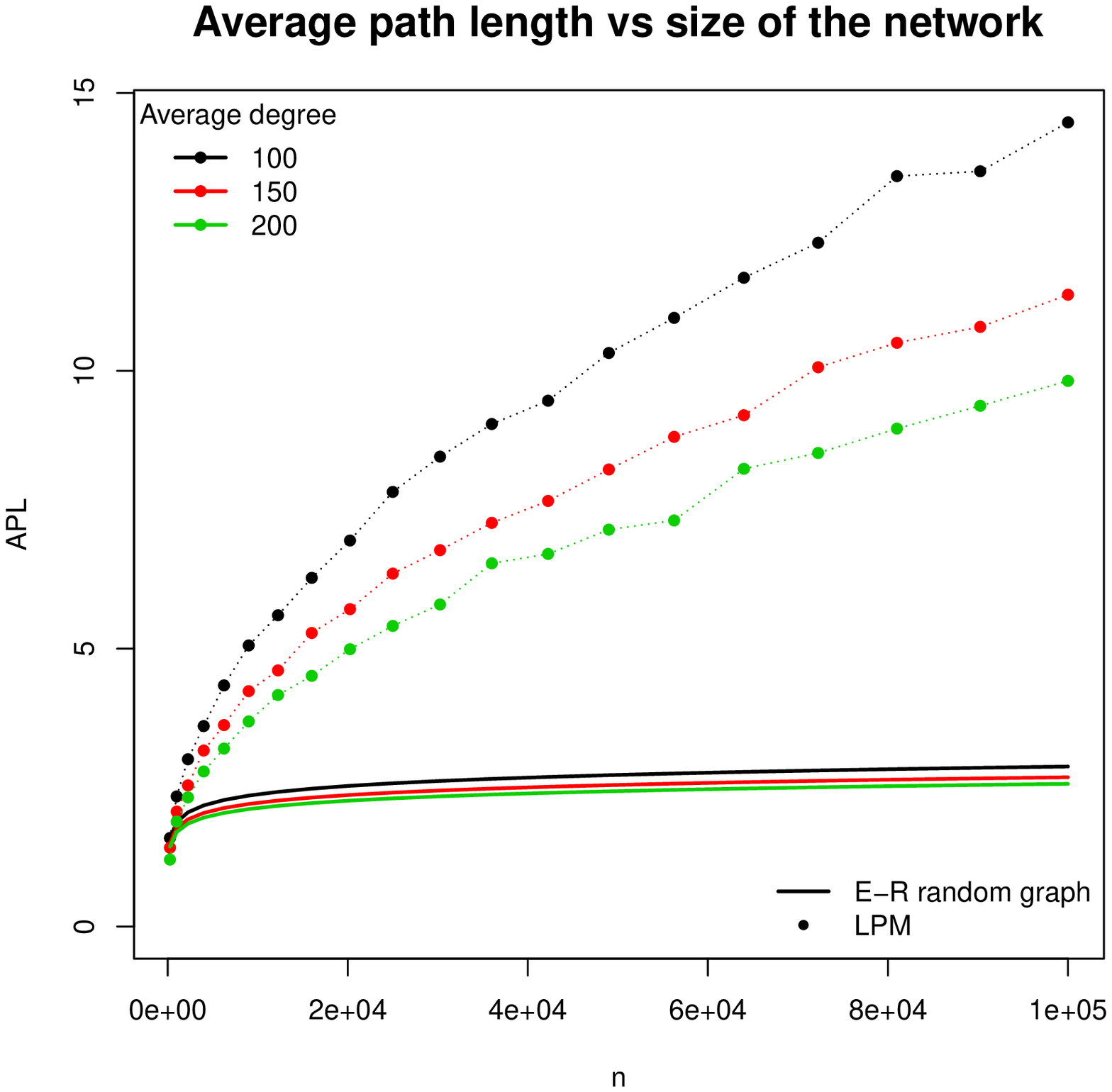}
\caption{\textbf{Left}: APL against the average degree of a $1000$ nodes network, compared with the corresponding Erd\H{o}s-R\'enyi random graph. The two behaviours 
diverge for sparse graphs, in which case Gaussian LPMs exhibit a larger APL. \textbf{Right}: Asymptotic behaviour for the APL is shown. Average degree of the network 
is kept constant while the size $n$ is on the horizontal axis. The continuous lines represent the APL value for corresponding Erd\H{o}s-R\'enyi random graphs with 
same average degrees. APL is typically higher in LPM, and grows proportionally to a function which dominates the logarithm.}
\label{fig:SW2}
\end{figure}

Figure \ref{fig:SW2} shows that Gaussian LPMs typically have a higher APL than corresponding Erd\H{o}s-R\'enyi random graphs. 
In the left panel, the APL is plotted
against the average degree of the network. It appears that the sparser the network, the more marked the difference with Erd\H{o}s-R\'enyi random graphs is. 
Instead, as the network gets denser, Gaussian LPMs tend to behave more and more similarly to Erd\H{o}s-R\'enyi random graphs.
In the right panel of Figure \ref{fig:SW2}, APL values are shown for larger Gaussian LPMs networks. In this case the average degree is kept constant, 
highlighting the asymptotic behaviour of the statistic. 

APL values for the corresponding Erd\H{o}s-R\'enyi random graphs are also 
shown in Figure \ref{fig:SW2}. The Gaussian LPM 
networks typically have a higher APL, which grows faster than the logarithm 
of the size of the network. 


\begin{figure}[htb]
\centering
\includegraphics[width=0.49\textwidth]{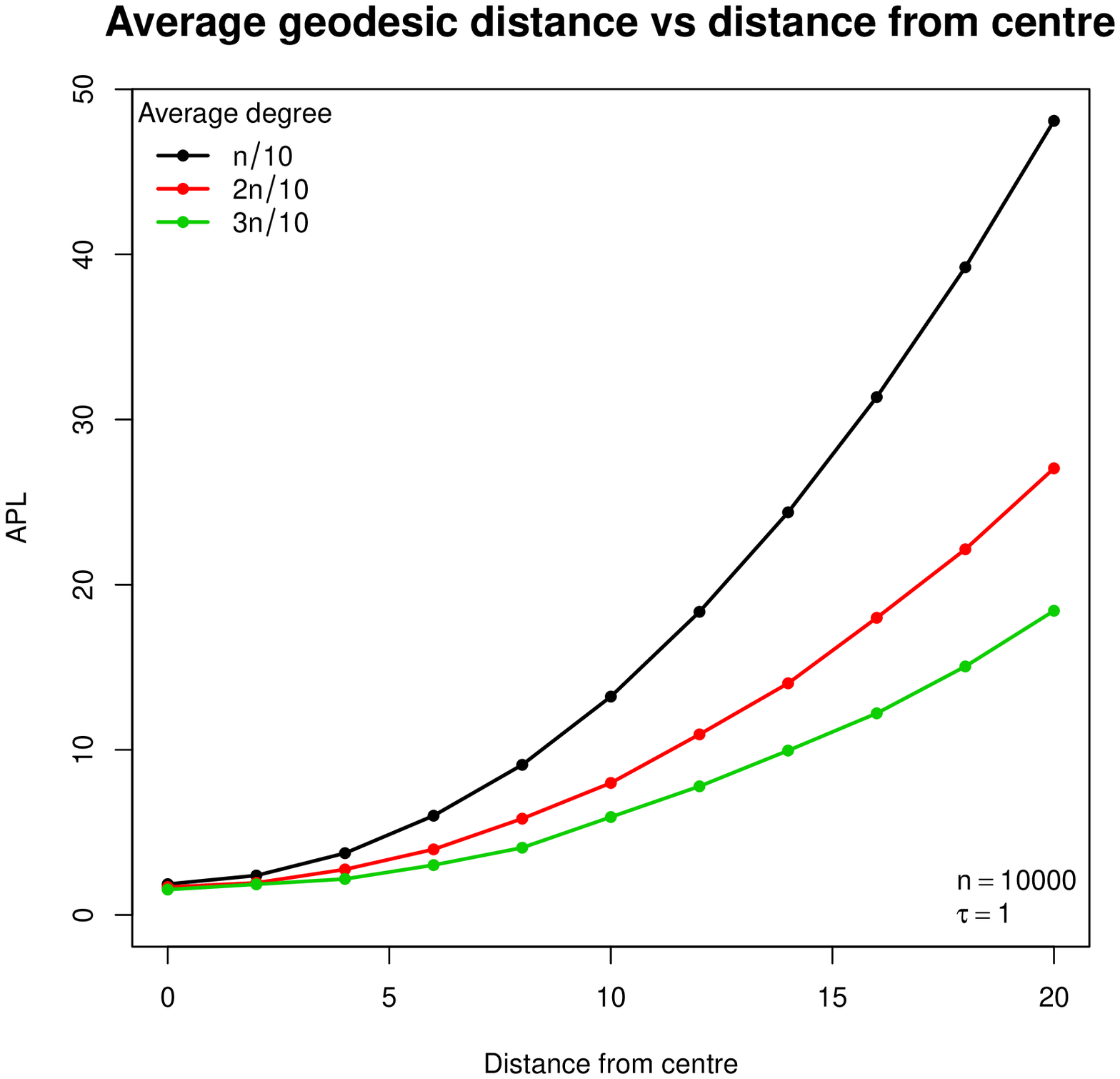}
\caption{Average geodesic distance from a node as a function of its distance from the centre of the latent space. The network is composed of $10000$ nodes, with $\tau=1$. 
Clearly, nodes which are closer to the centre will be better positioned to reach easily many other nodes, thus having a smaller APL index. Such heterogeneity in the connectivity structure
characterises Gaussian LPMs and separates them from Erd\H{o}s-R\'enyi random graphs, justifying the larger values for global APL.}
\label{fig:SW3}
\end{figure}

Figure \ref{fig:SW3} illustrates a possible reason for this behaviour. 
The distance from a node to the centre of the latent space 
is plotted versus its geodesic distance to a second node picked at random.
There is clear heterogeneity, in contrast with the behaviour of 
Erd\H{o}s-R\'enyi random graphs.
Clearly, when averaging over all the possible positions of the second randomly chosen node, important contributions are given by distant isolated nodes,
thereby increasing the APL value.

\section{Advantages of random effects models}\label{sec:LPMRE}
In the previous section, we have shown that, although the Gaussian LPM can capture degree heterogeneity, it cannot represent the power-law behaviour of many observed degree distributions.
In addition, the model has shortcomings in representing the small-world behaviour, in that the APL grows faster than the log of the number of actors.

In the Logistic LPM context, \textcite{krivitsky2009representing} 
addressed similar issues by adding node-specific
random effects to represent different levels of social involvement.
Here, we propose an extension of the Gaussian LPM (namely the Gaussian LPMRE of Table \ref{table:models}) following the same reasoning.

In the Gaussian LPMRE, the connectivity parameter $\varphi$ becomes node dependent, and is a realisation of an Inverse Gamma distribution with parameters $\beta_0$ and $\beta_1$. 
Essentially, an increase in $\varphi$ will mainly affect how prone the corresponding actor is to creating long-range connections, rather than short-range ones. 
This behaviour is in line with typical scenarios in large social networks,
where hubs differ from ordinary nodes in that 
they entail connections between distant areas (or communities) of the graph, decreasing the average path length \parencite{watts1998collective}.

We can approximate \eqref{thetaLPMRE} and \eqref{Gder2} and characterise the factorial moments 
of the degree of a random node as a function of the model parameters $\tau,\gamma,\beta_0,\beta_1$, 
allowing an assessment of the extent to which such models can represent heavy tails.
Since the value of $\tau$ makes no difference here, we fix it to $1$.

Table \ref{tables:avgdeg} shows that the variance of random effects
does not have much influence on the average degree of the network.  
This is relevant for studying heavy tails, since sparser networks 
will naturally have a higher skewness index.  Hence, if we keep the 
mean of the random effect constant and change the variance, 
not much of the change in the skewness index will be due to the network becoming sparser. 

\begin{table}[htb] 
\centering 
\vspace{0.4cm}
\begin{tabular}{l c c c c c c c} 
\toprule 
& \multicolumn{7}{c}{Variance} \\ 
\cmidrule(l){2-8} 
Mean& 0.0001 & 0.1 & 1 & 10 & 100 & 1000 & 100000\\ 
\midrule 
0.1 & $\phantom{0}1.95$ & $\phantom{0}2.88$ & $\phantom{0}2.73$ & $\phantom{0}2.91$ & $\phantom{0}2.85$ & $\phantom{0}2.81$ & $\phantom{0}2.83$ \\
0.2 & $\phantom{0}7.34$ & $\phantom{0}8.30$ & $\phantom{0}8.25$ & $\phantom{0}8.20$ & $\phantom{0}8.30$ & $\phantom{0}8.21$ & $\phantom{0}8.17$ \\
0.3 & $14.97$ & $15.19$ & $14.83$ & $14.35$ & $14.40$ & $14.33$ & $14.38$ \\
0.4 & $24.11$ & $23.28$ & $21.14$ & $20.49$ & $20.73$ & $20.60$ & $20.37$ \\
\bottomrule 
\end{tabular}
\caption{Average degree of a network of $100$ actors for different values of mean and variance of the nodal random effects. 
The variance has very little impact on the overall average degree of the network. This is an important property which is 
needed to state that any increase of skewness is not due to the network getting sparser.} 
\label{tables:avgdeg} 
\end{table}

\begin{figure}[htb]
\centering
\includegraphics[width=0.49\textwidth]{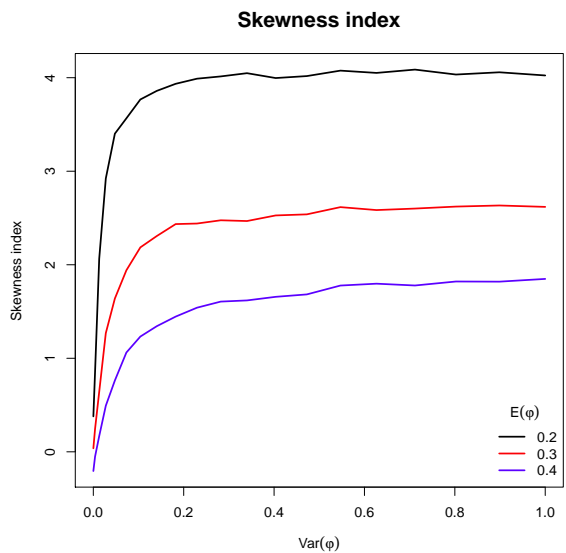}
\caption{Skewness index versus variance of nodal random effects. An increase in the variance of the random effects leads to an increase of the skewness index,
corresponding to heavier tails.}
\label{fig:avgdegskew}
\end{figure}

Figure \ref{fig:avgdegskew} shows that an increase in the variance of the 
random effects does yield an increase in the skewness index, 
corresponding to a right-skewed heavy-tailed shape.  
Therefore, these two results indicate that the heaviness of the tails 
can be controlled by changing the variance of the random effects, 
without changing the average degree of the network by much.  
The smallest skewness index is obtained with a null variance 
for random effects, which corresponds to the Gaussian LPM.

\begin{figure}[htb]
\centering
\includegraphics[width=0.42\textwidth]{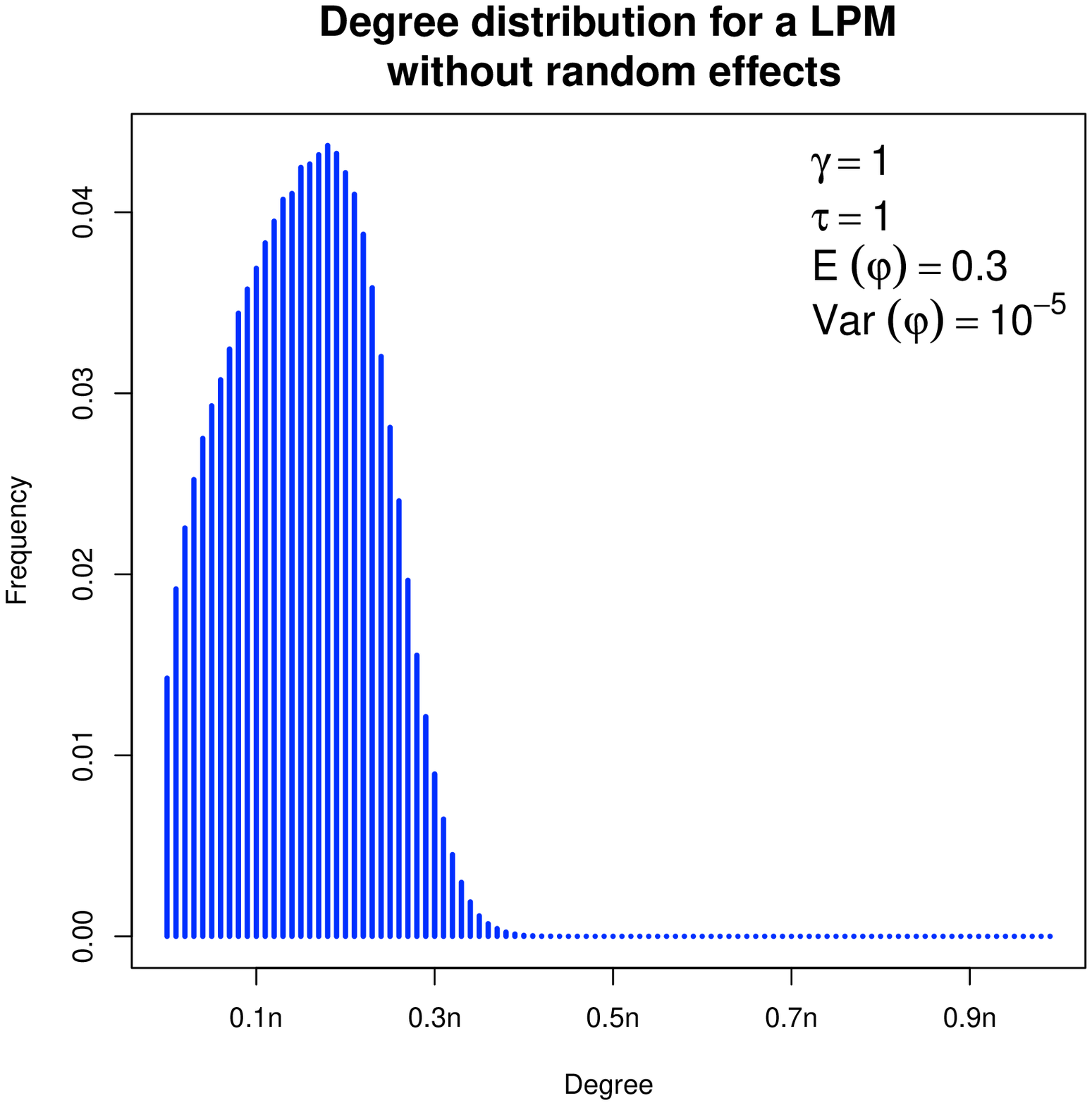}
\includegraphics[width=0.42\textwidth]{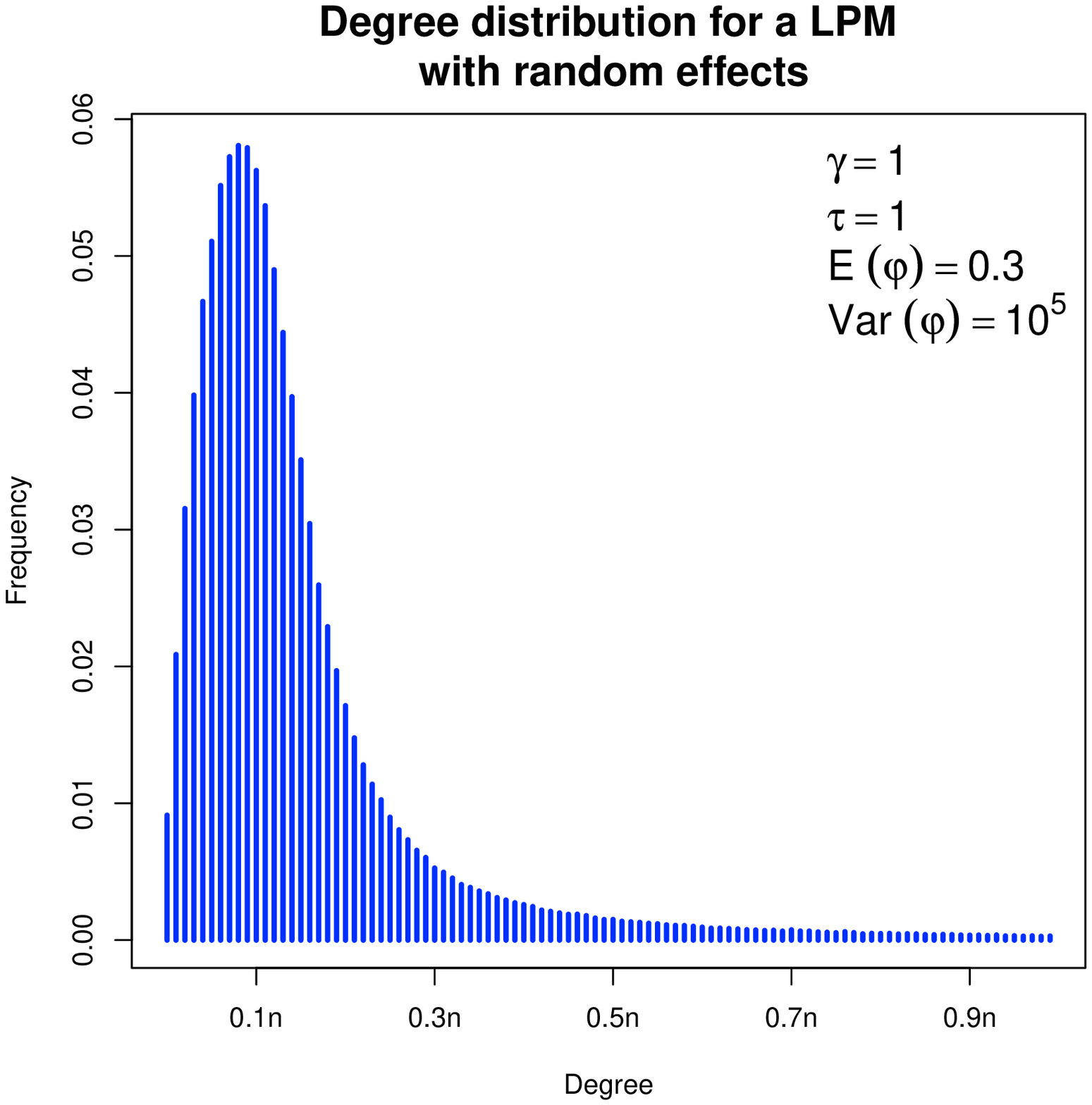}
\includegraphics[width=0.42\textwidth]{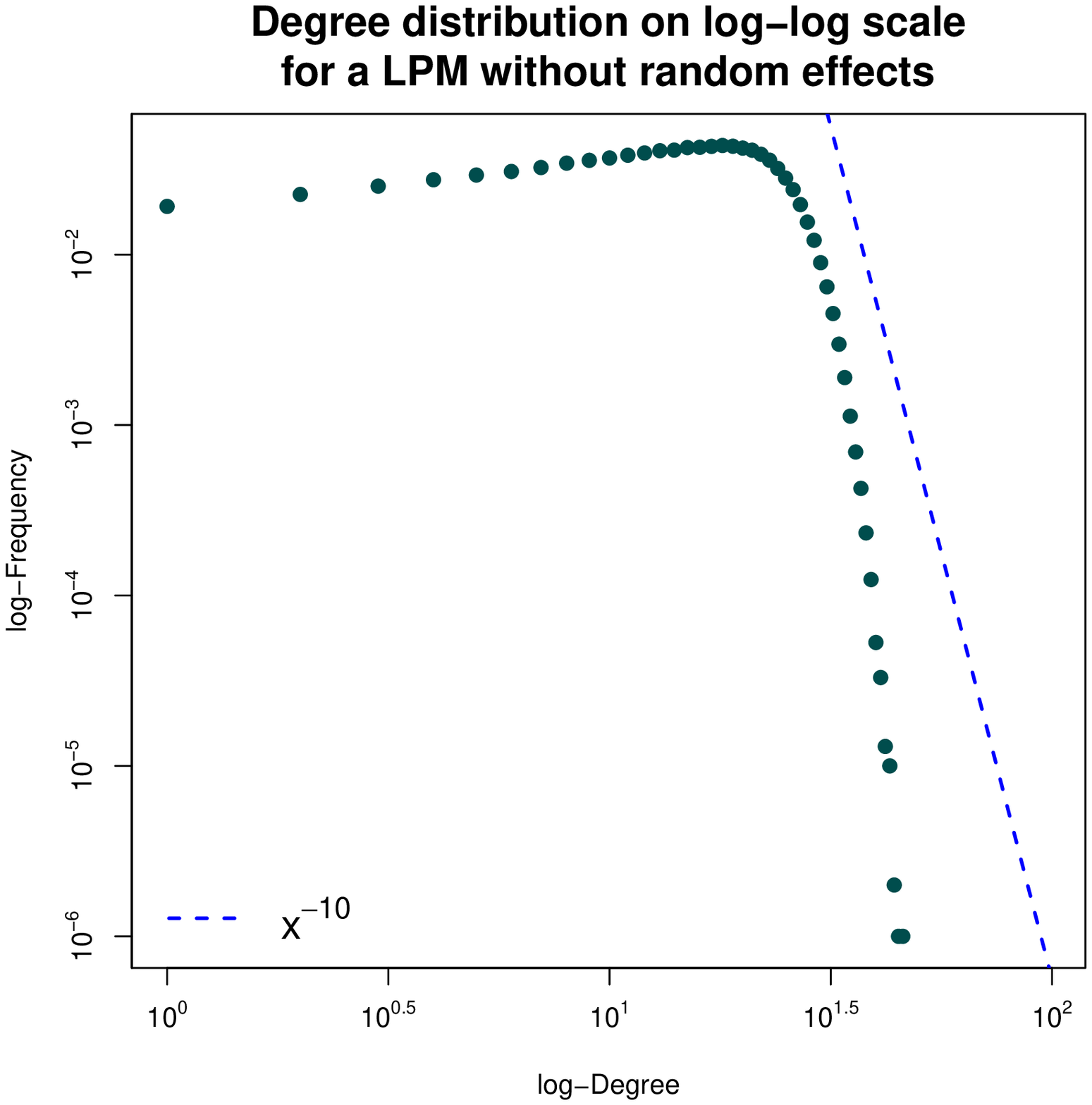}
\includegraphics[width=0.42\textwidth]{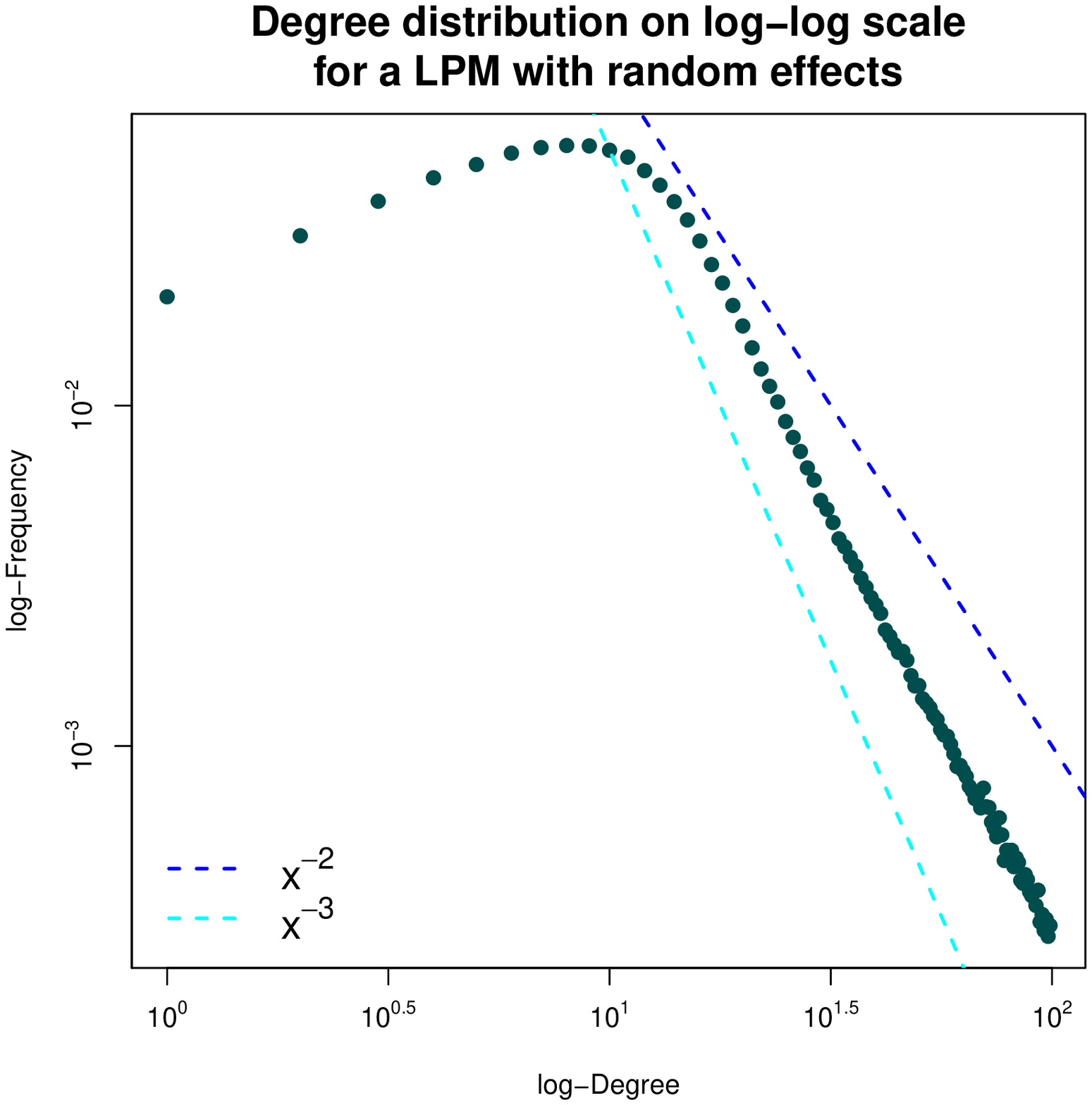}
\caption{\textbf{Top}: degree distributions for Gaussian LPMREs with null-variance random effects (\textbf{left}) and large-variance random effects (\textbf{right}).
\textbf{Bottom}: corresponding degree distribution on the log-log scale. An increase in the variance of the random effects results in a heavier power-law tailed degree 
distribution. The average degrees are: $0.151n$ and $0.144n$ for the case on left and right respectively, while skewness indexes are $-0.07$ and $2.53$ respectively.}
\label{fig:examp}
\end{figure}

But how heavy are the tails corresponding to a given positive skewness? 
Figure \ref{fig:examp} shows the empirical degree frequencies obtained through simulations of Gaussian LPMREs.
The two panels on the left side of Figure \ref{fig:examp} show
the degree distribution for a LPM (on both standard and log-log scale), where the variance of random effects is set to a very small value. 
The right-hand panels are obtained with
the same parameters, except for the variance of the random effect, which is increased to $10^5$. The average degrees for the two cases are: 
$0.151n$ and $0.144n$ respectively
and the skewness indexes are $-0.07$ and $2.53$ respectively.
The log-log scale plots are represented to show that the decay switches from a high-order power-law (reasonably comparable to a Poissonian tail) to a power-law with an exponent 
which falls between $2$ and $3$.

The results shown confirm that random effects can extend the family of networks represented using LPMs. However, other features of interest are non-trivially influenced.
Hence, we propose an empirical study to explore how random effects affect the asymptotic behaviour of LPMRE with respect to small-world behaviour and transitivity.
Simulations of LPMREs are very inefficient, so the results are rather limited. 
However, such a procedure is the only feasible one,
since theoretical results on the LPMRE are not available. In fact, we are currently investigating alternative ways to approach this analysis using more rigorous theoretical frameworks.

\begin{figure}[htb]
\centering
\includegraphics[width=0.42\textwidth]{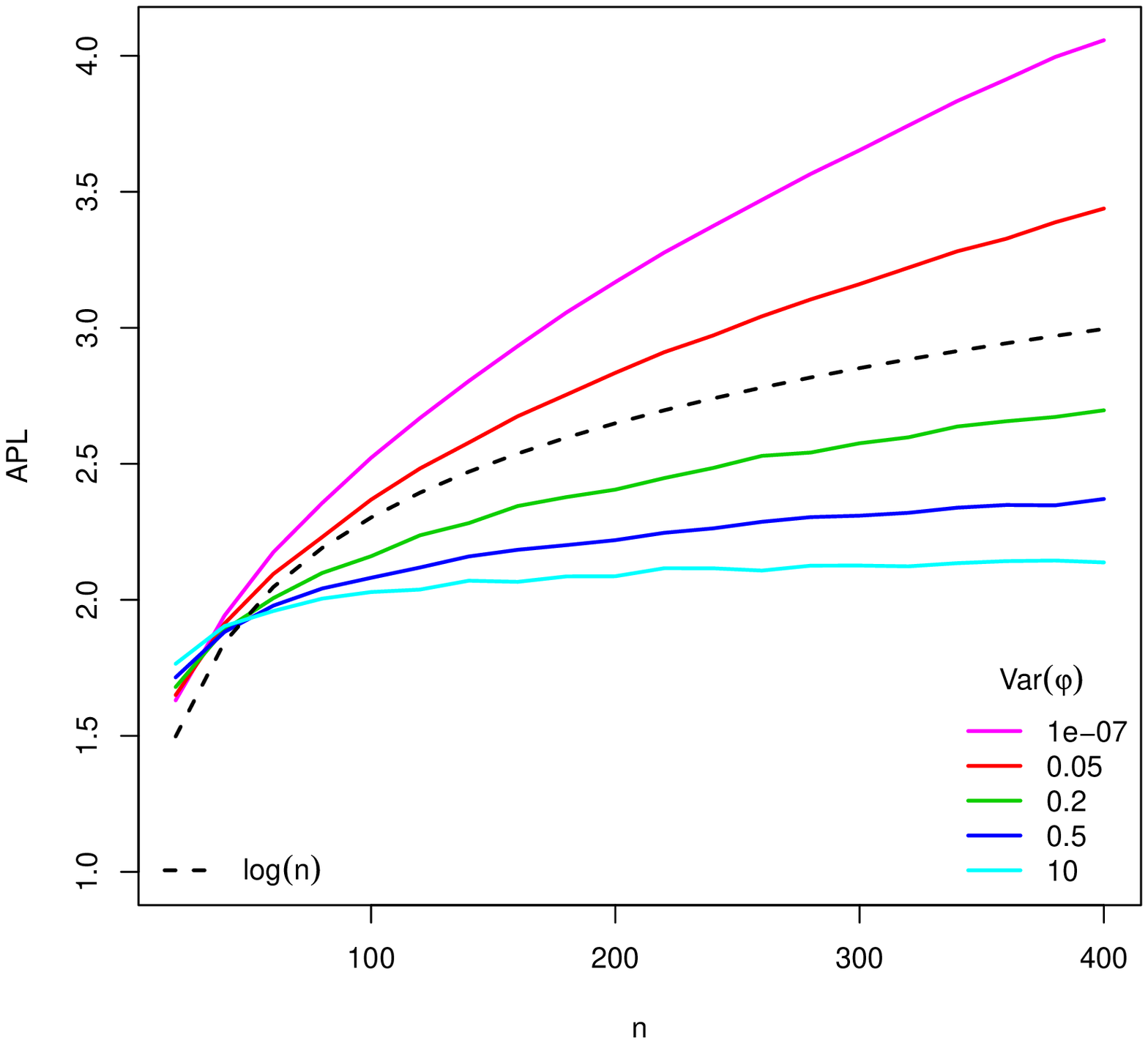}
\includegraphics[width=0.42\textwidth]{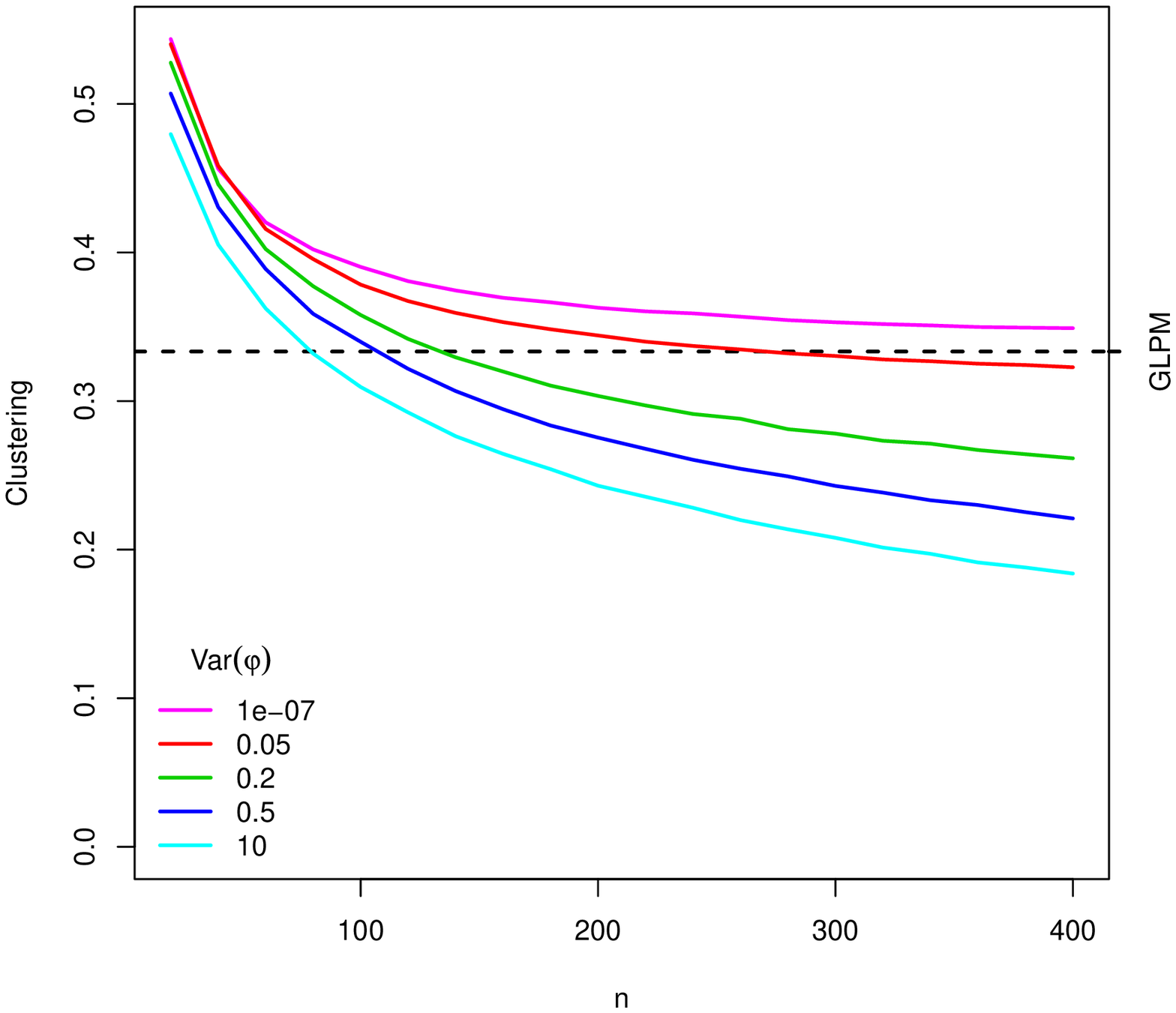}
\caption{APL (\textbf{left}) and clustering coefficient (\textbf{right}) as a function of $n$, holding an approximately constant average degree. The remaining model parameters are 
$\tau=1$, $\mathbb{E}[\varphi]=0.6$ and $\gamma=0.05(n-1)$. The number of networks generated for each value of $n$ is $1000$. The dashed black lines represent the log function 
and the asymptotic value for $\mathcal{C}$ under the Gaussian LPM for the left and right panel respectively.}
\label{fig:LPMREemp}
\end{figure}

In this experiment, we have selected a particular set of model parameters, generated a sequence of IID networks and studied the average features exhibited. Since we are interested in the 
asymptotic behaviour of APL and $\mathcal{C}$, we have held the average degree approximately constant by imposing $\gamma\propto n$, with $n$ increasing.
Figure \ref{fig:LPMREemp} illustrates the results. 
The left panel shows that an increase in the variance of the random effects results in a smaller APL. Furthermore, the APL growth as a
function of $n$ becomes slower than the log function, exhibiting the small-world behaviour. 
The right panel represents instead the empirical asymptotic clustering coefficient. 
Here, it appears that $\mathcal{C}$ tends to stabilise to a non-zero limiting value, which clearly depends on the variance of the
random effects. Such interaction between the presence of hubs and the clustering coefficient could be somehow expected, since for an extreme case, 
the $n$-nodes star, $\mathcal{C}$ is equal to zero.

Considering the results shown in this Section, random effects can be regarded as a useful addition to LPMs to capture several important features that arise in large social networks.

\section{Real data examples}\label{sec:Real data examples}
We have characterised the models introduced by showing how some 
important statistics of realised networks depend on the parameters of LPMs. 
We now show that several well known real social networks have statistics 
that can be well captured by a fitted LPM, using the following datasets:
\begin{itemize}[noitemsep] 
 \item \textbf{Dolphins}: This is a social network of frequent associations between $62$ dolphins in a community living off Doubtful Sound, New Zealand \parencite{lusseau2003bottlenose}.
 \item \textbf{Monks}: This describes the interpersonal relations among $18$ monks in a monastery \parencite{sampson1968novitiate}. 
 \item \textbf{Florentine}: This describes the connections by marriage between $16$ noble families in Florence during Renaissance \parencite{padgett1994marriage}.
 \item \textbf{Prison}: Data collected in the 1950s by John Gagnon from $67$ prison inmates, each one being asked to specify his preferences among other participants \parencite{macrae1960direct}.
 \item \textbf{High-tech}: This network contains the friendship ties among $36$ employees of a hi-tech company, which were gathered by means of the question: who do you consider to be a personal friend? \parencite{krackhardt1999ties}.
 \item \textbf{Math method}: $38$ school superintendents were asked to indicate their friendship ties with other superintendents in the county with the following question: among the chief school administrators in Allegheny County (PA, USA), 
 who are your three best friends? \parencite{carlson1965adoption}.
 \item \textbf{Sawmill}: $36$ employees of a sawmill were asked to quantify the time they spent discussing work matters with each of their colleagues \parencite{michael1997modeling}.
 \item \textbf{San Juan}: Study carried out in a rural area in Costa Rica. Edges represent visiting frequencies between $75$ families living in farms in a neighbourhood called San Juan Sur \parencite{de2011exploratory}.
 \item \textbf{Network sciences} ($1589$ nodes): Coauthorship network of scientists working on network theory and experiment \parencite{newman2006finding}.
 \item \textbf{Geometry} ($7343$ nodes): Coauthorship network of scientists working on computational geometry \parencite{jonescomputational}.
 \item \textbf{Condensed Matter} ($16726$ nodes): Coauthorships between scientists posting preprints on the Condensed Matter E-Print Archive \parencite{newman2001structure}.
 \item \textbf{High energy} ($27770$ nodes): Coauthorships between scientists posting preprints on the High-Energy Theory E-Print Archive \parencite{newman2001structure}.
\end{itemize}
Where necessary, the datasets have been transformed into binary undirected (no self-edges) graphs, using standard reasonable procedures. 

We can obtain the following network statistics for the Gaussian LPM using
Theorem \ref{thm1}:
the average degree $\bar{k}$, the clustering coefficient $\mathcal{C}$, 
the average path length APL and the skewness index $S$. 
Table \ref{table:realexamples1} shows their observed and 
theoretical values for the smaller datasets.

\begin{table}[htb]
\footnotesize
\begin{center}

\begin{tabular}{| c | c |}
    \hline
    \multicolumn{2}{|c|}{\textbf{Parameters}} \\ \hline
    $\tau$ & 0.810\\ \hline
    $\varphi/\gamma$ & 0.232\\ \hline
  \end{tabular}
  \begin{tabular}{|p{4.6cm}|p{1cm}|p{1cm}|p{1cm}|p{1cm}|}
    \hline
    \textbf{Dolphins (n=62)} & $\bar{k}$ &$\mathcal{C}$& S & APL \\ \hline
    Observed & 5.129 & 0.309 & 0.292 & 3.357 \\ \hline
    Theoretical & 5.129 & 0.309 & 0.461 & 3.282 \\ \hline
  \end{tabular}
\vspace{0.2cm}

\begin{tabular}{| c | c |}
    \hline
    \multicolumn{2}{|c|}{\textbf{Parameters}} \\ \hline
    $\tau$ & 0.763\\ \hline
    $\varphi/\gamma$ & 2.115\\ \hline
  \end{tabular}
  \begin{tabular}{|p{4.6cm}|p{1cm}|p{1cm}|p{1cm}|p{1cm}|}
    \hline
    \textbf{Monks (n=18)} & $\bar{k}$ &$\mathcal{C}$& S & APL \\ \hline
    Observed & 6.667 & 0.465 & 0.877 & 1.68 \\ \hline
    Theoretical & 6.667 & 0.465 & -0.05 & 1.724 \\ \hline
  \end{tabular}
\vspace{0.2cm}

\begin{tabular}{| c | c |}
    \hline
    \multicolumn{2}{|c|}{\textbf{Parameters}} \\ \hline
    $\tau$ & 0.302\\ \hline
    $\varphi/\gamma$ & 2.460\\ \hline
  \end{tabular}
  \begin{tabular}{|p{4.6cm}|p{1cm}|p{1cm}|p{1cm}|p{1cm}|}
    \hline
    \textbf{Florentine (n=16)} & $\bar{k}$ &$\mathcal{C}$& S & APL \\ \hline
    Observed & 2.5 & 0.191 & 0.424 & 2.486 \\ \hline
    Theoretical & 2.5 & 0.191 & 0.503 & 2.827 \\ \hline
\end{tabular}
\vspace{0.2cm}

\begin{tabular}{| c | c |}
    \hline
    \multicolumn{2}{|c|}{\textbf{Parameters}} \\ \hline
    $\tau$ & 0.776\\ \hline
    $\varphi/\gamma$ & 0.180\\ \hline
  \end{tabular}
  \begin{tabular}{|p{4.6cm}|p{1cm}|p{1cm}|p{1cm}|p{1cm}|}
    \hline
    \textbf{Prison (n=67)} & $\bar{k}$ &$\mathcal{C}$& S & APL \\ \hline
    Observed & 4.239 & 0.288 & 0.855 & 3.355 \\ \hline
    Theoretical & 4.239 & 0.288 & 0.562 & 3.831 \\ \hline
\end{tabular}
\vspace{0.2cm}

\begin{tabular}{| c | c |}
    \hline
    \multicolumn{2}{|c|}{\textbf{Parameters}} \\ \hline
    $\tau$ & 0.913\\ \hline
    $\varphi/\gamma$ & 0.376\\ \hline
  \end{tabular}
  \begin{tabular}{|p{4.6cm}|p{1cm}|p{1cm}|p{1cm}|p{1cm}|}
    \hline
    \textbf{High-tech (n=36)} & $\bar{k}$ &$\mathcal{C}$& S & APL \\ \hline
    Observed & 5.056 & 0.372 & 0.785 & 2.360 \\ \hline
    Theoretical & 5.056 & 0.372 & 0.376 & 2.749 \\ \hline
\end{tabular}
\vspace{0.2cm}

\begin{tabular}{| c | c |}
    \hline
    \multicolumn{2}{|c|}{\textbf{Parameters}} \\ \hline
    $\tau$ & 0.616\\ \hline
    $\varphi/\gamma$ & 0.328\\ \hline
  \end{tabular}
  \begin{tabular}{|p{4.6cm}|p{1cm}|p{1cm}|p{1cm}|p{1cm}|}
    \hline
    \textbf{Math method (n=38)} & $\bar{k}$ &$\mathcal{C}$& S & APL \\ \hline
    Observed & 3.211 & 0.246 & 0.654 & 2.644 \\ \hline
    Theoretical & 3.211 & 0.246 & 0.612 & 3.480 \\ \hline
\end{tabular}
\vspace{0.2cm}

\begin{tabular}{| c | c |}
    \hline
    \multicolumn{2}{|c|}{\textbf{Parameters}} \\ \hline
    $\tau$ & 0.550\\ \hline
    $\varphi/\gamma$ & 0.436\\ \hline
  \end{tabular}
  \begin{tabular}{|p{4.6cm}|p{1cm}|p{1cm}|p{1cm}|p{1cm}|}
    \hline
    \textbf{Sawmill (n=36)} & $\bar{k}$ &$\mathcal{C}$& S & APL \\ \hline
    Observed & 3.444 & 0.230 & 2.290 & 3.138 \\ \hline
    Theoretical & 3.444 & 0.230 & 0.558 & 3.210 \\ \hline
\end{tabular}
\vspace{0.2cm}

\begin{tabular}{| c | c |}
    \hline
    \multicolumn{2}{|c|}{\textbf{Parameters}} \\ \hline
    $\tau$ & 0.657\\ \hline
    $\varphi/\gamma$ & 0.186\\ \hline
  \end{tabular}
  \begin{tabular}{|p{4.6cm}|p{1cm}|p{1cm}|p{1cm}|p{1cm}|}
    \hline
    \textbf{San Juan (n=75)} & $\bar{k}$ &$\mathcal{C}$& S & APL \\ \hline
    Observed & 4.133 & 0.245 & 1.622 & 3.485 \\ \hline
    Theoretical & 4.133 & 0.245 & 0.579 & 3.883 \\ \hline
\end{tabular}
\vspace{0.2cm}

\caption{Theoretical and observed statistics for small-sized social networks. Statistics shown are the average degree $\bar{k}$, the clustering coefficient $\mathcal{C}$, 
the average path length APL and the skewness index $S$. Following the criterion described, the average degree and the clustering coefficient are matched exactly in every case, while
the corresponding skewness index and average path length are fairly close to the observed counterparts.}
\label{table:realexamples1}
\end{center}
\end{table}

\begin{figure}[htb]
\centering
\includegraphics[width=0.244\textwidth]{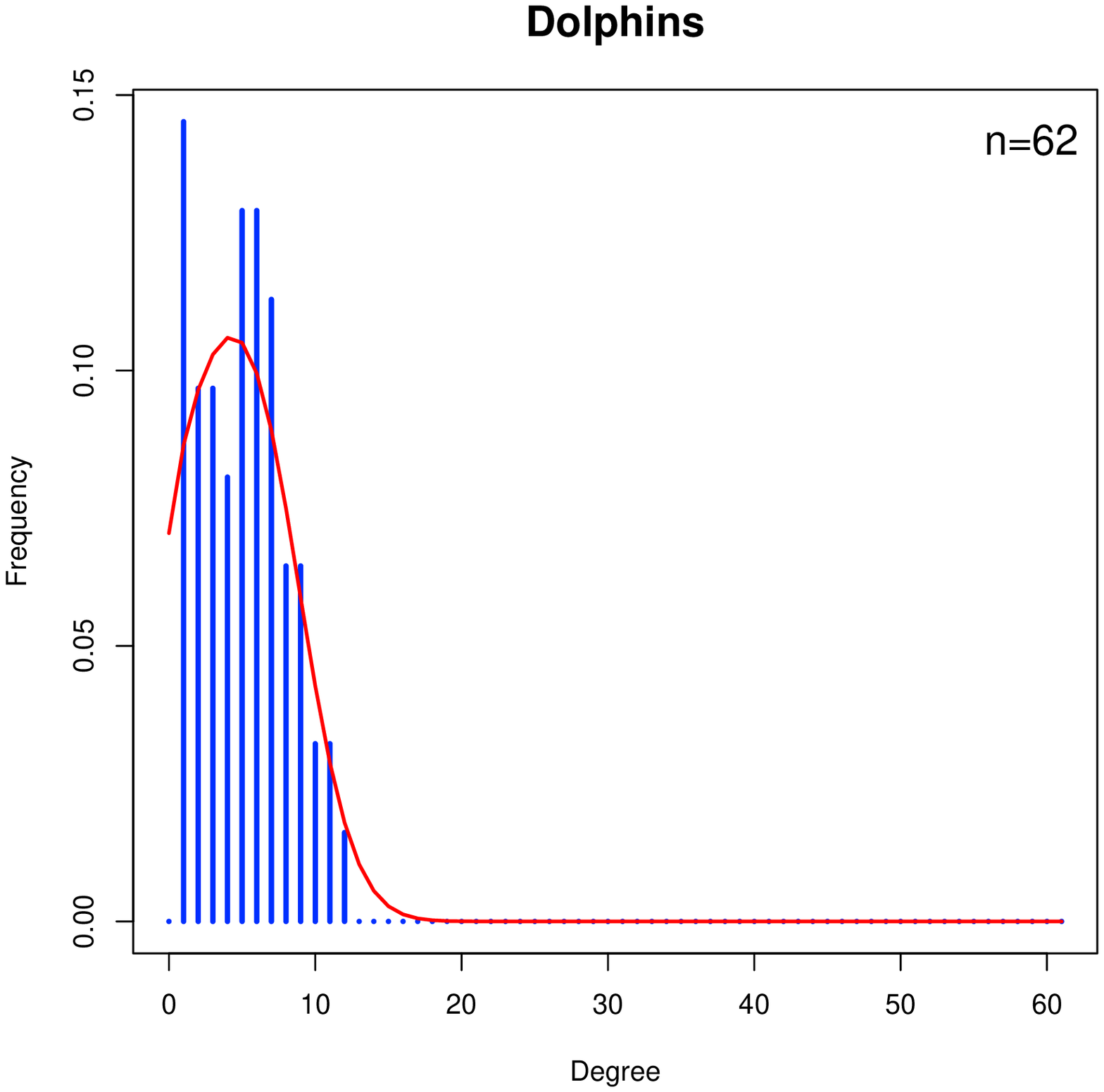}
\includegraphics[width=0.244\textwidth]{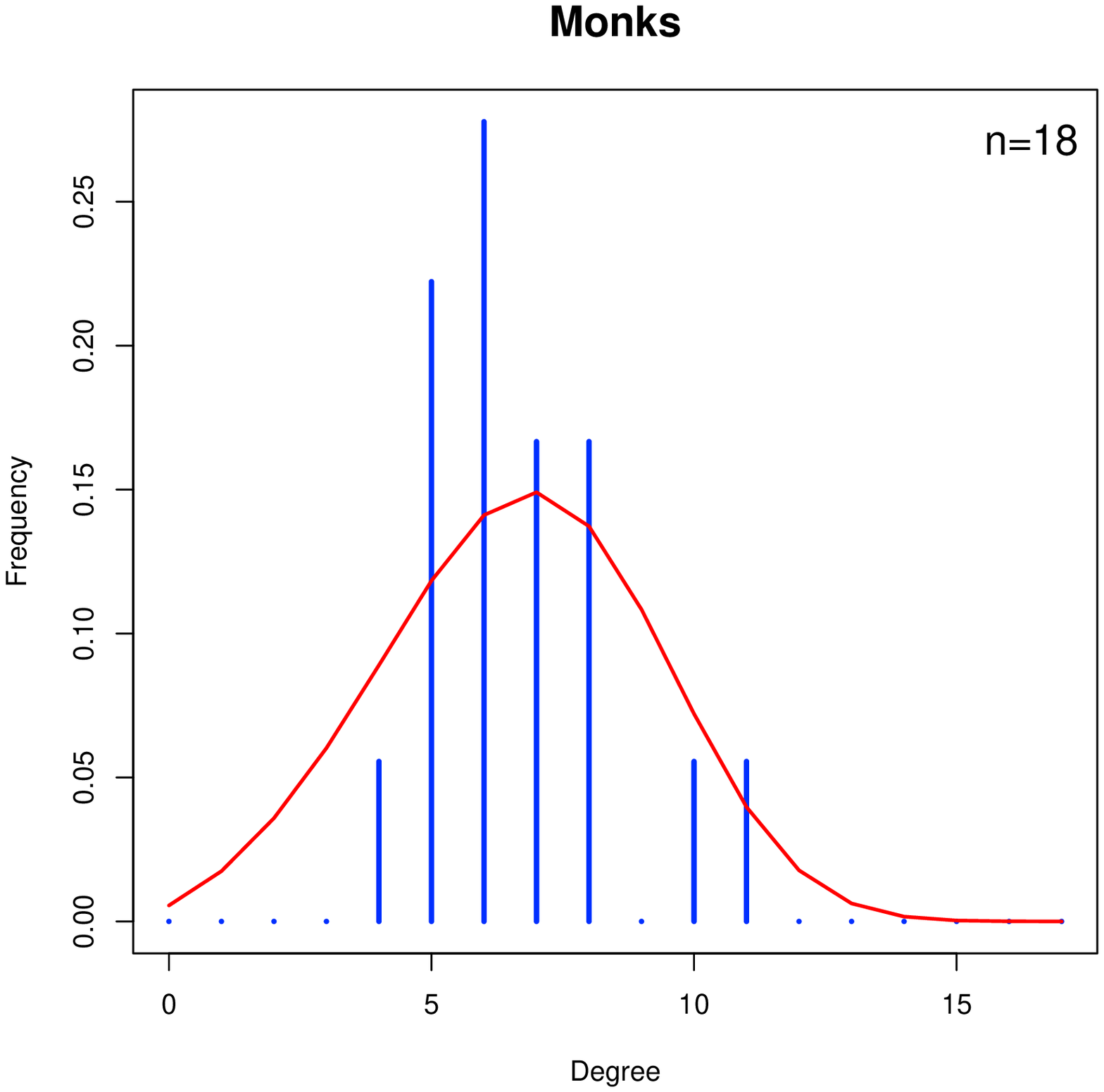}
\includegraphics[width=0.244\textwidth]{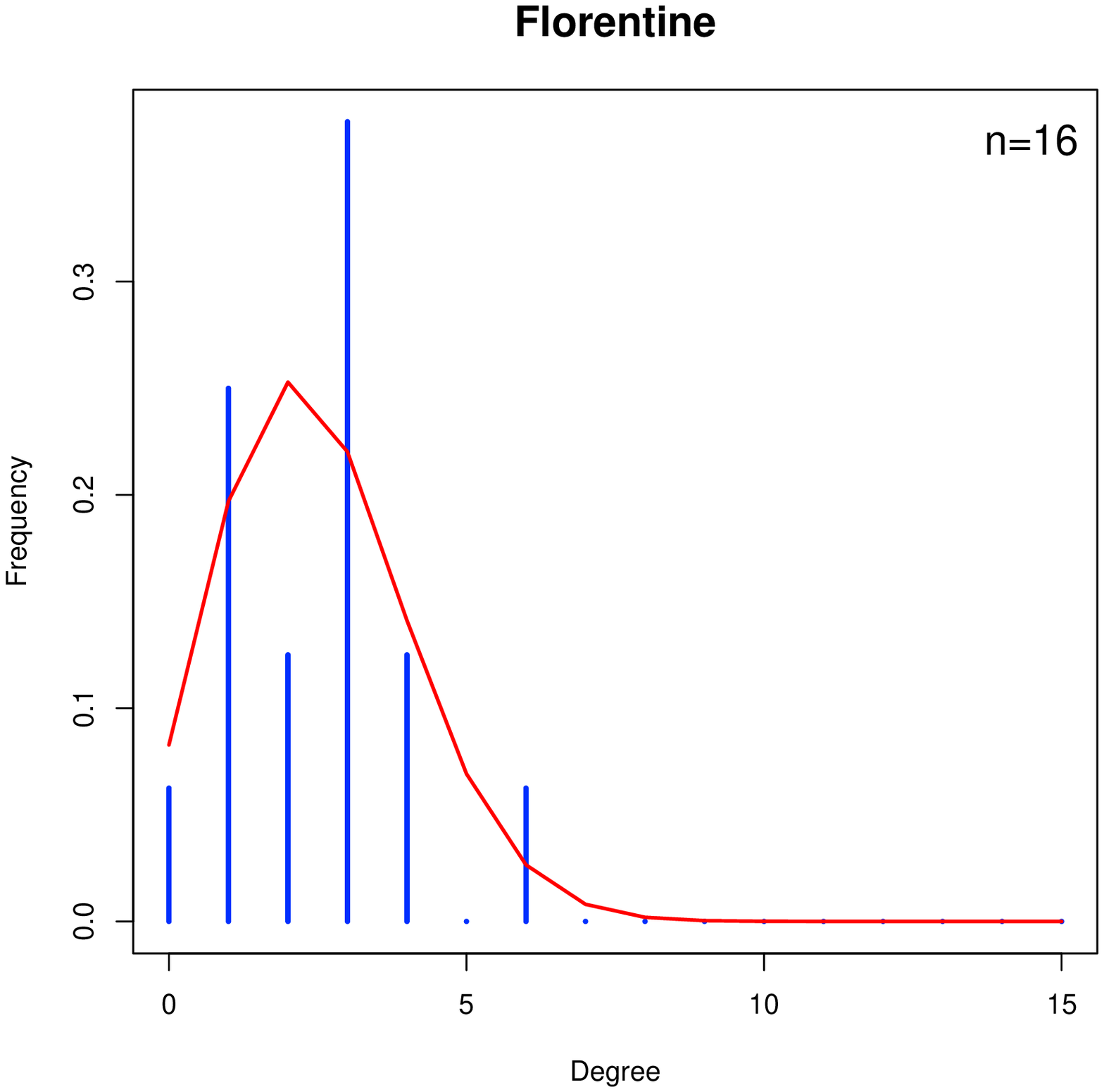}
\includegraphics[width=0.244\textwidth]{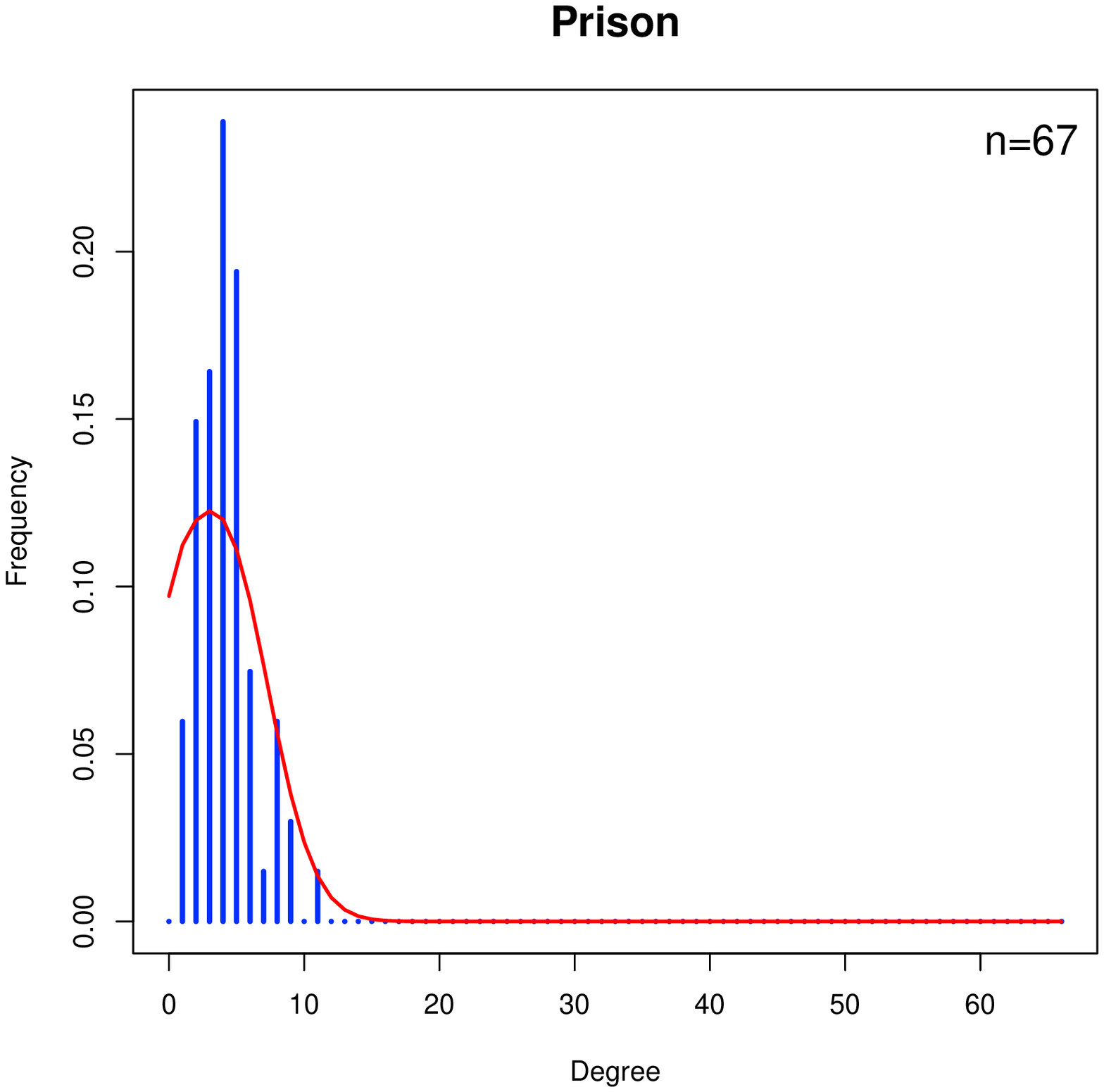}
\includegraphics[width=0.244\textwidth]{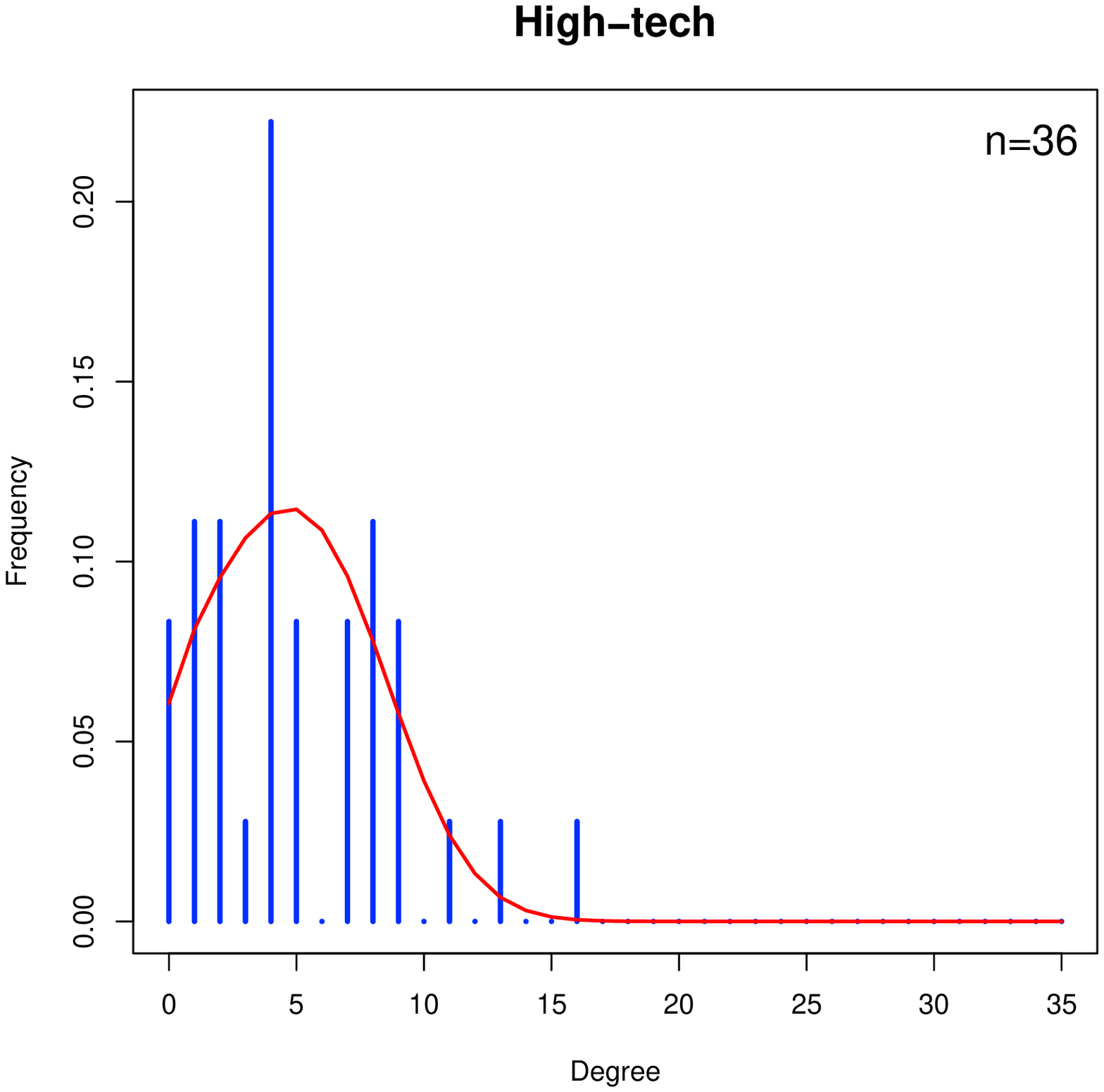}
\includegraphics[width=0.244\textwidth]{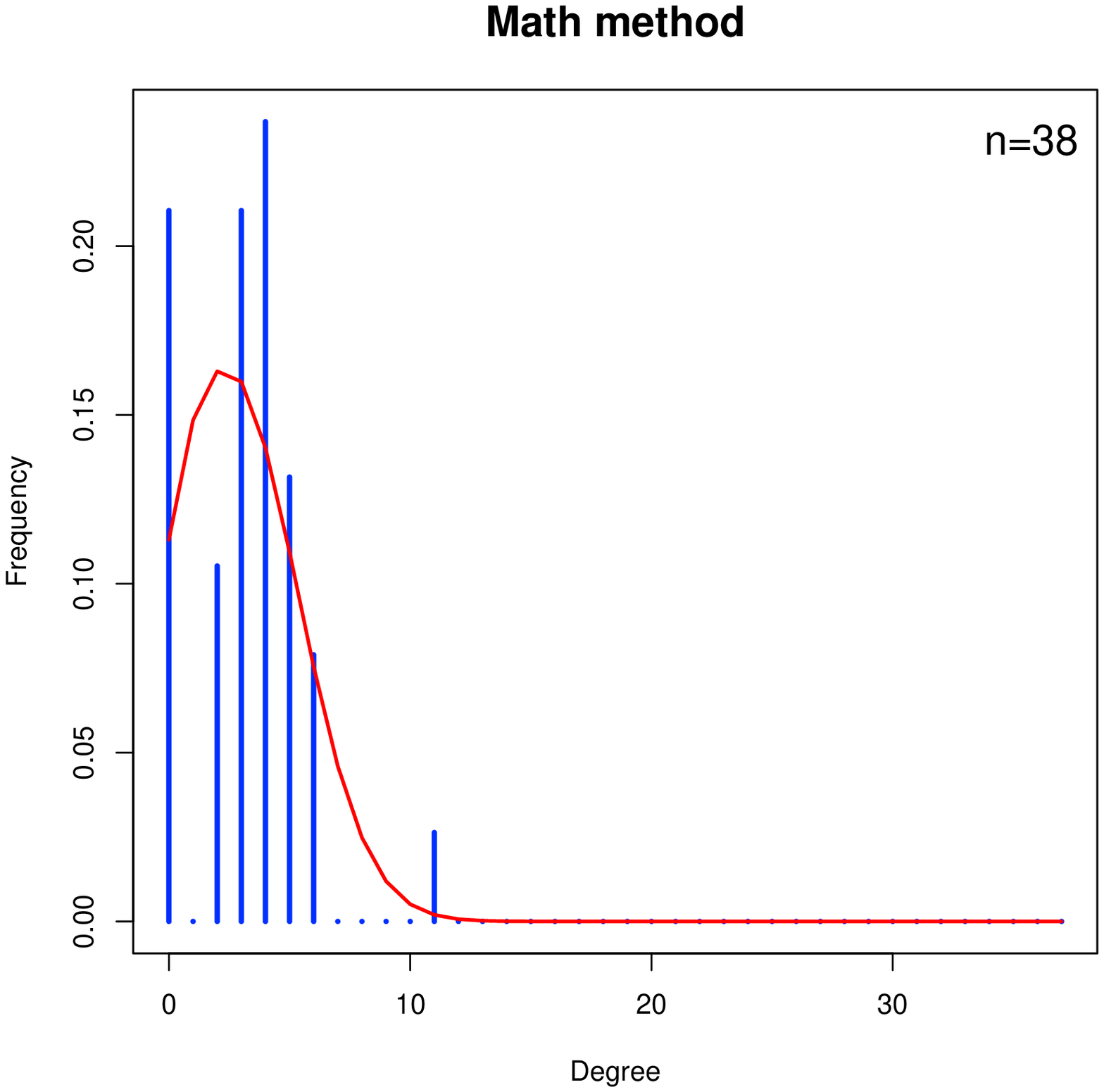}
\includegraphics[width=0.244\textwidth]{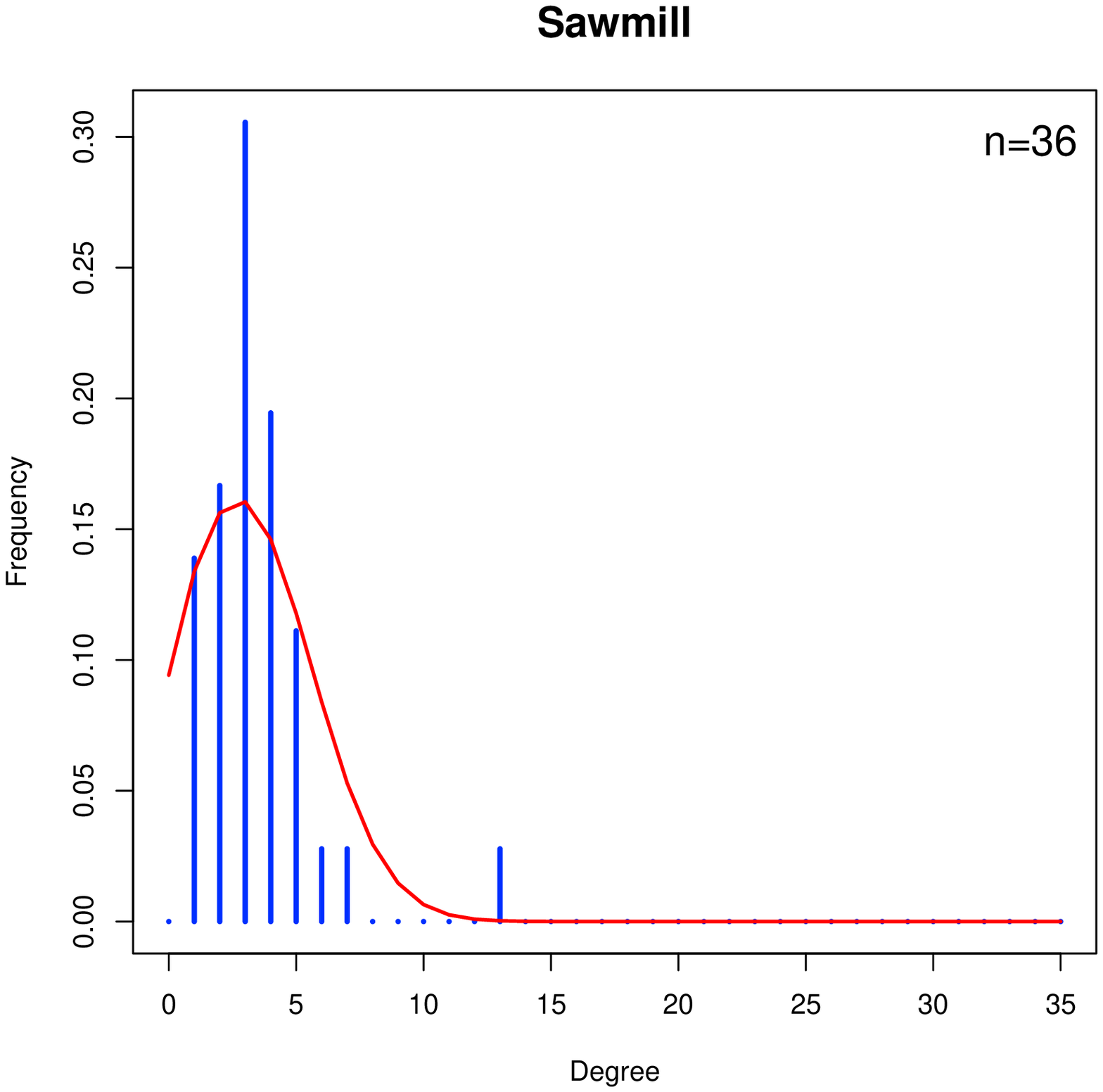}
\includegraphics[width=0.244\textwidth]{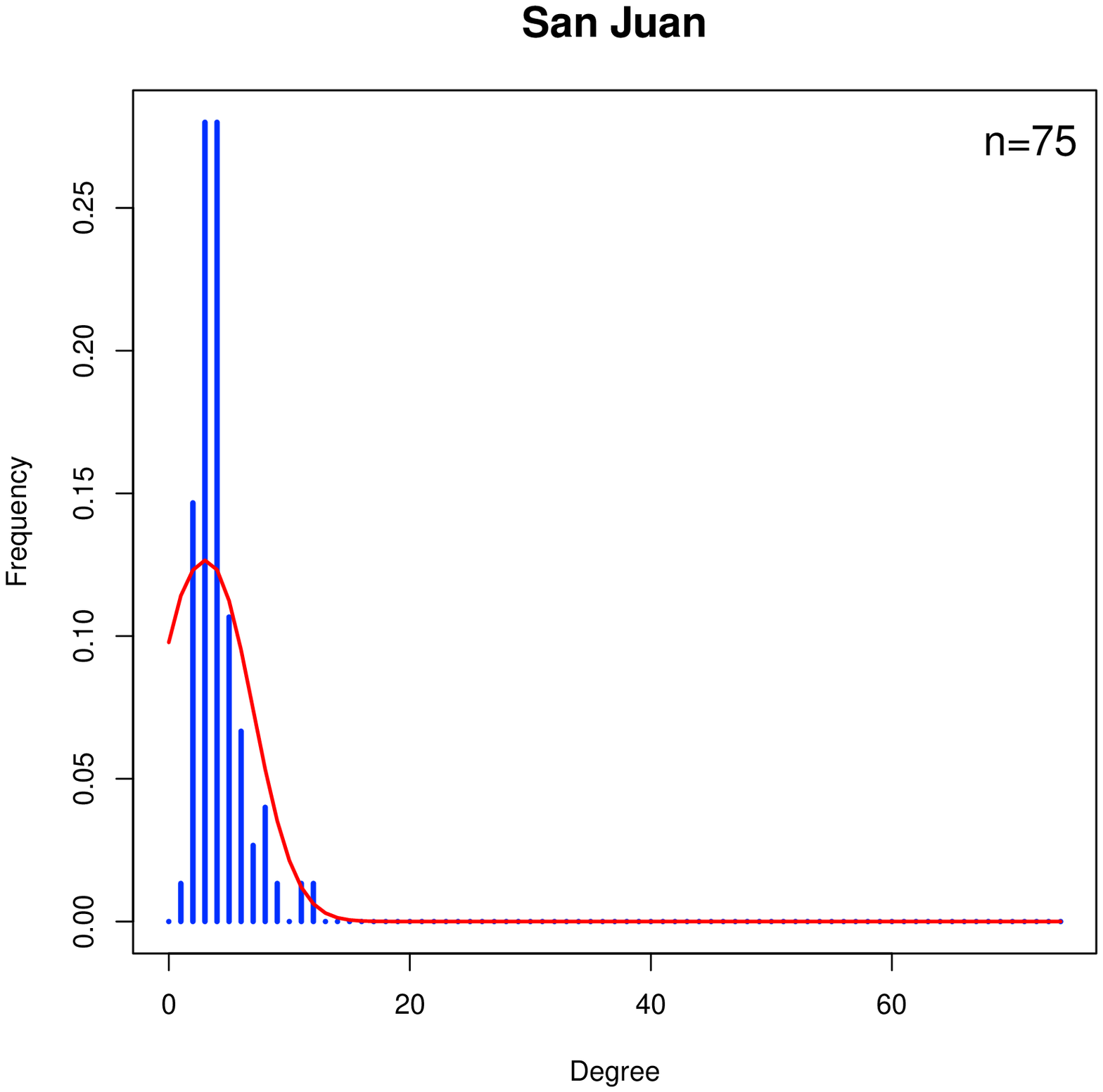}
\caption{Comparison between the observed degree distributions (blue bars) and the theoretical ones (red lines) for several small-size real social networks. Datasets used (from top left by row): Dolphins, Monks, Florentine, Prison, High-tech,
Math method, Sawmill, San Juan.}
 \label{fig:realexamples2}
\end{figure}

The theoretical values shown in Table \ref{table:realexamples1} 
correspond to model parameters chosen to match the
observed with the theoretical $\bar{k}$ and $\mathcal{C}$.
This simple criterion performs well for the networks presented, as indicated by Figure \ref{fig:realexamples2}, which shows theoretical and observed degree distributions.

\begin{figure}[htb]
\centering
\includegraphics[width=0.42\textwidth]{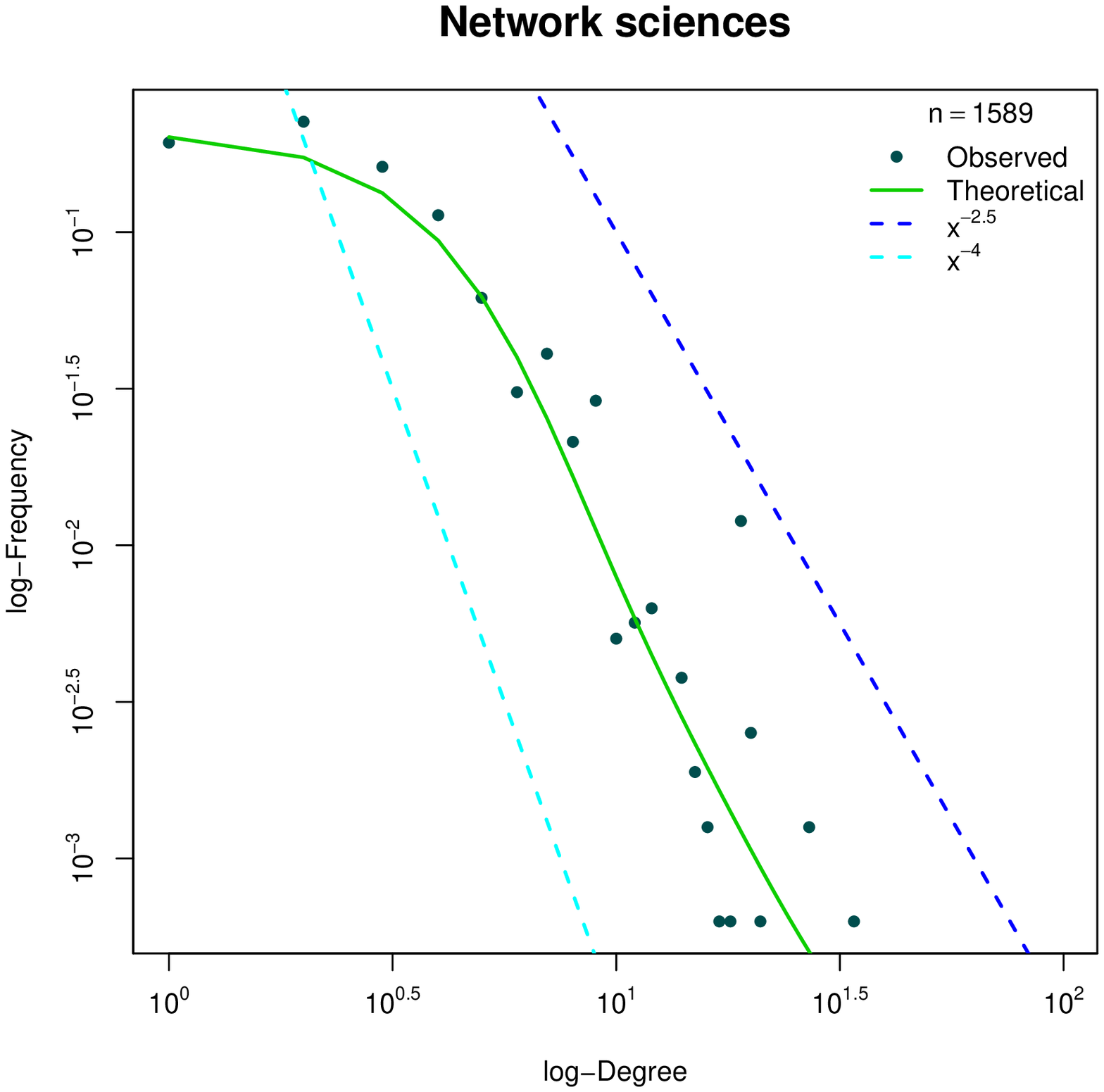}
\includegraphics[width=0.42\textwidth]{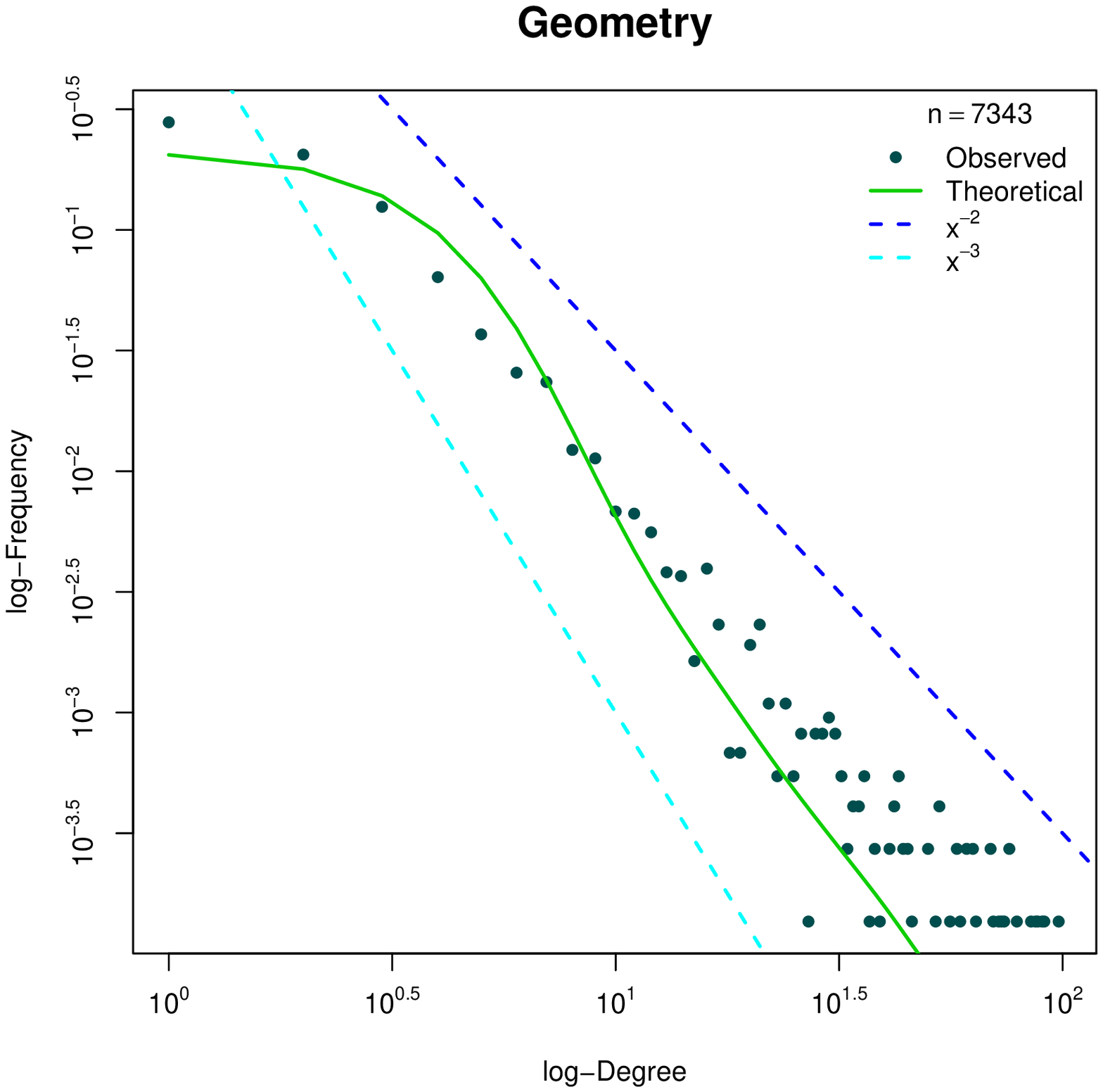}
\includegraphics[width=0.42\textwidth]{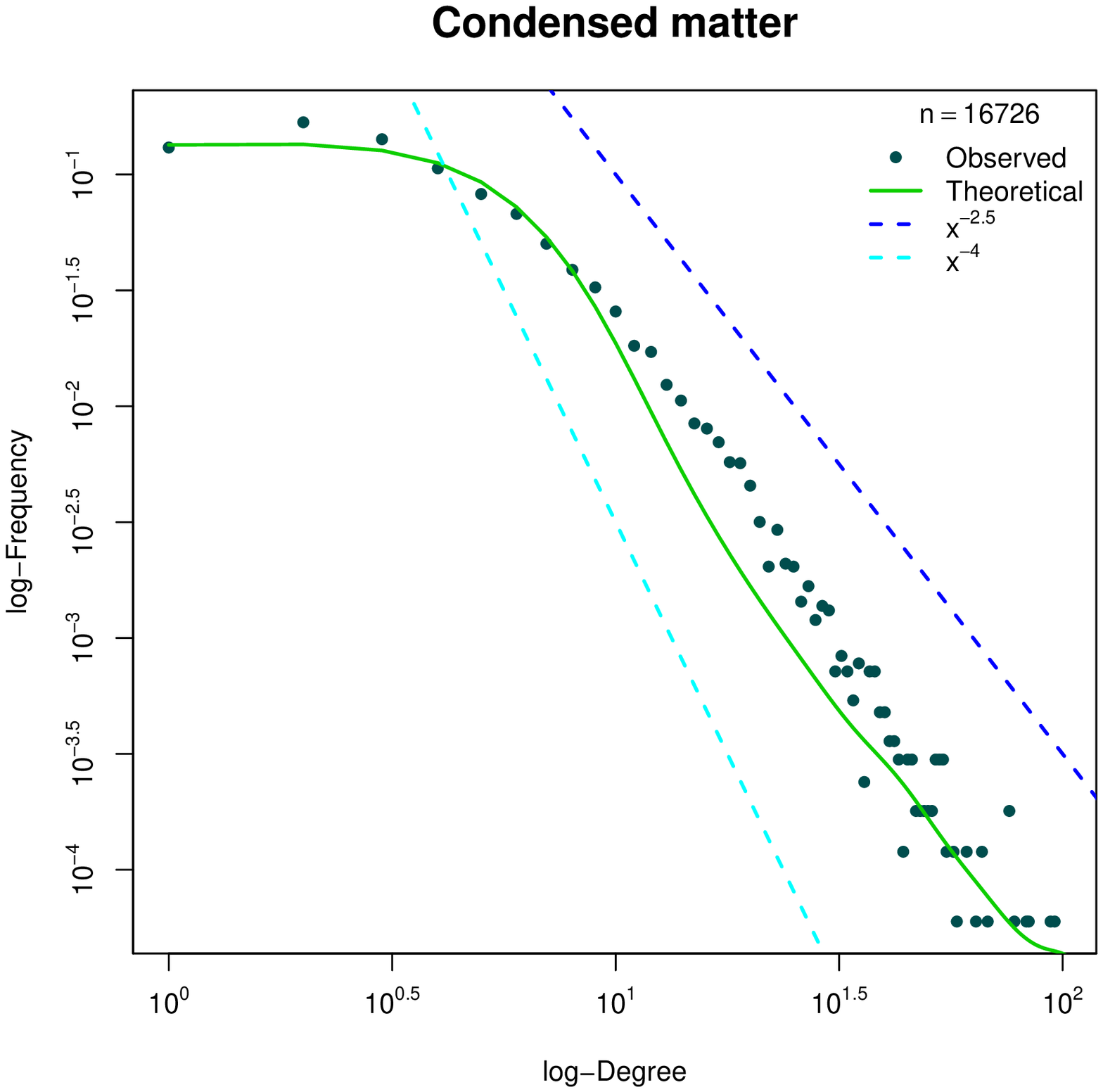}
\includegraphics[width=0.42\textwidth]{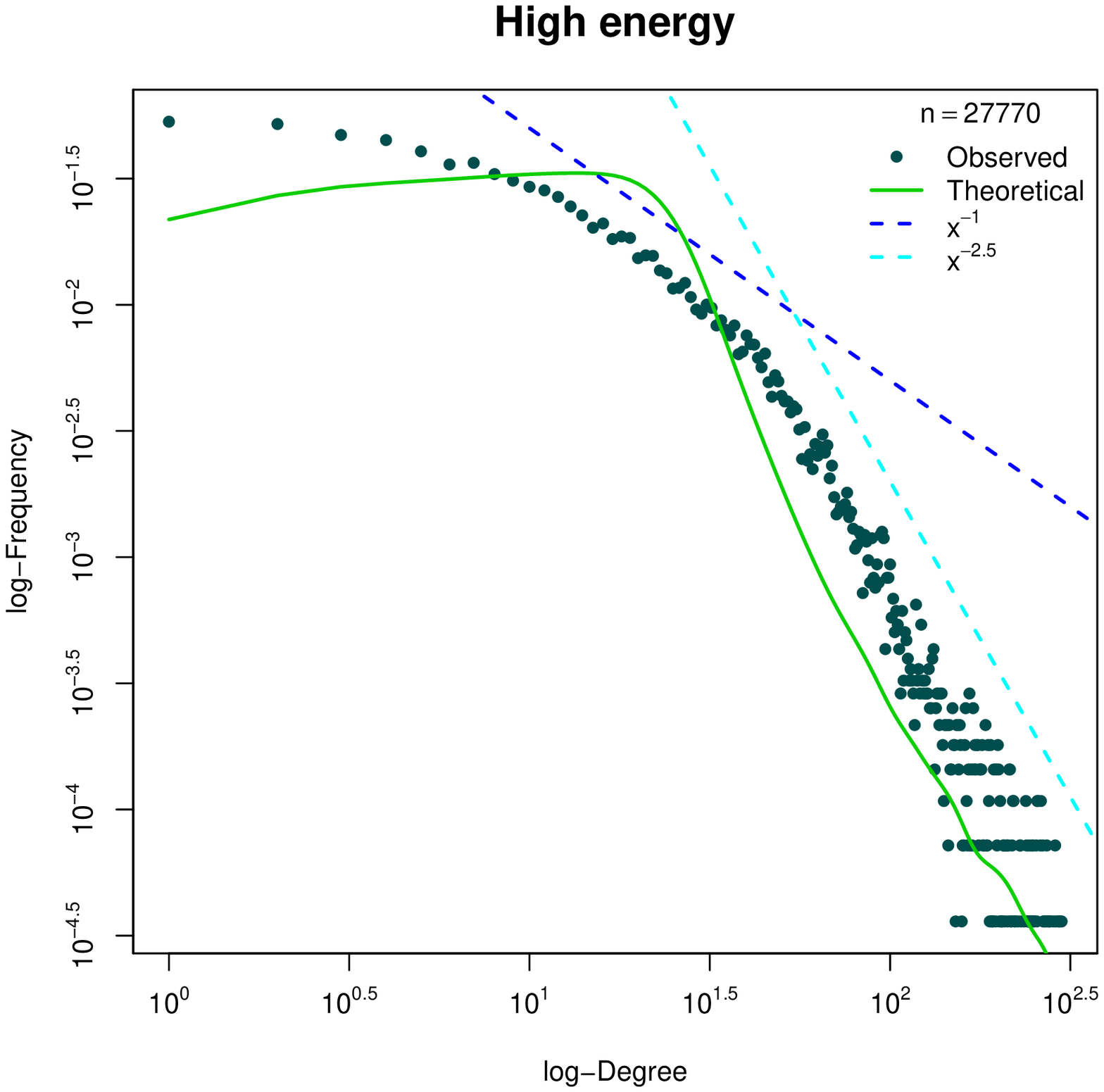}
\caption{Empirical (blue dots) and theoretical (green line) degree distributions on log-log scale for various large citation networks. 
The datasets exhibit different asymptotic power-law orders. Gaussian LPMREs reasonably represent the asymptotic tendency of the degree distributions in every case. 
Datasets used: Network sciences (top left), Geometry (top right), Condensed matter (bottom left), High energy (bottom right).}
 \label{fig:realexamples3}
\end{figure}

A slightly different study was carried out for the larger datasets,
to assess to what extent the Gaussian LPMRE
can represent the asymptotic scale-free
decay of the degree distribution, for different orders of the power-law. 
We consider several collaboration networks where nodes correspond to 
authors and two nodes are linked if the corresponding scientists published 
a paper as coauthors. 
All the networks shown exhibit a power-law degree distribution, with different slopes, which vary in the range $1$ to $4$. 
Figure \ref{fig:realexamples3} shows the theoretical and observed degree distributions on the log-log scale, indicating that the asymptotic behaviour is 
reasonably well represented in all the cases.

\section{Conclusions}\label{sec:Conclusions}
The main contribution of this paper is to advance our 
understanding of Latent Position Models for networks by  
providing several probabilistic results.
Our main results describe features of realised Latent Position networks, characterising their degree distribution, the mixing properties of the degrees, the 
clustering coefficient and the path lengths' distribution.
Although this work deals only with undirected graphs, the same results can be extended in a similar fashion to directed ones.

Gaussian LPMs have been shown not to be appropriate for modelling
scale-free networks, since the average degree frequencies exhibit a left-skewed and truncated shape. 
However, modifying the basic LPM to include nodal random effects resulted in the ability of the model to represent power-law degree distributions of different slopes in both simulated and real networks. 

It has been also shown that Gaussian LPMs have an asymptotically strictly positive clustering coefficient, in contrast to other well known models, such as 
Erd\H{o}s-R\'enyi and Exponential Random Graph models, whose clustering
coefficient is asymptotically zero.
This result suggests that LPMs can generate highly clustered networks and that 
they can capture the persistent clustering behaviour of large social networks. 

The average degree of the closest neighbours to a node has been characterised, 
showing that positive degree correlations arise in LPM networks. 
This is in line with observed social networks, 
where assortative mixing in the nodal degrees frequently occurs.

It has also been shown how the distribution of geodesic distances can be efficiently approximated, yielding an analysis of the asymptotic behaviour of the average path length. 
It appears that dense LPM networks have the same behaviour of Erd\H{o}s-R\'enyi random graphs, while sparser LPM networks do not exhibit the small-world effect. 

Through simulations, important advantages of using nodal random effects have been outlined, suggesting that the Gaussian LPMRE has properties that makes it suitable for modelling
large social networks. An important extension of this work would be to develop new strategies to study analytically the LPMRE and LPCM.


\section*{Acknowledgements}\label{sec:Acknowledgements}
The Insight Centre for Data Analytics is supported by Science Foundation Ireland under Grant Number SFI/12/RC/2289.
Nial Friel and Riccardo Rastelli's research was also supported by a Science Foundation Ireland grant: 12/IP/1424.
Adrian Raftery's research was supported by the Eunice Kennedy Shriver National 
Institute of Child Health and Development through NIH grants nos.~R01
HD054511 and R01 HD070936, by Science Foundation Ireland grant 11/W.1/I2079 and by National Institutes of Health grant U54-HL127624.

\newpage

\appendix
\section{Appendix: proofs}
\subsection{Theorem \ref{thm1}}\label{proof1}
\paragraph{D1.} This is straightforward since $\forall \textbf{z}_s\in \mathbb{R}^d:$
\begin{equation}
 \theta\left( \textbf{z}_s \right) = Pr\left( y_{sj}=1\middle\vert \textbf{z}_s \right) = \int_{\mathcal{Z}} p\left( \textbf{z}_j \right)r\left( \textbf{z}_s,\textbf{z}_j \right)d\textbf{z}_j.
\end{equation}

\paragraph{D2.}
\begin{equation}
\begin{split}
 G\left( x \right) &= \sum_{k=0}^{n-1}x^kp_k =\sum_{k=0}^{n-1}x^kPr\left( D_s=k \right)\\
 &=\sum_{k=0}^{n-1}x^k\int_{\mathcal{Z}}\cdots\int_{\mathcal{Z}}p\left( \textbf{z}_1\right)\cdots p\left( \textbf{z}_n \right)Pr\left( D_s=k\middle\vert P \right)d\textbf{z}_1\cdots d\textbf{z}_n\\
 &=\int_{\mathcal{Z}}\cdots\int_{\mathcal{Z}}\left[\prod_{j=1}^{n}p\left( \textbf{z}_j\right)\right]\mathbb{E}\left[x^{D_s}\middle\vert P\right]d\textbf{z}_1\cdots d\textbf{z}_n\\
 &=\int_{\mathcal{Z}}\cdots\int_{\mathcal{Z}}\left[\prod_{j=1}^{n}p\left( \textbf{z}_j\right)\right]\left\{\prod_{j=1}^{n}\mathbb{E}\left[x^{Y_{sj}}\middle\vert P\right]\right\}d\textbf{z}_1\cdots d\textbf{z}_n\\
 &=\int_{\mathcal{Z}}\cdots\int_{\mathcal{Z}}\left\{\prod_{j=1}^{n}p\left( \textbf{z}_j\right)\left[xr\left( \textbf{z}_s,\textbf{z}_j \right)+1-r\left( \textbf{z}_s,\textbf{z}_j \right)\right]\right\}d\textbf{z}_1\cdots d\textbf{z}_n\\
 &=\int_{\mathcal{Z}}p\left( \textbf{z}_s \right)\left\{\int_{\mathcal{Z}}p\left( \textbf{z}_j\right)\left[xr\left( \textbf{z}_s,\textbf{z}_j \right)+1-r\left( \textbf{z}_s,\textbf{z}_j \right)\right]d\textbf{z}_j\right\}^{n-1}d\textbf{z}_s\\
 &=\int_{\mathcal{Z}}p\left( \textbf{z}_s \right)\left\{x\int_{\mathcal{Z}}p\left( \textbf{z}_j\right)r\left( \textbf{z}_s,\textbf{z}_j \right)d\textbf{z}_j+1-\int_{\mathcal{Z}}p\left( \textbf{z}_j\right)r\left( \textbf{z}_s,\textbf{z}_j \right)d\textbf{z}_j\right\}^{n-1}d\textbf{z}_s\\
 &=\int_{\mathcal{Z}}p\left(\textbf{z}_s\right)\left[x\theta(\textbf{z}_s) + 1 - \theta(\textbf{z}_s)\right]^{n-1}d\textbf{z}_s.
\end{split}
\end{equation}

\paragraph{D3.} The $r$-th factorial moment of $D_s$ corresponds to the $r$-th derivative of $G$ evaluated in $1$:
\begin{equation}\label{GderProof}
\begin{split}
 \frac{\partial{}^rG}{\partial{x}^r}(x) &= \int_{\mathcal{Z}}p\left(\textbf{z}_s\right)\frac{\partial{}^r}{\partial{x}^r}\left[x\theta(\textbf{z}_s) + 1 - \theta(\textbf{z}_s)\right]^{n-1}d\textbf{z}_s\\
 &= \int_{\mathcal{Z}}p\left(\textbf{z}_s\right)\left( n-1 \right)\cdots\left( n-r \right)\theta\left( \textbf{z}_s \right)^r\left[x\theta(\textbf{z}_s) + 1 - \theta(\textbf{z}_s)\right]^{n-r-1}d\textbf{z}_s\\
 &= \frac{(n-1)!}{(n-r-1)!}\int_{\mathcal{Z}}p\left(\textbf{z}_s\right)\theta\left( \textbf{z}_s \right)^r\left[x\theta(\textbf{z}_s) + 1 - \theta(\textbf{z}_s)\right]^{n-r-1}d\textbf{z}_s;\\
\end{split}
\end{equation}
and the final formula evaluated in $x=1$ gives \eqref{Gder}.
\paragraph{D4.} The average degree is the first factorial moment, thus:
\begin{equation}
 \bar{k} = G'(1) = \frac{\left( n-1 \right)!}{\left( n-2 \right)!}\int_{\mathcal{Z}}p\left(\textbf{z}_s\right)\theta\left( \textbf{z}_s \right)d\textbf{z}_s = \left( n-1 \right)\int_{\mathcal{Z}}p\left(\textbf{z}_s\right)\theta\left( \textbf{z}_s \right)d\textbf{z}_s.
\end{equation}

\paragraph{D5.} The distribution of the degree of a random node can be recovered by differentiating $G$ as well. Indeed, using \eqref{GderProof}, for every $k$:
\begin{equation}
 p_k = \frac{1}{k!} \frac{\partial{}^rG}{\partial{x}^r}(0) = \binom{n-1}{k}\int_{\mathcal{Z}}p\left(\textbf{z}_s\right)\theta\left( \textbf{z}_s \right)\left[ 1-\theta\left( \textbf{z}_s \right)\right]^{n-k-1}d\textbf{z}_s.
\end{equation}

\paragraph{D6.} Define the PGF for the degree of a random node once its latent information is fixed to $\textbf{z}_s$:
\begin{equation}\label{Kzs}
 \begin{split}
  \tilde{G}\left( x;\textbf{z}_s \right) &= \sum_{k=0}^{n-1}x^kPr\left( D_s=k\middle\vert \textbf{z}_s \right)\\
  &=\int_{\mathcal{Z}}\cdots\int_{\mathcal{Z}}\left[\prod_{\substack{j=1 \\ j\neq s}}^{n}p\left( \textbf{z}_j\right)\right]\mathbb{E}\left[x^{D_s}\middle\vert P\right]d\textbf{z}_{-s}\\
  &=\left\{\int_{\mathcal{Z}}p\left( \textbf{z}_j\right)\left[xr\left( \textbf{z}_s,\textbf{z}_j \right)+1-r\left( \textbf{z}_s,\textbf{z}_j \right)\right]d\textbf{z}_j\right\}^{n-1}\\
  &=\left\{x\theta\left( \textbf{z}_s \right)+1-\theta\left( \textbf{z}_s \right)\right\}^{n-1};\\
 \end{split}
\end{equation}
which is simply the PGF of a binomial random variable with parameters $n-1$ and $\theta\left( \textbf{z}_s \right)$. 
Hence its average degree is $\bar{k}\left( \textbf{z}_s \right) = \left( n-1 \right)\theta\left( \textbf{z}_s \right)$.
Note that $d\textbf{z}_{-s}=\prod_{j\neq s}d\textbf{z}_j$.

\paragraph{D7.} We now write down the PGF for the degree of a random neighbour of a node located in $\textbf{z}_s$.
\begin{equation}
 \begin{split}
  H\left( x;\textbf{z}_s \right) &= \sum_{k=0}^{n-1}x^kPr\left( D_j=k\middle\vert y_{sj}=1,\textbf{z}_s \right)\\
  &=\int_{\mathcal{Z}}p\left( \textbf{z}_j \middle\vert y_{sj}=1,\textbf{z}_s\right)\sum_{k=0}^{n-1}x^kPr\left( D_j=k\middle\vert y_{sj}=1,\textbf{z}_s,\textbf{z}_j \right)d\textbf{z}_j\\
  &=\int_{\mathcal{Z}}p\left( \textbf{z}_j \middle\vert y_{sj}=1,\textbf{z}_s\right)\mathbb{E}\left[x^{D_j}\middle\vert y_{sj}=1,\textbf{z}_s,\textbf{z}_j\right]d\textbf{z}_j.\\
 \end{split}
\end{equation}
Note that $\mathbb{E}\left[x^{D_j}\middle\vert y_{sj}=1,\textbf{z}_s,\textbf{z}_j\right]$ corresponds to the PGF for the so called excess degree \parencite{newman2001random},
i.e. the degree of a node at one extreme of an edge picked at random. Hence, such PGF is equal to $\frac{x\tilde{G}'(x;\textbf{z}_j)}{\tilde{G}\left( 1;\textbf{z}_j \right)}$,
where $\tilde{G}$ has been defined in \eqref{Kzs}. Then:
\begin{equation}
 \begin{split}
  H\left( x;\textbf{z}_s \right) &=\int_{\mathcal{Z}}p\left( \textbf{z}_j \middle\vert y_{sj}=1,\textbf{z}_s\right)\frac{x\tilde{G}'(x;\textbf{z}_j)}{\tilde{G}\left( 1;\textbf{z}_j \right)}d\textbf{z}_j\\
  &=\int_{\mathcal{Z}}\frac{Pr\left( y_{sj}=1 \middle\vert \textbf{z}_j,\textbf{z}_s\right)p\left( \textbf{z}_j\right)}{Pr\left( y_{sj}=1 \middle\vert \textbf{z}_s\right)}\left\{x\left[x\theta\left( \textbf{z}_j+1-\theta\left( \textbf{z}_j \right) \right)\right]^{n-2}\right\}d\textbf{z}_j\\
  &=\frac{1}{\theta\left( \textbf{z}_s \right)}\int_{\mathcal{Z}}p\left( \textbf{z}_j\right)r\left( \textbf{z}_j,\textbf{z}_s\right)\left\{x\left[x\theta\left( \textbf{z}_j+1-\theta\left( \textbf{z}_j \right) \right)\right]^{n-2}\right\}d\textbf{z}_j.\\
 \end{split}
\end{equation}
Its average degree is then given by:
\begin{equation}
\begin{split}
 \bar{k}_{nn}\left( \textbf{z}_s \right) &= H'\left( 1;\textbf{z}_s \right) =\frac{1}{\theta\left( \textbf{z}_s \right)}\int_{\mathcal{Z}}p\left( \textbf{z}_j\right)r\left( \textbf{z}_j,\textbf{z}_s\right)\left\{1+\left( n-2 \right)\theta\left( \textbf{z}_j \right)\right\}d\textbf{z}_j\\
 &=1+\frac{\left( n-2 \right)}{\theta\left( \textbf{z}_s \right)}\int_{\mathcal{Z}}p\left( \textbf{z}_j\right)r\left( \textbf{z}_j,\textbf{z}_s\right)\theta\left( \textbf{z}_j \right)d\textbf{z}_j.
\end{split}
\end{equation}

\paragraph{D8.} The PGF for the degree of a neighbour of a node with degree $k$ is given by:
\begin{equation}
 \begin{split}
  \tilde{H}\left( x;k \right) &= \sum_{r=0}^{n-1} x^rPr\left( D_j=r\middle\vert D_s=k,y_{sj}=1 \right)\\
  &=\sum_{r=0}^{n-1} x^r\int_\mathcal{Z}p\left( \textbf{z}_s\middle\vert D_s=k \right)Pr\left( D_j=r\middle\vert \textbf{z}_s,y_{sj}=1 \right)d\textbf{z}_s\\
  &=\frac{1}{p_k}\int_\mathcal{Z}p\left( \textbf{z}_s \right)Pr\left( D_s=k\middle\vert \textbf{z}_s \right)H\left( x;\textbf{z}_s \right)d\textbf{z}_s\\
  &=\frac{1}{p_k}\int_\mathcal{Z}p\left( \textbf{z}_s \right)\left[\frac{\partial{}^k}{\partial{x}^k}\tilde{G}\left( 0;\textbf{z}_s \right)\right]H\left( x;\textbf{z}_s \right)d\textbf{z}_s\\
  &=\frac{1}{p_k}\int_\mathcal{Z}p\left( \textbf{z}_s \right)\binom{n-1}{k}\theta(\textbf{z}_s)^k\left[1-\theta(\textbf{z}_s)\right]^{n-k-1}H\left( x;\textbf{z}_s \right)d\textbf{z}_s;
 \end{split}
\end{equation}
and its first derivative evaluated in $x=1$ yields:
\begin{equation}
 \bar{k}_{nn}(k) = \frac{1}{p_k}\int_{\mathcal{Z}} p(\textbf{z}_s)\binom{n-1}{k}\theta(\textbf{z}_s)^k\left[1-\theta(\textbf{z}_s)\right]^{n-k-1} \bar{k}_{nn}(\textbf{z}_s)d\textbf{z}_s.
\end{equation}

\subsubsection{Proof for Corollary \ref{cor1}}\label{proof:cor1}
Recall that a convolution of two Gaussian densities is still a Gaussian density:
\begin{equation}
 \int_{\mathbb{R}^d} f_d\left( \textbf{z}_i; \boldsymbol{\mu}_1, \gamma_1 \right)f_d\left( \textbf{z}_j-\textbf{z}_i; \boldsymbol{\mu}_2, \gamma_2 \right)d\textbf{z}_i = f_d\left( \textbf{z}_j; \boldsymbol{\mu}_1+\boldsymbol{\mu}_2, \gamma_1+\gamma_2 \right),
\end{equation}
for every $\textbf{z}_i, \textbf{z}_j, \boldsymbol{\mu}_1, \boldsymbol{\mu}_2$ in $\mathbb{R}^d$ and every positive real numbers $\gamma_1$ and $\gamma_2$.

That being said:
\paragraph{D1.}
\begin{equation}
 \begin{split}
\theta(\textbf{z}_s) &= \int_{\mathbb{R}^d}f_d\left( \textbf{z}_j;\textbf{0},\gamma \right)\tau\left( 2\pi\varphi \right)^{\frac{d}{2}}f_d\left( \textbf{z}_s-\textbf{z}_j;\textbf{0},\varphi \right)d\textbf{z}_j\\
&= \tau\left( 2\pi\varphi \right)^{\frac{d}{2}}f_d\left( \textbf{z}_s;\textbf{0},\gamma+\varphi \right)\\
&=\tau\left(\frac{\varphi}{\gamma+\varphi}\right)^{\frac{d}{2}}\exp\left\{-\frac{1}{2(\gamma+\varphi)}\textbf{z}_s^t\textbf{z}_s\right\}.
\end{split}
\end{equation}
\paragraph{D3.}
\begin{equation}
 \begin{split}
\frac{\partial{}^rG}{\partial{x}^r}(1) &= \frac{(n-1)!}{(n-r-1)!}\int_{\mathbb{R}^d}f_d\left( \textbf{z}_s;\textbf{0},\gamma \right)\theta(\textbf{z}_s)^rd\textbf{z}_s\\
&= \frac{(n-1)!}{(n-r-1)!}\tau^r\left( \frac{\varphi}{\gamma+\varphi} \right)^{\frac{rd}{2}}\int_{\mathbb{R}^d}f_d\left( \textbf{z}_s;\textbf{0},\gamma \right)\exp\left\{- \frac{r}{2\left( \gamma+\varphi \right)}\textbf{z}_s^t\textbf{z}_s\right\}d\textbf{z}_s\\
&= \frac{(n-1)!}{(n-r-1)!}\tau^r\left( \frac{\varphi}{\gamma+\varphi} \right)^{\frac{rd}{2}}\left\{2\pi\frac{\left( \gamma+\varphi \right)}{r}\right\}^{\frac{d}{2}} \times\\
&\hspace{2cm}\times\int_{\mathbb{R}^d}f_d\left( \textbf{z}_s;\textbf{0},\gamma \right)f_d\left( \textbf{z}_s;\textbf{0},\frac{\gamma+\varphi}{r} \right)d\textbf{z}_s\\
&= \frac{(n-1)!}{(n-r-1)!}\tau^r\left( \frac{\varphi}{\gamma+\varphi} \right)^{\frac{rd}{2}}\left\{2\pi\frac{\left( \gamma+\varphi \right)}{r}\right\}^{\frac{d}{2}}\left\{2\pi\frac{\left[\left( r+1 \right)\gamma+\varphi\right]}{r}\right\}^{-\frac{d}{2}}\\
&=\frac{(n-1)!}{(n-r-1)!}\tau^r\left\{\frac{\varphi^r}{\left(\gamma+\varphi\right)^{r-1}\left[(r+1)\gamma+\varphi\right]}\right\}^\frac{d}{2}\\
\end{split}
\end{equation}
\paragraph{D4.}
\begin{equation}
 \begin{split}
\bar{k} = G'(1) = (n-1)\tau\left\{\frac{\varphi}{2\gamma+\varphi}\right\}^\frac{d}{2}\\
\end{split}
\end{equation}
\paragraph{D7.}
\begin{equation}
 \begin{split}
\bar{k}_{nn}(\textbf{z}_s) &= 1+\frac{(n-2)}{\theta\left( \textbf{z}_s \right)} \int_{\mathbb{R}^d} p\left( \textbf{z}_j \right)r\left( \textbf{z}_s,\textbf{z}_j \right)\theta\left( \textbf{z}_j \right)d\textbf{z}_j\\
&= 1+\frac{(n-2)}{\theta\left( \textbf{z}_s \right)}\tau^2\left( 2\pi\varphi \right)^d \times\\
&\hspace{2cm}\times\int_{\mathbb{R}^d} f_d\left( \textbf{z}_j;\textbf{0},\gamma \right)f_d\left( \textbf{z}_j;\textbf{0},\gamma+\varphi \right)f_d\left( \textbf{z}_s-\textbf{z}_j;\textbf{0},\varphi \right)d\textbf{z}_j\\
&= 1+\frac{(n-2)}{\theta\left( \textbf{z}_s \right)}\tau^2\left( 2\pi\varphi \right)^d \left\{2\pi\left( 2\gamma+\varphi \right)\right\}^{-\frac{d}{2}}\times\\
&\hspace{2cm}\times\int_{\mathbb{R}^d} f_d\left( \textbf{z}_j;\textbf{0},\frac{\gamma\left( \gamma+\varphi \right)}{2\gamma+\varphi} \right)f_d\left( \textbf{z}_s-\textbf{z}_j;\textbf{0},\varphi \right)d\textbf{z}_j\\
&= 1+(n-2)\tau\left( \frac{\varphi}{2\gamma+\varphi} \right)^{\frac{d}{2}} \frac{f_d\left( \textbf{z}_s;\textbf{0},\varphi+\frac{\gamma\left( \gamma+\varphi \right)}{2\gamma+\varphi}\right)}{f_d\left( \textbf{z}_s;\textbf{0},\gamma+\varphi \right)}\\
&=1+\bar{k}\left(\frac{n-2}{n-1}\right)\frac{f_d\left( \textbf{z}_s;\textbf{0},\frac{\gamma^2+3\gamma\varphi+\varphi^2}{2\gamma+\varphi} \right)}{f_d\left( \textbf{z}_s;\textbf{0},\gamma+\varphi \right)}.\\
\end{split}
\end{equation}

\subsubsection{Proof for Corollary \ref{cor2}}\label{proof:cor2}
\paragraph{D1.}
\begin{equation}
 \begin{split}
\theta(\textbf{z}_s) &= \int_{\mathbb{R}^d}\sum_{g=1}^{G}\pi_gf_d\left( \textbf{z}_j;\boldsymbol{\mu}_g,\gamma_g \right)\tau\left( 2\pi\varphi \right)^{\frac{d}{2}}f_d\left( \textbf{z}_s-\textbf{z}_j;\textbf{0},\varphi \right)d\textbf{z}_j\\
&= \tau\left( 2\pi\varphi \right)^{\frac{d}{2}}\sum_{g=1}^{G}\pi_g\int_{\mathbb{R}^d}f_d\left( \textbf{z}_j;\boldsymbol{\mu}_g,\gamma_g \right)f_d\left( \textbf{z}_s-\textbf{z}_j;\textbf{0},\varphi \right)d\textbf{z}_j\\
&= \tau\left( 2\pi\varphi \right)^{\frac{d}{2}}\sum_{g=1}^{G}\pi_gf_d\left( \textbf{z}_s;\boldsymbol{\mu}_g,\gamma_g+\varphi \right).\\
\end{split}
\end{equation}
\paragraph{D4.}
\begin{equation}
 \begin{split}
\bar{k} &= (n-1)\int_{\mathbb{R}^d}\sum_{g=1}^{G}\pi_gf_d\left( \textbf{z}_s;\boldsymbol{\mu}_g,\gamma_g \right)\tau\left( 2\pi\varphi \right)^{\frac{d}{2}}\sum_{h=1}^{G}\pi_hf_d\left( \textbf{z}_s;\boldsymbol{\mu}_h,\gamma_h+\varphi \right)d\textbf{z}_s\\
&=(n-1)\tau\left( 2\pi\varphi \right)^{\frac{d}{2}}\sum_{g=1}^{G}\sum_{h=1}^{G}\pi_g\pi_h\int_{\mathbb{R}^d}f_d\left( \textbf{z}_s;\boldsymbol{\mu}_g,\gamma_g \right)f_d\left( \textbf{z}_s;\boldsymbol{\mu}_h,\gamma_h+\varphi \right)d\textbf{z}_s\\
&=(n-1)\tau\left( 2\pi\varphi \right)^{\frac{d}{2}}\sum_{g=1}^{G}\sum_{h=1}^{G}\pi_g\pi_hf_d\left( \boldsymbol{\mu}_g-\boldsymbol{\mu}_h;\textbf{0},\gamma_g+\gamma_h+\varphi \right).
\end{split}
\end{equation}
While \textbf{D7} is straightforward from \eqref{degree1}.

\subsection{Proof of Proposition \ref{thm3}}\label{thm3proof}
First, we recall a few properties of the Gaussian distribution through a Lemma:
\begin{lemma}\label{lemma1}
Let $f_d\left( \cdot; \boldsymbol{\mu},\gamma \right)$ denote the $d$-dimensional Gaussian density centred in $\boldsymbol{\mu}$, with covariance matrix $\gamma\textbf{I}_d$. 
Let also $\textbf{x},\textbf{u},\textbf{v}\in\mathbb{R}^d$ and $a,b,\alpha\in\mathbb{R}^+$. Then:
\begin{align}
 f_d\left( \textbf{x};\textbf{u},a \right)f_d\left( \textbf{x};\textbf{v},b \right) &= f_d\left( \textbf{u}-\textbf{v};\textbf{0},a+b \right)f_d\left( \textbf{x};\frac{b\textbf{u}+a\textbf{v}}{a+b},\frac{ab}{a+b} \right)\label{lemma1eq1};\\
 f_d\left( \alpha\textbf{x};\textbf{u},a \right) &= \alpha^{-d}f_d\left( \textbf{x}; \frac{\textbf{u}}{\alpha}, \frac{a}{\alpha^2} \right).
\end{align}
\end{lemma}

Here follows the proof of Proposition \ref{thm3} by mathematical induction on $k$.
If $k=1$, then:
\begin{equation}
 I_{1}(\textbf{z}_i,\textbf{z}_j) = h_1f_d\left( \textbf{z}_j-\alpha_1\textbf{z}_i;\textbf{0},\omega_1 \right) = \tau\left( 2\pi\varphi \right)^{\frac{d}{2}}f_d\left( \textbf{z}_j-\textbf{z}_i;\textbf{0},\varphi \right) = r\left( \textbf{z}_i,\textbf{z}_j \right).
\end{equation}
Now assume that $I_{k}(\textbf{z}_i,\textbf{z}_j) = h_kf_d\left( \textbf{z}_j-\alpha_k\textbf{z}_i;\textbf{0},\omega_k \right)$, then we need to prove that 
$$I_{k+1}(\textbf{z}_i,\textbf{z}_j) = h_{k+1}f_d\left( \textbf{z}_j-\alpha_{k+1}\textbf{z}_i;\textbf{0},\omega_{k+1} \right),$$ where
$h_{k+1},\alpha_{k+1},\omega_{k+1}$ are defined recursively by \eqref{recurrence}.
\begin{equation}
 \begin{split}
  I_{k+1}(\textbf{z}_i,\textbf{z}_j) &= \int_{\mathcal{Z}}\cdots\int_{\mathcal{Z}} p(\textbf{z}_1)\dots p(\textbf{z}_{k})r\left( \textbf{z}_i,\textbf{z}_1 \right)\cdots r\left( \textbf{z}_{k},\textbf{z}_j \right)d\textbf{z}_1\cdots d\textbf{z}_{k}\\
  &= \int_{\mathcal{Z}}p\left( \textbf{z}_k \right)r\left( \textbf{z}_k,\textbf{z}_j\right)\int_{\mathcal{Z}}\cdots\int_{\mathcal{Z}} p(\textbf{z}_1)\dots p(\textbf{z}_{k-1})\times\\
  &\hspace{0.5cm}\times r\left( \textbf{z}_i,\textbf{z}_1 \right)\cdots r\left( \textbf{z}_{k-1},\textbf{z}_k \right)d\textbf{z}_1\cdots d\textbf{z}_{k}\\
  &= \int_{\mathcal{Z}}p\left( \textbf{z}_k \right)r\left( \textbf{z}_k,\textbf{z}_j\right)I_k\left( \textbf{z}_i,\textbf{z}_k \right)d\textbf{z}_{k}\\
  &= \int_{\mathcal{Z}}p\left( \textbf{x} \right)r\left( \textbf{x},\textbf{z}_j\right)I_k\left( \textbf{z}_i,\textbf{x} \right)d\textbf{x}.
 \end{split}
\end{equation}
Now, we introduce the Gaussian LPM assumptions and use the results of the Lemma \ref{lemma1}:
\begin{equation}
 \begin{split}
  I_{k+1}(\textbf{z}_i,\textbf{z}_j) &= \tau\left( 2\pi\varphi \right)^{\frac{d}{2}}h_k\int_{\mathbb{R}^d}f_d\left( \textbf{x};\textbf{0},\gamma \right)f_d\left( \textbf{x}-\textbf{z}_j;\textbf{0},\varphi \right)f_d\left( \textbf{x}-\alpha_k\textbf{z}_i;\textbf{0},\omega_k \right)d\textbf{x}\\
  &= \tau\left( 2\pi\varphi \right)^{\frac{d}{2}}h_k\times\\
  &\hspace{0.5cm}\times\int_{\mathbb{R}^d}f_d\left( \textbf{x}-\textbf{z}_j;\textbf{0},\varphi \right)f_d\left( -\alpha_k\textbf{z}_i;\textbf{0},\omega_k+\gamma \right)f_d\left( \textbf{x};\frac{\gamma\alpha_k\textbf{z}_i}{\omega_k+\gamma},\frac{\omega_k\gamma}{\omega_k+\gamma} \right)d\textbf{x}\\  
  &= \tau\left( 2\pi\varphi \right)^{\frac{d}{2}}h_k\alpha^{-d}f_d\left( \textbf{z}_i;\textbf{0},\frac{\omega_k+\gamma}{\alpha_k^2} \right)\times\\
  &\hspace{0.5cm}\times\int_{\mathbb{R}^d}f_d\left( \textbf{x}-\textbf{z}_j;\textbf{0},\varphi \right)f_d\left( \textbf{x};\frac{\gamma\alpha_k\textbf{z}_i}{\omega_k+\gamma},\frac{\omega_k\gamma}{\omega_k+\gamma} \right)d\textbf{x}\\  
  &= h_{k+1}f_d\left( \textbf{z}_j; \frac{\gamma\alpha_k\textbf{z}_i}{\omega_k+\gamma},\frac{\omega_k\gamma+\omega_k\varphi+\varphi\gamma}{\omega_k+\gamma} \right)\\
  &= h_{k+1}f_d\left( \textbf{z}_j-\alpha_{k+1}\textbf{z}_i;\textbf{0}.\omega_{k+1} \right).
 \end{split}
\end{equation}

\subsection{Proof of Corollary \ref{cor:dispersion}}\label{proof:dispersion}
Let $G$ be the PGF of the random variable $D$, denoting the degree of a node picked at random. Then the $r$-th derivative of $G$ evaluated in $1$ is equal to the $r$-th factorial moment of $D$, denoted here $c_r$:
\begin{equation}
 c_r = \frac{\partial{}^rG}{\partial{x}^r}(1) = \mathbb{E}\left[D\left( D-1 \right)\cdots\left( D-r+1 \right)\right].
\end{equation}
In particular:
\begin{align}
c_1 &= \mathbb{E}\left[D\right] = m_1\\
c_2 &= \mathbb{E}\left[D\left( D-1 \right)\right] = \mathbb{E}\left[D^2\right] - \mathbb{E}\left[D\right] = m_2 - m_1\\
&\hspace{1cm}\Longrightarrow m_2 = c_1+c_2,
\end{align}
where $m_1$ and $m_2$ denote the first two non-central moments of $D$.
That being said, using Corollary \ref{cor1} the dispersion index can be evaluated exactly:
\begin{equation}
 \begin{split}
  \mathcal{D} &= \frac{\mathbb{E}\left[ \left( D-m_1 \right)^2\right]}{m_1}= \frac{m_2-m_1^2}{m_1}= \frac{m_2}{m_1} - m_1= 1+\frac{c_2}{c_1} - c_1\\
  &= 1+\frac{\left( n-1 \right)\left( n-2 \right)\tau^2\left\{ \frac{\varphi^2}{\left( \gamma+\varphi \right)\left( 3\gamma+\varphi \right)}\right\}^{\frac{d}{2}}}{\left( n-1 \right)\tau\left\{ \frac{\varphi}{2\gamma+\varphi }\right\}^{\frac{d}{2}}} - \left( n-1 \right)\tau\left\{ \frac{\varphi}{ 2\gamma+\varphi}\right\}^{\frac{d}{2}}\\
  &= 1+\left( n-2 \right)\tau\left\{ \frac{\varphi\left( 2\gamma+\varphi \right)}{\left( \gamma+\varphi \right)\left( 3\gamma+\varphi \right)}\right\}^{\frac{d}{2}}- \left( n-1 \right)\tau\left\{ \frac{\varphi}{ 2\gamma+\varphi}\right\}^{\frac{d}{2}},\\
 \end{split}
\end{equation}
which proves the corollary. Also, when $d=2$, the threshold between underdispersion and overdispersion is given by:
\begin{equation}
 \frac{\left( n-2 \right)\left( 2\gamma+\varphi \right)}{\left( \gamma+\varphi \right)\left( 3\gamma+\varphi \right)} - \frac{\left( n-1 \right)}{\left( 2\gamma+\varphi \right)} = 0.
\end{equation}
Now, recalling that $\varphi>0$ and $\gamma>0$, this is equivalent to:
\begin{equation}
\begin{split}
 &\left( n-2 \right)\left( 2\gamma+\varphi \right)^2 - \left( n-1 \right)\left( \gamma+\varphi \right)\left( 3\gamma+\varphi \right)=0\\
 \Rightarrow&\varphi^2+4\gamma\varphi+5\gamma^2-n\gamma^2=0\\
 \Rightarrow&\varphi=\gamma\left( -2\pm \sqrt{n-1}\right).
\end{split}
\end{equation}
One solution is negative thus not feasible, then the threshold is given by:
$$\varphi=\gamma\left( \sqrt{n-1}-2 \right).$$

\subsection{Proof of Proposition \ref{thm2}}\label{proof:clustering}
Formula in \eqref{clustering1} is straightforward since it is obtained by conditioning on the latent information. We now show how to obtain the exact formula \eqref{trans} under the Gaussian LPM.
We solve the numerator $\mathcal{C}_N$ and the denominator $\mathcal{C}_D$ independently.
\begin{equation}
 \begin{split}
  \mathcal{C}_D &= \int_{\mathbb{R}^d}\int_{\mathbb{R}^d}\int_{\mathbb{R}^d} p(\textbf{z}_i)p(\textbf{z}_k)p(\textbf{z}_j)r\left( \textbf{z}_i,\textbf{z}_k\right)r\left(\textbf{z}_k,\textbf{z}_j \right)d\textbf{z}_kd\textbf{z}_id\textbf{z}_j\\
  &= \int_{\mathbb{R}^d} p(\textbf{z}_k)\left\{\int_{\mathbb{R}^d} p(\textbf{z}_i)r\left( \textbf{z}_i,\textbf{z}_k \right)d\textbf{z}_i\right\} \left\{\int_{\mathbb{R}^d} p(\textbf{z}_j)r\left( \textbf{z}_k,\textbf{z}_j \right)d\textbf{z}_j\right\}d\textbf{z}_k\\
  &= \int_{\mathbb{R}^d} p(\textbf{z}_k)\theta\left( \textbf{z}_k \right)^2d\textbf{z}_k\\
  &= \frac{G''(1)}{\left( n-1 \right)\left( n-2 \right)}\\
  &= \tau^2\left\{\frac{\varphi^2}{\left( \gamma+\varphi \right)\left( 3\gamma+\varphi \right)}\right\}^{\frac{d}{2}}
 \end{split}
\end{equation}
Now we solve the numerator. 
\begin{equation}
 \begin{split}
  \mathcal{C}_N &= \int_{\mathbb{R}^d}\int_{\mathbb{R}^d}\int_{\mathbb{R}^d} p(\textbf{z}_i)p(\textbf{z}_k)p(\textbf{z}_j)r\left( \textbf{z}_i,\textbf{z}_k \right)r\left( \textbf{z}_k,\textbf{z}_j \right)r\left( \textbf{z}_j,\textbf{z}_i \right)d\textbf{z}_id\textbf{z}_kd\textbf{z}_j\\
  &= \int_{\mathbb{R}^d}p(\textbf{z}_i)\int_{\mathbb{R}^d}p(\textbf{z}_j)r\left( \textbf{z}_j,\textbf{z}_i \right)\left\{\int_{\mathbb{R}^d} p(\textbf{z}_k)r\left( \textbf{z}_i,\textbf{z}_k \right)r\left( \textbf{z}_k,\textbf{z}_j \right)d\textbf{z}_k\right\}d\textbf{z}_jd\textbf{z}_i\\
  &= \int_{\mathbb{R}^d}p(\textbf{z}_i)\int_{\mathbb{R}^d}p(\textbf{z}_j)r\left( \textbf{z}_j,\textbf{z}_i \right) I_2\left( \textbf{z}_i,\textbf{z}_j \right)d\textbf{z}_jd\textbf{z}_i\\
  &= \int_{\mathbb{R}^d}p(\textbf{z}_i)I_{3}(\textbf{z}_i,\textbf{z}_i)d\textbf{z}_i\\
 \end{split}
\end{equation}
where $I_{k}(\textbf{z}_i,\textbf{z}_j)$ is defined in \ref{WalkIntegral1} for every $k\in \mathbb{N}^0$, $\textbf{z}_i\in\mathbb{R}^d$ and $\textbf{z}_j\in\mathbb{R}^d$.

For more clarity, we define the recurring quantity 
\begin{equation}
 \lambda=\varphi^2+3\gamma\varphi+\gamma^2.
\end{equation}
We first discover the quantities needed to write $I_{3}(\textbf{z}_i,\textbf{z}_i)$ explicitly:
\begin{equation}
 \begin{cases}
 \alpha_1&=1\\
 \omega_1&=\varphi\\
 h_1&=\tau\left( 2\pi\varphi \right)^{\frac{d}{2}}\\
\end{cases};\hspace{2cm}
\begin{cases}
 \alpha_2&=\frac{\gamma}{\gamma+\varphi}\\
 \omega_2&=\frac{\varphi\left( 2\gamma+\varphi \right)}{\gamma+\varphi}\\
 h_2&=\tau^2\left( 2\pi\varphi \right)^df_d\left( \textbf{z}_i;\textbf{0},\gamma+\varphi \right)\\
\end{cases};
\end{equation}
\begin{align}
 \alpha_3&=\frac{\alpha_2\gamma}{\omega_2+\gamma} = \frac{\gamma^2}{\lambda};\\
 \omega_3&=\frac{\omega_2\varphi + \omega_2\gamma + \gamma\varphi}{\omega_2+\gamma} = \frac{\varphi\left( \gamma+\varphi \right)\left( 3\gamma+\varphi \right)}{\lambda};\\
 h_3&=\tau^3\left( 2\pi\varphi \right)^{\frac{3}{2}d}f_d\left( \textbf{z}_i;\textbf{0},\gamma+\varphi \right)\left( \frac{\gamma+\varphi}{\gamma} \right)^df_d\left( \textbf{z}_i;\textbf{0},\frac{\lambda\left( \gamma+\varphi \right)}{\gamma^2} \right).
\end{align} 
Now, for $h_3$, we use Lemma \ref{lemma1} and join the two Gaussian densities:
\begin{equation}
\begin{split}
  h_3&=\tau^3\left( 2\pi\varphi \right)^{\frac{3}{2}d}\left( \frac{\gamma+\varphi}{\gamma} \right)^d \left\{2\pi\frac{\left( \gamma+\varphi \right)^2\left( 2\gamma+\varphi \right)}{\gamma^2}\right\}^{-\frac{d}{2}}f_d\left( \textbf{z}_i;\textbf{0},\frac{\lambda}{2\gamma+\varphi}\right)\\
  &= \tau^3\left( 2\pi\varphi \right)^{d}\left\{\frac{\varphi}{2\gamma+\varphi}\right\}^{\frac{d}{2}}f_d\left( \textbf{z}_i;\textbf{0},\frac{\lambda}{2\gamma+\varphi}\right).\\
\end{split}
\end{equation}
Also:
\begin{align}
 &\left( 1-\alpha_3 \right) = \frac{\varphi\left( 3\gamma+\varphi \right)}{\lambda}\\
 &\frac{\omega_3}{\left( 1-\alpha_3 \right)^2} = \frac{\lambda\left( \gamma+\varphi \right)}{\varphi\left( 3\gamma+\varphi \right)}\\
\end{align}
Then, it follows:
\begin{equation}
 \begin{split}
  I_{3}(\textbf{z}_i,\textbf{z}_i) &= h_3\left( 1-\alpha_3 \right)^{-d}f_d\left( \textbf{z}_i;\textbf{0},\frac{\omega_3}{\left( 1-\alpha_3 \right)^2}\right)\\
  &= \tau^3\left( 2\pi\varphi \right)^{d}\left\{\frac{\varphi}{2\gamma+\varphi}\right\}^{\frac{d}{2}}f_d\left( \textbf{z}_i;\textbf{0},\frac{\lambda}{2\gamma+\varphi}\right)\times\\
  &\hspace{1cm}\times\left\{\frac{\lambda}{\varphi\left( 3\gamma+\varphi \right)}\right\}^df_d\left( \textbf{z}_i;\textbf{0},\frac{\lambda\left( \gamma+\varphi \right)}{\varphi\left( 3\gamma+\varphi \right)}\right).
 \end{split}
\end{equation}
Collapsing again the Gaussian densities:
\begin{equation}
  I_{3}(\textbf{z}_i,\textbf{z}_i) =  \tau^3\left\{\frac{2\pi\varphi^2}{2\left( 3\gamma+\varphi \right)}\right\}^{\frac{d}{2}}f_d\left( \textbf{z}_i;\textbf{0},\frac{\gamma+\varphi}{2} \right)
\end{equation}
We can now obtain the final result for the numerator:
\begin{equation}
 \begin{split}
  \mathcal{C}_N &= \int_{\mathbb{R}^d}p(\textbf{z}_i)I_{3}(\textbf{z}_i,\textbf{z}_i)d\textbf{z}_i\\
  &= \tau^3\left\{\frac{2\pi\varphi^2}{2\left( 3\gamma+\varphi \right)}\right\}^{\frac{d}{2}}\int_{\mathbb{R}^d}f_d\left( \textbf{z}_i;\textbf{0},\gamma \right)f_d\left( \textbf{z}_i;\textbf{0},\frac{\gamma+\varphi}{2} \right)d\textbf{z}_i\\
  &= \tau^3\left\{\frac{\varphi^2}{\left( 3\gamma+\varphi \right)^2}\right\}^{\frac{d}{2}}
 \end{split}
\end{equation}
The final formula for the clustering coefficient follows:
\begin{equation}
 \mathcal{C} = \frac{\mathcal{C}_N}{\mathcal{C}_D} = \frac{\tau^3\left\{\frac{\varphi^2}{\left( 3\gamma+\varphi \right)^2}\right\}^{\frac{d}{2}}}{\tau^2\left\{\frac{\varphi^2}{\left( \gamma+\varphi \right)\left( 3\gamma+\varphi \right)}\right\}^{\frac{d}{2}}}
 = \tau\left(\frac{\gamma+\varphi}{3\gamma+\varphi}\right)^{\frac{d}{2}}.
\end{equation}

\printbibliography

\end{document}